\documentclass[12pt]{article}
\usepackage{latexsym}
\usepackage{amsmath,amsfonts}
\usepackage{times}
\allowdisplaybreaks[4]

\catcode`\@=11

\newcount\hour
\newcount\minute
\newtoks\amorpm \hour=\time\divide\hour by 60\minute
=\time{\multiply\hour by 60 \global\advance\minute by-\hour}
\edef\standardtime{{\ifnum\hour<12 \global\amorpm={am}%
        \else\global\amorpm={pm}\advance\hour by-12 \fi
        \ifnum\hour=0 \hour=12 \fi
        \number\hour:\ifnum\minute<10
        0\fi\number\minute\the\amorpm}}
\edef\militarytime{\number\hour:\ifnum\minute<10
0\fi\number\minute}

\def\draftlabel#1{{\@bsphack\if@filesw {\let\thepage\relax
   \xdef\@gtempa{\write\@auxout{\string
      \newlabel{#1}{{\@currentlabel}{\thepage}}}}}\@gtempa
   \if@nobreak \ifvmode\nobreak\fi\fi\fi\@esphack}
        \gdef\@eqnlabel{#1}}
\def\@eqnlabel{}
\def\@vacuum{}
\def\marginnote#1{}
\def\draftmarginnote#1{\marginpar{\raggedright\scriptsize\tt#1}}
\def\draft{
        \pagestyle{plain}
        \overfullrule=2pt
        \oddsidemargin -.5truein
        \def\@oddhead{\sl \phantom{\today\quad\militarytime} \hfil
        \smash{\Large\sl DRAFT} \hfil \today\quad\militarytime}
        \let\@evenhead\@oddhead
        \let\label=\draftlabel
        \let\marginnote=\draftmarginnote
        \def\ps@empty{\let\@mkboth\@gobbletwo
        \def\@oddfoot{\hfil \smash{\Large\sl DRAFT} \hfil}
        \let\@evenfoot\@oddhead}
        \def\@eqnnum{(\theequation)\rlap{\kern\marginparsep\tt\@eqnlabel}%
        \global\let\@eqnlabel\@vacuum}  }

\newcommand{\rf}[1]{(\ref{#1})}
\renewcommand{\theequation}{\thesection.\arabic{equation}}
\renewcommand{\thefootnote}{\fnsymbol{footnote}}
\newcommand{\newsection}{    
\setcounter{equation}{0}\section}

\def\appendix#1{\addtocounter{section}{1}\setcounter{equation}{0}
\renewcommand{\thesection}{\Alph{section}}
\section*{Appendix \thesection\protect\indent \parbox[t]{11.15cm}{#1}}
\addcontentsline{toc}{section}{Appendix \thesection\ \ \ #1}}

\def\nline{\,\nabla\kern -0.7em\raise0.2ex\hbox{/}\,\,}
\def\yline{\,y\kern -0.47em /}
\def\aline{\,a\kern -0.49em /}
\def\parline{\,\partial\kern -0.55em /\,\,}

\newcommand{\Mo}{\mathbb{M}}
\newcommand{\No}{\mathbb{N}}
\newcommand{\Po}{\mathbb{P}}

\newcommand{\Zo}{\mathbb{Z}}

\def\be{\begin{equation}}
\def\ee{\end{equation}}
\def\beq{\begin{eqnarray}}
\def\eeq{\end{eqnarray}}

\def\Rsm{{\scriptscriptstyle R}}
\def\Lsm{{\scriptscriptstyle L}}

\def\smCSF{{\scriptscriptstyle CSF}}

\def\smone{{\scriptscriptstyle (1)}}

\def\smpt{{\scriptscriptstyle [2]}}
\def\smp3{{\scriptscriptstyle [3]}}

\def\smpn{{\scriptscriptstyle [n]}}

\def\Jbf{{\bf J}}
\def\Lbf{{\bf L}}
\def\Mbf{{\bf M}}
\def\Pbf{{\bf P}}

\def\Xbf{{\bf X}}

\def\ibf{{\bf i}}
\def\iibf{{\bf ii}}
\def\iiibf{{\bf iii}}
\def\ivbf{{\bf iv}}
\def\vbf{{\bf v}}
\def\vibf{{\bf vi}}
\def\viibf{{\bf vii}}

\def\half{\frac{1}{2}}

\def\alphab{\bar{\alpha}}
\def\upsilonb{\bar{\upsilon}}
\def\zetab{\bar{\zeta}}

\def\Vb{{\bar{V}}}

\def\gb{{\bar{g}}}

\def\phik{|\phi\rangle}

\def\irm{{\rm i}}

\def\dyn{{\rm dyn}}

\def\betach{\check{\beta}}

\begin{document}


\begin{flushright}
FIAN-TD-2017-19 \ \ \ \ \ \  \\
arXiv: 1709.08596 V2
\end{flushright}

\vspace{1cm}

\begin{center}

{\Large \bf Cubic interaction vertices for continuous-spin fields

\medskip
and arbitrary spin massive fields}

\vspace{2.5cm}

R.R. Metsaev\footnote{ E-mail: metsaev@lpi.ru }

\vspace{1cm}

{\it Department of Theoretical Physics, P.N. Lebedev Physical
Institute, \\ Leninsky prospect 53,  Moscow 119991, Russia }

\vspace{3cm}

{\bf Abstract}

\end{center}

Light-cone gauge formulation of relativistic dynamics of a continuous-spin field propagating in the flat space is developed. Cubic interaction vertices of continuous-spin massless fields
and totally symmetric arbitrary spin massive fields are studied. We consider parity invariant cubic vertices that involve one continuous-spin massless field and two arbitrary spin massive fields and parity invariant cubic vertices that involve two continuous-spin massless fields and one arbitrary spin massive field. We construct the complete list of such vertices explicitly. Also we demonstrate that there are no cubic vertices describing consistent interaction of continuous-spin massless fields with arbitrary spin massless fields.

\vspace{3cm}


\newpage
\renewcommand{\thefootnote}{\arabic{footnote}}
\setcounter{footnote}{0}

\section{ \large Introduction}

In view of certain remarkable features, the theory of continuous-spin gauge field has attracted some interest  recently \cite{Bekaert:2005in}-\cite{Najafizadeh:2017tin}. An extensive list of references devoted to various aspects of this topic may be found in Refs.\cite{Bekaert:2006py,Schuster:2014hca,Bekaert:2017khg,Brink:2002zx}.
We note interesting discussions about possible interrelations between the string theory and  continuous-spin field theory in Refs.\cite{Savvidy:2003fx}. Also, we note that, it turns out that a continuous-spin field can be decomposed into an infinite chain of coupled scalar, vector, and totally symmetric tensor fields which consists of every spin just once. This property of a continuous-spin field triggered our interest in this topic because a similar infinite chain of scalar, vector and totally symmetric fields enters the theory of higher-spin gauge field in AdS space \cite{Vasiliev:1990en} and we expect therefore some interesting interrelations between continuous-spin gauge field theory and higher-spin gauge theory.

In this paper, we study interacting continuous-spin fields.%
\footnote{ Continuous-spin field is an infinite component field. Finite component fields with arbitrary but fixed integer values of spin are referred to as arbitrary spin fields in this paper.}
Namely, our major aim in this paper is to study interaction of continuous-spin fields with arbitrary spin fields which propagate in flat space. To this end we use a light-cone gauge formulation of relativistic dynamics of fields propagating in flat space. The light-cone formulation discovered in Ref.\cite{Dirac:1949cp}
offers conceptual and technical simplifications of approaches  to
many problems of string theories and modern quantum field theory. Though this
formulation hides Lorentz symmetries but eventually turns out to be effective. For example, we mention the light-cone gauge formulation of superstring field theories \cite{Green:1983hw}.
Recently studied light-cone gauge superspace formulations of some supersymmetric field theories may be found in Refs.\cite{Metsaev:2004wv,Ananth:2005vg}. Various interesting applications of the light-cone formalism to field theories such as QCD are discussed in Refs.\cite{Brodsky:2013dca}.

In this paper, we apply a light-cone formalism for studying
vertices describing interaction of continuous-spin fields with arbitrary spin fields.
To this end, first, we develop the light-cone gauge $so(d-2)$ covariant formulation of free continuous-spin fields propagating in the flat space $R^{d-1,1}$ with arbitrary $d\geq 4$.%
\footnote{ In the framework of light-cone gauge helicity formalism,
free continuous-spin massless field in $R^{3,1}$ was discussed in Ref.\cite{Brink:2002zx}, while  all cubic interaction vertices for arbitrary spin massless fields in $R^{3,1}$ were obtained in Refs.\cite{Bengtsson:1983pd}.}
Second, using such formulation and adopting the method for constructing cubic interaction
vertices developed in Refs.\cite{Metsaev:1993ap}-\cite{Metsaev:2007rn} for arbitrary spin massive and massless fields, we find cubic vertices describing interaction of continuous-spin fields with arbitrary spin massive fields propagating in $R^{d-1,1}$, $d\geq 4$.
Namely, we find parity invariant cubic vertices describing interaction of one continuous-spin massless field with two arbitrary spin massive fields and parity invariant cubic vertices describing interaction of two continuous-spin massless fields with one arbitrary spin massive field. We provide the complete classification for such vertices. Also, we analyse equations for parity invariant cubic vertices describing interaction of one continuous-spin massless field with two arbitrary spin massless fields and equations for parity invariant cubic vertices describing interaction of two continuous-spin massless fields with one arbitrary spin massless field. We show that such equations do not have solutions. In other words, we demonstrate that, in the framework of $so(d-2)$ covariant light-cone gauge formalism, there are no parity invariant cubic vertices describing consistent interaction of continuous-spin massless fields with arbitrary spin massless fields.

The long term motivation for our study of parity invariant cubic vertices by using the $so(d-2)$ covariant light-cone gauge formulation  is related to the fact that it is the parity invariant light-cone gauge vertices that can be cast into BRST gauge invariant form in a relatively straightforwardly way. For more discussion of this theme, see Conclusions.

This paper is organized as follows.

In Sec.\ref{sec-02}, we
introduce our notation and describe the manifestly $so(d-2)$
covariant light-cone gauge formulation of free continuous-spin field propagating in $R^{d-1,1}$ space.
Also we recall the well-known light-cone gauge formulation of arbitrary spin
massless and massive fields.

In Sec.\ref{sec-03},  we study restrictions imposed by the Poincar\'e algebra symmetries on  arbitrary $n$-point interaction vertices.
After that we restrict our attention to cubic vertices. We formulate the light-cone gauge dynamical principle and discuss restrictions imposed by this principle on cubic vertices. In other words, we find complete system of equations imposed on cubic vertices by the Poincar\'e algebra symmetries and the light-cone gauge dynamical principle.

In Sec.\ref{sec-04}, we present solution to equations
for parity invariant cubic vertices describing interaction of one continuous
spin massless field with two arbitrary spin massive fields having the same masses,
while, in Sec.\ref{sec-05}, we present solution to equations
for parity invariant cubic interaction vertices for one continuous
spin massless field and two arbitrary spin massive fields having different masses.
We provide the complete classification of the just mentioned cubic interaction vertices.

In Sec.\ref{sec-06}, we present solution to equations for parity invariant cubic  vertices describing interaction of two continuous-spin massless fields with one arbitrary spin massive field.
Using our solution, we provide the complete classification of such cubic interaction vertices.

In Sec.\ref{sec-07}, we summarize our conclusions and suggest directions for future research.

In Appendix A, we describe the basic notation and conventions we use in this paper.

In Appendix B, we discuss light-cone gauge formulation of continuous-spin field propagating in $R^{3,1}$ by using  realization of physical fields in the helicity basis.

In appendix C, we outline the procedure of derivation of cubic vertices describing interaction of one continuous-spin massless field with two arbitrary spin massive fields, while, in Appendix D, we outline the procedure of derivation of cubic vertices for two continuous-spin massless fields and one arbitrary spin  massive field.

In Append E, we discuss equations for cubic vertices describing interaction of one continuous-spin massless fields with two arbitrary spin massless fields, while, in Appendix F, we study equations for cubic vertices describing interaction of two continuous-spin massless fields with one arbitrary spin massless field. We demonstrate that such equations do not have consistent solution.

\newsection{ \large Free light-cone gauge continuous-spin fields and arbitrary spin massive and massless fields}\label{sec-02}

\noindent {\bf Poincar\'e algebra in light-cone frame}. Light cone gauge method developed in Ref.\cite{Dirac:1949cp} reduces the problem of finding a new dynamical system  to a problem
of finding a new (light cone gauge) solution for commutation relations of a basic symmetry algebra.
For continuous-spin field and arbitrary spin massive and massless fields that propagate in the flat space $R^{d-1,1}$, basic symmetries are associated with the Poincar\'e algebra  $iso(d-1,1)$.
We start therefore with a description of a realization of the Poincar\'e algebra symmetries on a space of continuous-spin field and arbitrary spin massive and massless fields. In this section, we discuss free light-cone gauge fields.

The Poincar\'e algebra $iso(d-1,1)$ is spanned by the translation generators $P^\mu$ and rotation generators $J^{\mu\nu}$ which are generators of
the Lorentz algebra $so(d-1,1)$. The commutation relations of the Poincar\'e algebra  we use are given by%
\footnote{ Indices $\mu,\nu,\rho,\sigma = 0,1,\ldots,d-1$ are vector indices of the Lorentz algebra $so(d-1,1)$.}
\be \label{17082017-man02-01}
{} [P^\mu,\,J^{\nu\rho}]=\eta^{\mu\nu} P^\rho - \eta^{\mu\rho} P^\nu\,,
\qquad {} [J^{\mu\nu},\,J^{\rho\sigma}] = \eta^{\nu\rho} J^{\mu\sigma} + 3\hbox{ terms}\,,
\ee
where $\eta^{\mu\nu}$ stands for the mostly positive flat metric tensor.
The translation generators $P^\mu$ are considered to be hermitian, while the Lorentz algebra generators  $J^{\mu\nu}$ are taken to be anti-hermitian.

In order to discuss the light-cone formulation, we introduce, in place of
the Lorentz basis coordinates $x^\mu$, the light-cone basis
coordinates $x^\pm$, $x^i$ which are defined by the relations
\be \label{17082017-man02-02}
x^\pm \equiv \frac{1}{\sqrt{2}}(x^{d-1}  \pm x^0)\,,\qquad
x^i\,,\quad i=1,\ldots, d-2\,.
\ee
In what follows, the coordinate $x^+$ is treated as an evolution parameter. Using notation in \rf{17082017-man02-02}, we note then that the $so(d-1,1)$ Lorentz algebra vector $X^\mu$ is decomposed as $X^+,X^-,X^i$, while scalar product of the $so(d-1,1)$ Lorentz algebra vectors $X^\mu$ and $Y^\mu$ is decomposed as
\be \label{19082017-man02-01}
\eta_{\mu\nu}X^\mu Y^\nu = X^+Y^- + X^-Y^+ + X^i Y^i\,.
\ee
From \rf{19082017-man02-01}, we see that in light-cone frame, non vanishing elements of the flat metric are given by $\eta_{+-} = \eta_{-+}=1$, $\eta_{ij} = \delta_{ij}$, i.e., for the covariant and contravariant components of vectors we have the relations $X^+=X_-$, $X^-=X_+$, $X^i=X_i$. In
light-cone approach, generators of the Poincar\'e algebra are
separated into the following two groups:
\beq
\label{19082017-man02-02} &&
P^+,\quad
P^i,\quad
J^{+i},\quad
J^{+-},\quad
J^{ij}, \qquad
\hbox{ kinematical generators};
\\
\label{19082017-man02-03}  &&
P^-,\quad
J^{-i}\,, \hspace{4.3cm}
\hbox{ dynamical generators}.
\eeq
For $x^+=0$, in the field theoretical realization, kinematical generators \rf{19082017-man02-02} are
quadratic in fields%
\footnote{For arbitrary $x^+ \ne 0 $, dynamical generators \rf{19082017-man02-03} can be presented as
$G= G_1 + x^+ G_2$, where a functional $G_1$ is quadratic in fields, while a functional $G_2$
involves quadratic and higher order terms in fields.},
while, dynamical generators \rf{19082017-man02-03} involve quadratic and higher order terms in fields..

In light-cone frame, commutators of the Poincar\'e algebra generators \rf{19082017-man02-02},\rf{19082017-man02-03} are
obtained from the ones in \rf{17082017-man02-01} by using the
non vanishing elements of the flat metric, $\eta^{+-}=\eta^{-+}=1$,
$\eta^{ij} = \delta^{ij}$. We assume the following hermitian conjugation rules for the generators of the Poincar\'e algebra,
\be \label{19082017-man02-04}
P^{\pm \dagger} = P^\pm, \quad
P^{i\dagger} = P^i, \quad
J^{ij\dagger} = - J^{ij}\,,\quad
J^{+-\dagger}=-J^{+-}, \quad
J^{\pm i\dagger} = -J^{\pm i}\,.
\ee
In order to provide a field theoretical realization of the Poincar\'e algebra generators on a space of continuous-spin fields and arbitrary spin  massive and massless fields we exploit a light-cone gauge description of the fields. We discuss continuous-spin field and arbitrary spin massive and massless fields in turn.

{\bf Continuous-spin massless/massive field}. To discuss the light-cone gauge description of a continuous-spin massless/massive field, we introduce the following set of scalar, vector and traceless tensor fields of the $so(d-2)$ algebra:
\be \label{19082017-man02-05}
\phi^{i_1\ldots i_n}\,, \qquad n=0,1,2,\ldots,\infty\,.
\ee
In \rf{19082017-man02-05}, fields with $n=0$ and $n=1$ are the respective scalar and vector fields of the $so(d-2)$ algebra, while fields with $n\geq 2$ are traceless tensor fields of the $so(d-2)$ algebra,
\be \label{19082017-man02-06}
\phi^{iii_3\ldots i_n} = 0 \,, \qquad n =2,3,\ldots, \infty\,.
\ee

In order to discuss the light-cone gauge formulation of a continuous-spin field in an easy-to-use form we introduce the creation operators $\alpha^i$, $\upsilon$ and the respective annihilation operators $\alphab^i$, $\upsilonb$,
\be  \label{19082017-man02-07}
[\alphab^i,\alpha^j] = \delta^{ij}\,,  \quad [\upsilonb,\upsilon]=1\,, \quad \upsilonb |0\rangle = 0\,, \quad \alphab^i|0\rangle = 0\,, \quad \alpha^{i\dagger} = \alphab^i\,, \quad \upsilon^\dagger = \upsilonb\,.
\ee
Throughout this paper, the creation and annihilation operators will be referred to as oscillators. We note that the oscillators $\alpha^i$, $\alphab^i$ and $\upsilon$, $\upsilonb$ transform in the respective vector and scalar representations of the $so(d-2)$ algebra. Using the oscillators  $\alpha^i$, $\upsilon$, we collect all fields \rf{19082017-man02-05} into a ket-vector $\phik$ defined as
\be  \label{19082017-man02-08}
|\phi(p,\alpha)\rangle = \sum_{n=0}^\infty
\frac{\upsilon^n}{n!\sqrt{n!}}
\alpha^{i_1} \ldots \alpha^{i_n}
\phi^{i_1\ldots i_n}(p) |0\rangle\,,
\ee
where the argument $\alpha$ in \rf{19082017-man02-08} stands for the oscillators $\alpha^i$, $\upsilon$, while the argument $p$ stands for the momenta $p^i$, $\beta$.
Ket-vector \rf{19082017-man02-08} satisfies the following algebraic constraints
\beq
\label{19082017-man02-09} && (N_\alpha - N_\upsilon)\phik = 0 \,,  \qquad N_\alpha =\alpha^i\alphab^i\,, \qquad N_\upsilon = \upsilon\upsilonb\,,
\\
\label{19082017-man02-10} && \alphab^2 \phik = 0\,.
\eeq
We note that constraint \rf{19082017-man02-10} amounts to tracelessness constraints \rf{19082017-man02-06}.

\noindent {\bf Arbitrary spin massive fields}.  To discuss light-cone gauge description of an arbitrary spin-$s$ massive field, we introduce the following set of scalar, vector, and tensor fields of the $so(d-2)$ algebra:
\be \label{19082017-man02-14}
\phi^{i_1\ldots i_n}\,, \qquad n=0,1,2,\ldots,s\,.
\ee
In \rf{19082017-man02-14}, fields with $n=0$ and $n=1$ are the respective scalar and vector fields of the $so(d-2)$ algebra, while fields with $n\geq 2$ are totally symmetric tensor fields of the $so(d-2)$ algebra. Physical D.o.F of a massive field in flat space $R^{d-1,1}$ are described by irreps of the $so(d-1)$ algebra. For
the fields \rf{19082017-man02-14} to be associated with irreps of the $so(d-1)$ algebra, we should impose a constraint on fields \rf{19082017-man02-14}.
To simplify the presentation of the constraint we use the vector oscillators $\alpha^i$, $\alphab^i$ \rf{19082017-man02-07} and scalar oscillators $\zeta$, $\zetab$ defined by the relations
\be  \label{19082017-man02-15}
[\zetab,\zeta] = 1\,, \qquad \zetab |0\rangle = 0\,, \qquad \zeta^\dagger = \zetab\,.
\ee
Using the oscillators  $\alpha^i$, $\zeta$, we collect all fields \rf{19082017-man02-14} into a ket-vector $\phik$ defined as
\be  \label{19082017-man02-16}
|\phi_s(p,\alpha)\rangle = \sum_{n=0}^s
\frac{\zeta^{s-n}}{n!\sqrt{(s-n)!}}
\alpha^{i_1} \ldots \alpha^{i_n}
\phi^{i_1\ldots i_n}(p) |0\rangle\,,
\ee
where the argument $\alpha$ in \rf{19082017-man02-16} stands for the oscillators $\alpha^i$, $\zeta$, while the argument $p$ stands for the momenta $p^i$, $\beta$.
Ket-vector \rf{19082017-man02-16} satisfies the algebraic constraints
\beq
\label{19082017-man02-17} &&  (N_\alpha + N_\zeta -s) |\phi_s\rangle  = 0\,,
\\
\label{19082017-man02-18} && (\alphab^2 + \zetab^2) |\phi_s\rangle =0 \,.
\eeq
Constraint \rf{19082017-man02-17} tells us that ket-vector $\phik$ \rf{19082017-man02-16} is a degree-$s$ homogeneous polynomial in the oscillators $\alpha^i$, $\zeta$, while relation \rf{19082017-man02-18} is the constraint required for the fields \rf{19082017-man02-14} to be associated with irreps of the $so(d-1)$ algebra. Sometimes we prefer to use an infinite chain of massive fields which consists of every spin just once. Such chain of massive fields is described by the ket-vector
\be  \label{19082017-man02-16-a1}
|\phi(p,\alpha)\rangle = \sum_{s=0}^\infty |\phi_s(p,\alpha)\rangle\,,
\ee
where, in \rf{19082017-man02-16-a1}, the $|\phi_s(p,\alpha)\rangle$ stands for the ket-vector of spin-$s$ massive field given in \rf{19082017-man02-16}.

\noindent {\bf Arbitrary spin massless fields}.  To discuss light-cone gauge description of an arbitrary spin-$s$ massless field, we introduce a rank-$s$ totally symmetric traceless tensor field of the $so(d-2)$ algebra
\be \label{19082017-man02-19}
\phi^{i_1\ldots i_s}\,, \qquad \phi^{iii_3\ldots i_s} = 0 \,.
\ee
To simplify the presentation we use oscillators $\alpha^i$ \rf{19082017-man02-07} and  introduce the following ket-vector:
\be  \label{19082017-man02-20}
|\phi_s(p,\alpha)\rangle = \frac{1}{s!}
\alpha^{i_1} \ldots \alpha^{i_s}
\phi^{i_1\ldots i_s}(p) |0\rangle\,,
\ee
where the argument $\alpha$ in \rf{19082017-man02-20} stands for the oscillators $\alpha^i$, while the argument $p$ stands for the momenta $p^i$, $\beta$.
Ket-vector \rf{19082017-man02-20} satisfies the algebraic constraints
\beq
\label{19082017-man02-21} && (N_\alpha  -s ) |\phi_s\rangle  = 0\,,
\\
\label{19082017-man02-22} && \alphab^2 |\phi_s\rangle =0 \,.
\eeq
From \rf{19082017-man02-21}, we learn that ket-vector \rf{19082017-man02-20} is a degree-$s$ homogeneous polynomial in the oscillators $\alpha^i$, while the constraint for the ket-vector in \rf{19082017-man02-22} amounts to the tracelessness constraint for tensor fields in \rf{19082017-man02-19}. Sometimes it is convenient to use an infinite chain of massless fields which consists of every spin just once. Such chain of massless fields is described by the ket-vector
\be  \label{19082017-man02-20-a1}
|\phi(p,\alpha)\rangle = \sum_{s=0}^\infty |\phi_s(p,\alpha)\rangle\,,
\ee
where, in \rf{19082017-man02-20-a1}, the $|\phi_s(p,\alpha)\rangle$ stands for the ket-vector of spin-$s$ massless field given in \rf{19082017-man02-20}.

\noindent {\bf Field-theoretical realization of Poincar\'e algebra}. We now discuss a field theoretical realization of the Poincar\'e algebra on the space of continuous fields and arbitrary spin massive and massless fields. A realization of kinematical generators \rf{19082017-man02-02} and dynamical generators \rf{19082017-man02-03} in terms of differential operators acting on the ket-vector $|\phi\rangle$ is given by%
\footnote{ In this paper, without loss of generality, the generators of the Poincar\'e algebra are analysed for $x^+=0$.}
\beq
&& \hspace{-5cm} \hbox{\it Kinematical generators}:
\nonumber\\
\label{20082017-man02-01} && P^i=p^i\,, \qquad \qquad\quad
P^+=\beta\,,
\\
\label{20082017-man02-02} && J^{+i}=\partial_{p^i} \beta\,, \qquad \quad \
J^{+-} = \partial_\beta \beta\,,
\\
\label{20082017-man02-03} && J^{ij}=p^i \partial_{p^j} - p^j \partial_{p^i} + M^{ij}\,,
\\
&& \hspace{-5cm}  \hbox{\it Dynamical generators}:
\nonumber\\
\label{20082017-man02-04} && P^- =  -\frac{p^i p^i + m^2}{2\beta}\,,
\\
\label{20082017-man02-05} && J^{-i} = - \partial_\beta p^i + \partial_{p^i} P^- + \frac{1}{\beta}(M^{ij} p^j + M^i)\,,
\eeq
where we use the notation
\be \label{20082017-man02-06}
\beta\equiv p^+\,,\qquad
\partial_\beta\equiv \partial/\partial \beta\,, \quad
\partial_{p^i}\equiv \partial/\partial p^i\,.
\ee
In \rf{20082017-man02-03},\rf{20082017-man02-05} and below, the $M^{ij}$ stands for a spin operator of the $so(d-2)$ algebra,
\be \label{20082017-man02-07}
[M^{ij},M^{kl}] =
\delta^{jk} M^{il} + 3\hbox{ terms}.
\ee
On spaces of ket-vectors of continuous-spin \rf{19082017-man02-08}, arbitrary spin massive \rf{19082017-man02-16} and arbitrary spin massless \rf{19082017-man02-20} fields, the operator $M^{ij}$ is realized as
\be \label{20082017-man02-08}
M^{ij} = \alpha^i \alphab^j - \alpha^j \alphab^i\,.
\ee
In \rf{20082017-man02-04}, \rf{20082017-man02-05}, the $m$ is a mass parameter, while the $M^i$ is a spin operator. The $m$ and $M^i$ satisfy the commutation relations
\beq
\label{20082017-man02-09} && [M^i,M^{jk}] = \delta^{ij} M^k - \delta^{ik} M^j \,,
\\
\label{20082017-man02-10} && [M^i, M^j ] = - m^2 M^{ij}\,.
\eeq
We now see that all that remains to complete the description of the differential operators in \rf{20082017-man02-01}-\rf{20082017-man02-05} is to provide a realization of the spin operator $M^i$
on the ket-vectors of the fields under consideration. On spaces of ket-vectors of continuous-spin \rf{19082017-man02-08}, massive \rf{19082017-man02-16} and massless \rf{19082017-man02-20} fields, the operator $M^i$ is realized in the following way:

\noindent {\it Continuous-spin field (massless, $m^2=0$, and massive, $m^2< 0$)}:
\beq
\label{20082017-man02-14} && M^i = g \bar\alpha^i + A^i \gb\,,
\\
\label{20082017-man02-15} && A^i  =  \alpha^i - \alpha^2 \frac{1}{2N_\alpha+d-2}\alphab^i\,,
\\
\label{20082017-man02-16} && g =  g_\upsilon \upsilonb\,, \qquad  \gb  = - \upsilon g_\upsilon\,,
\\
\label{20082017-man02-17} && g_\upsilon  = \Bigl(\frac{1}{(N_\upsilon + 1) (2N_\upsilon + d - 2)} F_\upsilon\Bigr)^{1/2} \,,
\\
\label{20082017-man02-18} && F_\upsilon  =    \kappa^2 -  N_\upsilon (N_\upsilon + d - 3) m^2\,,
\\
\label{20082017-man02-19} && N_\alpha =  \alpha^i\alphab^i\,, \qquad  N_\upsilon =  \upsilon \upsilonb\,.
\eeq
\noindent {\it Arbitrary spin massive field, $m^2> 0$}:
\be \label{20082017-man02-20}
M^i = m (\zeta\alphab^i - \alpha^i \zetab).
\ee
\noindent {\it Arbitrary spin massless field, $m=0$}:
\be \label{20082017-man02-21}
M^i = 0\,.
\ee
In \rf{20082017-man02-18}, the $m$ stands for mass parameter of continuous-spin field, while $\kappa$ is a dimensionfull real-valued parameter, $\kappa^2>0$. We note that $\kappa^2$ is realized as eigenvalue of square of Pauli-Lubanski vector operator. Also we note that, for a continuous-spin massless field, we have $m =0$, while for a continuous-spin massive field, we assume $m^2 < 0$.

Realizations of the operator $M^i$ for massive field \rf{20082017-man02-20} and massless field \rf{20082017-man02-21} are well known from textbook \cite{Siegel:1988yz}. To our knowledge, the realization of the operator $M^i$ for continuous-spin field in $R^{d-1,1}$ with arbitrary $d$ and $m\ne 0$ given in \rf{20082017-man02-14}-\rf{20082017-man02-19} has not been discussed in earlier literature.%
\footnote{ For continuous-spin massless field in $R^{3,1}$ and $R^{4,1}$, the  discussion of the spin operator $M^i$ can be found in Sec.2 in Ref.\cite{Brink:2002zx}.}

Having found the realization of the Poincar\'e algebra generators
in terms of  differential operators in \rf{20082017-man02-01}-\rf{20082017-man02-21} we are ready to provide a field theoretical realization of the Poincar\'e algebra generators in terms of the ket-vectors $|\phi\rangle$.  At the quadratic level, a field theoretical realization of the kinematical generators  \rf{19082017-man02-02} and the dynamical generators \rf{19082017-man02-03} takes the form
\be \label{20082017-man02-22}
G_\smpt
=\int \beta d^{d-1}p\,
\langle\phi(p)| G |\phi(p)\rangle\,, \qquad
d^{d-1}p \equiv
d\beta d^{d-2}p\,,
\ee
where $G$ stands for the differential operators given in
\rf{20082017-man02-01}-\rf{20082017-man02-21}, while
$G_\smpt$ stands for the field theoretical generators.
The ket-vector $|\phi\rangle$ satisfies the well known Poisson-Dirac commutation relations
\be \label{20082017-man02-23}
[\,|\phi(p,\alpha)\rangle\,,\,|\phi(p^\prime\,,\alpha^\prime)\rangle\,]
\bigl|_{equal\, x^+}=\bigr.\frac{\delta^{d-1}(p+p^\prime)}{2\beta} \Pi\,,
\ee
where $\Pi$ stands for the projector on space of the respective ket-vectors of continuous \rf{19082017-man02-08}, massive \rf{19082017-man02-16} and massless \rf{19082017-man02-20} fields. Using \rf{20082017-man02-22}, \rf{20082017-man02-23}, we verify the standard commutation relation
\be \label{20082017-man02-24}
[ |\phi\rangle,G_\smpt\,]
= G
|\phi\rangle\,.
\ee

The following remarks are in order.

\noindent \ibf) In the framework of the Lagrangian approach, the light-cone gauge action is given by
\be \label{20082017-man02-25}
S = \int dx^+  d^{d-1} p\,\, \langle \phi(p)|{\rm
i}\, \beta
\partial^- |\phi(p)\rangle +\int dx^+ P^-\,,
\ee
where $P^-$ is the Hamiltonian. Representation for the
light-cone gauge action given in \rf{20082017-man02-25} is valid both for the free and interacting
fields. In the theory of free fields, the Hamiltonian is obtained from relations
\rf{20082017-man02-04},\rf{20082017-man02-22}.

\noindent \iibf) The light cone gauge formulation of free continuous-spin field we described in this section can be derived by using the Lorentz covariant and gauge invariant formulation of continuous-spin field in terms of the double-traceless gauge fields we developed in Ref.\cite{Metsaev:2016lhs} and by applying the standard method for the derivation of light-cone gauge formulation from the Lorentz covariant and gauge invariant formulation. For the case of totally symmetric fields in AdS, the pattern of such derivation  can be found in Section 3 in Ref.\cite{Metsaev:1999ui}.

\newsection{ \large Restrictions imposed on interaction vertices by Poincar\'e algebra symmetries and by light-cone gauge dynamical principle} \label{sec-03}

In theories of interacting fields, the dynamical generators given in \rf{19082017-man02-03} receive corrections which involve higher powers of fields. Dynamical generators \rf{19082017-man02-03} can be expanded in fields as
\be \label{20082017-man02-26}
G^\dyn
= \sum_{n=2}^\infty
G_\smpn^\dyn\,,
\ee
where we use the notation $G_\smpn^\dyn$ in \rf{20082017-man02-26} to denote a functional
that has $n$ powers of ket-vector $\phik$.
Problem of finding a dynamical system  of interacting fields amounts to a problem
of finding a non-trivial solution to dynamical generators $G_\smpn^\dyn$ for $n\geq 3$.
In this Section, we describe restrictions imposed on $G_\smpn^\dyn$ by the Poincar\'e algebra kinematical and dynamical symmetries. After that we discuss restrictions imposed on $G_\smp3$ by light-cone gauge dynamical principle. For the reader convenience, we start with a discussion of Poincar\'e algebra kinematical symmetries of dynamical generators $G_\smpn^\dyn$ for arbitrary $n \geq 3$.

\noindent {\bf Kinematical symmetries of dynamical generators $G_\smpn^\dyn$ for $n\geq 3$}. From the commutators of the dynamical generators \rf{19082017-man02-03}  with the kinematical generators $P^i$ and $P^+$, we find that the dynamical generators $G_\smpn^\dyn$ with $n\geq 3$ can be cast into the following form:
\beq
\label{20082017-man02-27} &&
P_\smpn^-
=  \int\!\! d\Gamma_n \,\,
\langle \Phi_\smpn ||p_\smpn^-\rangle \,,
\\
\label{20082017-man02-28} &&
J_\smpn^{-i}
= \int \!\! d\Gamma_n\,\,
\langle \Phi_\smpn | j_\smpn^{-i}\rangle
+ (\Xbf_\smpn^i \langle \Phi_\smpn |)|p_\smpn^-\rangle \,,
\eeq
where we use the notation
\beq
\label{20082017-man02-29} && \langle \Phi_\smpn| \equiv \prod_{a=1}^n \langle \phi(p_a,\alpha_a)|\,,\qquad\quad \langle \phi(p_a,\alpha_a)| \equiv |\phi(p_a,\alpha_a)\rangle^\dagger\,,
\\
\label{20082017-man02-30} && d\Gamma_n \equiv (2\pi)^{d-1} \delta^{d-1}(\sum_{a=1}^np_a)
\prod_{a=1}^n \frac{d^{d-1} p_a}{(2\pi)^{(d-1)/2}} \,,
\\
\label{20082017-man02-31} && \Xbf_\smpn^i \equiv - \frac{1}{n}\sum_{a=1}^n \partial_{p_a^i}\,.
\eeq
The $n$-point densities $|p_\smpn^-\rangle$ and $|j_\smpn^{-i}\rangle$ entering the respective generators $P_\smpn^-$ and $J_\smpn^{-i}$ \rf{20082017-man02-27}, \rf{20082017-man02-28} can be presented as
\beq
\label{20082017-man02-32} && |p_\smpn^-\rangle = p_\smpn^- (p_a,\beta_a, \alpha_a)|0\rangle\,,
\\
\label{20082017-man02-33} &&  |j_\smpn^{-i}\rangle = j_\smpn^{-i} (p_a,\beta_a, \alpha_a)|0\rangle\,.
\eeq
In this Section and below, we use the indices $a,b=1,\ldots,n$ to label
fields entering $n$-point interaction vertex. The Dirac $\delta$- functions in \rf{20082017-man02-30} respect the
conservation laws for the momenta $p_a^i$ and $\beta_a$. We note that argument $p_a$ in \rf{20082017-man02-29}, \rf{20082017-man02-30} stands for the momenta $p_a^i$, $\beta_a$. As seen from \rf{20082017-man02-32}, \rf{20082017-man02-33}, the densities $p_\smpn^-$ and $j_\smpn^{-i}$ depend on the momenta $p_a^i$, $\beta_a$, and the quantity $\alpha_a$ which is shortcut for spin variables. Namely, for continuous-spin field, the shortcut $\alpha_a$ stands for the set of oscillators $\alpha_a^i$, $\upsilon_a$, while for massive and massless fields, the shortcut $\alpha_a$ stands for the respective sets of oscillators $\alpha_a^i$, $\zeta_a$ and $\alpha_a^i$.
In this paper, the density $p_\smpn^-$ will often be referred to as an $n$-point interaction vertex. For $n=3$, the density $p_\smpn^-$ will be referred to as cubic interaction vertex.

\noindent {\bf $J^{+-}$-symmetry}. Commutators of the dynamical generators $P^-$, $J^{-i}$ with the kinematical generator $J^{+-}$ lead to the following equations for the densities:
\beq
\label{21082017-man02-01} && \sum_{a=1}^n \beta_a\partial_{\beta_a} \, |p_\smpn^-\rangle = 0 \,,
\\
\label{21082017-man02-02} && \sum_{a=1}^n \beta_a\partial_{\beta_a} \, |j_\smpn^{-i}\rangle = 0 \,.
\eeq

\noindent {\bf $J^{ij}$-symmetries}. Commutators of the dynamical generators $P^-$, $J^{-i}$ with the kinematical generators $J^{ij}$ lead to the following equations for the densities:
\beq
 \label{21082017-man02-03}&& \sum_{a=1}^n \bigl( p_a^i\partial_{p_a^j} - p_a^j\partial_{p_a^i}  + M_a^{ij}  \bigr) |p_\smpn^-\rangle =  0 \,,
\\
\label{21082017-man02-04}&& \sum_{a=1}^n \bigl( p_a^i\partial_{p_a^j} - p_a^j\partial_{p_a^i}  + M_a^{ij}  \bigr) |j_\smpn^{-k}\rangle = \delta^{jk} |j_\smpn^{-i}\rangle - \delta^{ik} |j_\smpn^{-j}\rangle\,.
\eeq

\noindent {\bf $J^{+i}$-symmetries}. From the commutators of the dynamical generators $P^-$, $J^{-i}$ with the kinematical generators $J^{+i}$, we learn that the densities $p_\smpn^-$ and $j_\smpn^{-i}$ depend on the momenta $p_a^i$ through the new momentum variables $\Po_{ab}^i$ defined by the relation
\be  \label{21082017-man02-05}
\Po_{ab}^i \equiv p_a^i \beta_b - p_b^i \beta_a\,.
\ee
Thus we see that the densities $p_\smpn^-$ and $j_\smpn^{-i}$ turn out to be functions of $\Po_{ab}^i$ in place of $p_a^i$,
\beq
\label{21082017-man02-06} && p_\smpn^- = p_\smpn^- (\Po_{ab},\beta_a, \alpha_a)\,, \qquad j_\smpn^{-i} = j_\smpn^{-i} (\Po_{ab},\beta_a, \alpha_a)\,.
\eeq

To summarize our study of restrictions imposed by the kinematical symmetries we note that the commutators  between the dynamical generators $P^-$, $J^{-i}$ and the kinematical
generators $J^{+-}$, $J^{ij}$ amount to equations given in \rf{21082017-man02-01}-\rf{21082017-man02-04}, while, from the commutators between the dynamical generators $P^-$, $J^{-i}$ and the kinematical generators $J^{+i}$, we learn that the $n$-point densities $p_\smpn^-$, $j_\smpn^{-i}$  turn out to be functions of the new momenta $\Po_{ab}^i$ in place of the generic momenta $p_a^i$.

Using definition of the new momenta $\Po_{ab}^i$ \rf{21082017-man02-05} and the conservation laws for the momenta $p_a^i$, $\beta_a$, we verify that there are only $n-2$ independent new momenta $\Po_{ab}^i$ . For example, for $n=3$, there is only one independent $\Po_{ab}^i$ (see relations \rf{21082017-man02-08} below). This simplifies study of restrictions imposed by kinematical symmetries on  dynamical generators. To demonstrate this we  consider kinematical symmetries for cubic densities $p_\smp3^-$ and $j_\smp3^{-i}$.

\noindent {\bf Kinematical symmetries of cubic densities}. Taking into account the momentum conservation laws
\be  \label{21082017-man02-07}
p_1^i + p_2^i + p_3^i = 0\,, \qquad \quad \beta_1 +\beta_2 +\beta_3 =0 \,,
\ee
it is easy to check that the momenta $\Po_{12}^i$, $\Po_{23}^i$, $\Po_{31}^i$ are expressed in terms of a new momentum $\Po^i$,
\be  \label{21082017-man02-08}
\Po_{12}^i =\Po_{23}^i = \Po_{31}^i = \Po^i \,,
\ee
where a new momentum $\Po^i$ is defined by the following relations:
\be \label{21082017-man02-09}
\Po^i \equiv \frac{1}{3}\sum_{a=1,2,3} \betach_a p_a^i\,, \qquad
\betach_a\equiv \beta_{a+1}-\beta_{a+2}\,, \quad \beta_a\equiv
\beta_{a+3}\,.
\ee
The use of the momentum $\Po^i$ \rf{21082017-man02-09} is preferable because this momentum is manifestly invariant under cyclic permutations of the external line indices
$1,2,3$. Thus the cubic densities $p_\smp3^-$ and $j_\smp3^{-i}$ are eventually a functions of
the momenta $\Po^i$, $\beta_a$ and the spin variables $\alpha_a$:
\be  \label{21082017-man02-10}
p_\smp3^- = p_\smp3^-(\Po,\beta_a, \alpha_a)\,, \qquad
j_\smp3^{-i} = j_\smp3^{-i}(\Po,\beta_a, \alpha_a)\,.
\ee
The fact that momenta $p_a^i$ enter cubic densities though the momentum $\Po^i$ allows us to simplify  the kinematical symmetry equations given in \rf{21082017-man02-01}-\rf{21082017-man02-04}. Namely, it is easy to check that, in terms of densities \rf{21082017-man02-10}, the kinematical symmetry equations \rf{21082017-man02-01}-\rf{21082017-man02-04} can be represented as follows.

\noindent {\bf $J^{+-}$-symmetry equations}:
\beq
\label{21082017-man02-14} && \Jbf^{+-} \,|p_\smp3^-\rangle  = 0\,,
\\
\label{21082017-man02-15} && \Jbf^{+-} |j_\smp3^{-i}\rangle  = 0\,,
\\
\label{21082017-man02-15-a1} && \hspace{1.8cm} \Jbf^{+-} \equiv \Po^j\partial_{\Po^j} + \sum_{a=1,2,3} \beta_a\partial_{\beta_a} \,.
\eeq
\noindent {\bf $J^{ij}$-symmetry equations}:
\beq
\label{21082017-man02-16} &&  \Jbf^{ij} |p_\smp3^-\rangle =0\,,
\\
\label{21082017-man02-17} &&  \Jbf^{ij} |j_\smp3^{-k}\rangle = \delta^{jk} |j_\smp3^{-i}\rangle - \delta^{ik} |j_\smp3^{-j}\rangle \,,
\\
\label{21082017-man02-18} && \hspace{1cm}  \Jbf^{ij} \equiv \Lbf^{ij}(\Po) + \Mbf^{ij}\,,
 \qquad  \Lbf^{ij}(\Po) \equiv \Po^i \partial_{\Po^j}  -
\Po^j \partial_{\Po^i} \,,\qquad  \Mbf^{ij} \equiv \sum_{a=1,2,3} M_a^{ij}\,.\qquad
\eeq

Obviously, the kinematical symmetries do not admit to fix vertices uniquely. Therefore we proceed with discussion of restrictions imposed by dynamical symmetries.

\noindent {\bf Dynamical symmetries of cubic densities}. In this paper, restrictions on the interaction vertices imposed by commutation relations between the dynamical generators will be referred to as dynamical symmetry restrictions.
We now discuss restrictions imposed on cubic interaction vertices by the dynamical symmetries of the Poincar\'e algebra. In other words, we consider the
commutators
\be  \label{21082017-man02-20}
[P^-,J^{-i}]=0\,,\qquad\quad [J^{-i},J^{-j}]=0\,.
\ee
In the cubic approximation, commutators \rf{21082017-man02-20} amount to the following commutators:
\beq
\label{21082017-man02-21} &&  [P_\smpt^- ,J_\smp3^{-i}] + [P_\smp3^-,J_\smpt^{-i}]=0\,,
\\
\label{21082017-man02-22} && [J_\smpt^{-i},J_\smp3^{-j}] + [J_\smp3^{-i},J_\smpt^{-j}]=0\,.
\eeq
From commutators \rf{21082017-man02-21}, we obtain the following equation for the cubic densities
$|p_\smp3^-(\Po,\beta_a,\alpha_a)\rangle$ and $|j_\smp3^{-i}(\Po,\beta_a,\alpha_a)\rangle$,
\be  \label{21082017-man02-23}
\Pbf^- |j_\smp3^{-i}\rangle = - \Jbf^{-i\dagger}
|p_\smp3^-\rangle\,,
\ee
where we use the notation
\beq
\label{21082017-man02-24} && \hspace{-0.8cm}   \Pbf^- \equiv \sum_{a=1,2,3} P_a^-\,, \hspace{1.1cm} P_a^- \equiv - \frac{p_a^i p_a^i + m_a^2}{2\beta_a}\,,
\\
\label{21082017-man02-25} && \hspace{-0.8cm} \Jbf^{-i\dagger} \equiv \sum_{a=1,2,3} J_a^{-i\dagger}  \,, \hspace{0.7cm} J_a^{-i\dagger} \equiv  p_a^i \partial_{\beta_a} - p_a^-\partial_{p_a^i} - \frac{1}{\beta_a} M_a^{ij}p_a^j + \frac{1}{\beta_a}
M_a^{i \dagger} \,.\qquad
\eeq
Quantities $\Pbf^-$ and $\Jbf^{-i\dagger}$ defined in \rf{21082017-man02-24}, \rf{21082017-man02-25}
can be expressed in terms of the momentum $\Po^i$ (see Appendix A in Ref.\cite{Metsaev:2005ar}):
\beq
\label{21082017-man02-26} \Pbf^- & = & \frac{\Po^i \Po^i}{2\beta} - \sum_{a=1,2,3} \frac{m_a^2}{2\beta_a}\,,
\\
\label{21082017-man02-27} \Jbf^{-i\dagger}  & = &   - \frac{\Po^i}{\beta} \No_\beta + \frac{1}{\beta} \Mo^{ij} \Po^j + \sum_{a=1,2,3}  \frac{\check\beta_a }{6\beta_a} m_a^2 \partial_{\Po^i}  + \frac{1}{\beta_a} M_a^{i\dagger}\,,
\\
\label{21082017-man02-28} && \beta  \equiv  \beta_1 \beta_2 \beta_3\,,
\\
\label{21082017-man02-29} && \No_\beta    \equiv  \frac{1}{3}\sum_{a=1,2,3} \check\beta_a \beta_a \partial_{\beta_a}\,, \qquad
\Mo^{ij}   \equiv \frac{1}{3}\sum_{a=1,2,3} \check\beta_a M_a^{ij}\,, \qquad \betach_a \equiv \beta_{a+1} - \beta_{a+2}\,.\qquad
\eeq
Equation \rf{21082017-man02-23} allows us to express the density $|j_\smp3^{-i}\rangle $ in terms of the vertex $|p_\smp3^-\rangle$,
\be \label{21082017-man02-30}
|j_\smp3^{-i}\rangle = -  (\Pbf^-)^{-1} \Jbf^{-i\dagger}|p_\smp3^-\rangle\,.
\ee
Plugging $|j_\smp3^{-i}\rangle$ \rf{21082017-man02-30} into \rf{21082017-man02-22}, we
check that commutators \rf{21082017-man02-22} are fulfilled. Plugging $|j_\smp3^{-i}\rangle$ \rf{21082017-man02-30} into kinematical symmetries equations \rf{21082017-man02-15}, \rf{21082017-man02-17} we verify that, if the vertex  $|p_\smp3^-\rangle$ satisfies Eqs.\rf{21082017-man02-14},\rf{21082017-man02-16} then Eqs.\rf{21082017-man02-15},\rf{21082017-man02-17} are also fulfilled. Thus, in the cubic approximation, we checked that Eqs.\rf{21082017-man02-14}, \rf{21082017-man02-16}, \rf{21082017-man02-30} provide the complete list of equations obtained from all commutation relations of the Poincar\'e algebra.

\noindent {\bf Light-cone gauge dynamical principle}. Equations \rf{21082017-man02-14}, \rf{21082017-man02-16}, \rf{21082017-man02-30} do not admit to fix the vertex $|p_\smp3^-\rangle$ uniquely. In order to fix the vertex $|p_\smp3^-\rangle$ uniquely we impose additional restrictions on the vertex $|p_\smp3^-\rangle$. These additional restrictions are referred to as light-cone gauge dynamical principle in this paper and they are formulated as follows.

\noindent \ibf) The densities $|p_\smp3^-\rangle$, $|j_\smp3^{-i}\rangle$ should be expandable in the momentum $\Po^i$;%
\footnote{ If a function $f(x)$ is expandable in power series in $x$, then we refer to such function as  function expandable in $x$.}

\noindent \iibf) The density  $|p_\smp3^-\rangle$ should satisfy the restriction
\be \label{21082017-man02-31}
|p_\smp3^-\rangle \ne {\bf P}^- |V\rangle\,, \quad |V\rangle \ \hbox{is expandable in } \Po^i\,,
\ee
where $\Pbf^-$ is given in \rf{21082017-man02-26}.

\noindent \iiibf) The densities  $|p_\smp3^-\rangle$, $|j_\smp3^{-i}\rangle$, and density $|V\rangle$  \rf{21082017-man02-31} should not involve $(\Pbf^-)^\gamma$-terms, $\gamma<0$.

We note that requirement \rf{21082017-man02-31} is
related to field redefinitions. Ignoring requirement \rf{21082017-man02-31} leads to vertices which can be removed by field redefinitions. As we are interested in the vertices
that cannot be removed by using field redefinitions, we impose the requirement in \rf{21082017-man02-31}. Also we note that the assumptions \ibf) and \iiibf) are the light-cone counterpart of locality condition commonly used in Lorentz covariant formulations.

\noindent {\bf Complete system of equations for cubic vertex}. To summarize the discussion in this section, we note that, for the cubic vertex given by
\be \label{21082017-man02-31-a1}
|p_\smp3^-\rangle = p_\smp3^-(\Po,\alpha_a,\beta_a)|0\rangle\,,
\ee
the complete system of equations which remains to be solved takes the form
\beq
\label{21082017-man02-32} && \Jbf^{+-} | p_\smp3^- \rangle =0 \,, \hspace{5.5cm} \hbox{kinematical } \  J^{+-}-\hbox{ symmetry};
\\
\label{21082017-man02-33} &&  \Jbf^{ij} |p_\smp3^-\rangle = 0\,, \hspace{5.7cm} \hbox{kinematical } \  J^{ij}-\hbox{ symmetries};
\\
\label{21082017-man02-34} && |j_\smp3^{-i}\rangle = - (\Pbf^-)^{-1} \Jbf^{-i\dagger}|p_\smp3^-\rangle\,, \qquad \hspace{2.2cm} \hbox{ dynamical } P^-, J^{-i} \hbox{ symmetries };\qquad
\\
&& \hspace{3cm} \hbox{ Light-cone gauge dynamical principle:}
\nonumber\\
\label{21082017-man02-35-a1} && |p_\smp3^-\rangle \hbox{ and } \ |j_\smp3^{-i}\rangle \hspace{0.5cm} \hbox{ are expandable in } \Po^i;
\\
\label{21082017-man02-35-a2} && |p_\smp3^-\rangle \ne \Pbf^- |V\rangle, \quad |V\rangle \ \hbox{is expandable in } \Po^i; \qquad
\\
\label{21082017-man02-35-a3} && |p_\smp3^-\rangle, |j_\smp3^{-i}\rangle, |V\rangle \quad \hbox{ do not involve } (\Pbf^-)^\gamma \hbox{-terms},\quad \gamma < 0 \,.
\eeq
Eqs.\rf{21082017-man02-32}-\rf{21082017-man02-35-a3}  constitute the complete
system of equations which admit to fix the cubic vertex $p_\smp3^-$ uniquely. Operators $\Jbf^{+-}$, $\Jbf^{ij}$, $\Pbf^-$, $\Jbf^{-i\dagger}$ entering equations \rf{21082017-man02-32}-\rf{21082017-man02-35-a3} are defined in \rf{21082017-man02-15-a1}, \rf{21082017-man02-18}, \rf{21082017-man02-26}, \rf{21082017-man02-27} respectively.
Let us remark that, if we consider the Yang-Mills and Einstein theories, then it can
be verified that Eqs.\rf{21082017-man02-32}-\rf{21082017-man02-34} and the light-cone gauge dynamical principle \rf{21082017-man02-35-a1}-\rf{21082017-man02-35-a3} admit to fix the cubic interaction vertices unambiguously (up to coupling constants). It seems then reasonable to use Eqs.\rf{21082017-man02-32}-\rf{21082017-man02-34} and the light-cone gauge dynamical principle \rf{21082017-man02-35-a1}-\rf{21082017-man02-35-a3} for studying the cubic interaction vertices of the continuous-spin field theory.

\subsection{ Equations for parity invariant cubic interaction vertices}

We recall that we study parity invariant cubic vertices for one continuous-spin massless field and two arbitrary spin massive fields and parity invariant cubic vertices for two continuous-spin massless fields and one arbitrary spin massive field. Namely, using the shortcut $(0,\kappa)_\smCSF$ for continuous-spin massless field and the shortcut $(m,s)$ for arbitrary but fixed spin-$s$ massive field with mass parameter $m$, we are going to consider cubic vertices for the following fields:
\beq
\label{23082017-man02-01} && \hspace{-1.3cm} (m_1,s_1)\hbox{-}(m_2,s_2)\hbox{-}(0,\kappa_3)_\smCSF \hspace{0.8cm} \hbox{\small two massive fields and one continuous-spin massless field}\quad
\\
\label{23082017-man02-02} &&  \hspace{-1.3cm} (0,\kappa_1)_\smCSF\hbox{-}(0,\kappa_2)_\smCSF\hbox{-}(m_3,s_3) \hspace{0.4cm}\hbox{ \small two  continuous-spin massless fields and one massive field}\quad
\eeq
Our notation in \rf{23082017-man02-01} implies that the mass-$m_1$, spin-$s_1$ and mass-$m_2$, spin-$s_2$ massive fields carry external line indices $a=1,2$, while the continuous-spin massless field corresponds to $a=3$. From our notation in \rf{23082017-man02-02}, we learn that continuous-spin massless fields carry external line indices $a=1,2$, while the mass-$m_3$, spin-$s_3$ massive field corresponds to $a=3$.

In general, besides the momentum variables $\Po^i$, $\beta_1$, $\beta_2$, $\beta_3$, vertex \rf{21082017-man02-31-a1} depends on oscillators that involved in the ket-vectors entering $P_\smp3^-$ \rf{20082017-man02-29}.  Taking this into account and recalling the definition of the ket-vectors for continuous-spin field \rf{19082017-man02-08} and massive field \rf{19082017-man02-16}, we note that cubic vertices describing interactions of the fields in \rf{23082017-man02-01} and \rf{23082017-man02-02} depend on the following respective set of the oscillators and the momenta:
\beq
\label{23082017-man02-03} && \Po^i, \qquad \alpha_1^i, \ \zeta_1, \ \beta_1, \qquad  \alpha_2^i, \ \zeta_2, \beta_2, \qquad \alpha_3^i, \  \upsilon_3, \   \beta_3;
\\
\label{23082017-man02-04} && \Po^i, \qquad \alpha_1^i, \ \upsilon_1, \ \beta_1, \qquad  \alpha_2^i, \ \upsilon_2, \beta_2, \qquad \alpha_3^i, \  \zeta_3, \   \beta_3\,.
\eeq
Taking into account variables in \rf{23082017-man02-03}, \rf{23082017-man02-04}, we now analyze restrictions \rf{21082017-man02-32}-\rf{21082017-man02-35-a3} in turn.

\noindent \ibf)  First, we analyze the restrictions imposed by the $J^{ij}$-symmetries
\rf{21082017-man02-33} which tell us that the vertex $p_\smp3^-$ \rf{21082017-man02-31-a1} depend on  invariants of the $so(d-2)$ algebra. The scalar oscillators $\zeta_a$, $\upsilon_a$ and momenta $\beta_a$ are invariants of the $so(d-2)$ algebra. The remaining invariants can be built by using the momentum $\Po^i$, the vector oscillators $\alpha_a^i$, the delta-Kroneker $\delta^{ij}$, and the Levi-Civita symbol $\epsilon^{ i_1\ldots i_{d-2} }$.
Vertices that do not involve the antisymmetric Levi-Civita symbol
are referred to as parity invariant vertices,
while vertices involving one antisymmetric Levi-Civita symbol are
referred to as parity non-invariant vertices. In this paper, we focus on the parity
invariant vertices. This implies that invariants of the $so(d-2)$ algebra that can be built by using
$\Po^i$, $\alpha_a^i$, and $\delta^{ij}$ are given by
\be \label{23082017-man02-05}
\Po^i\Po^i,\qquad \alpha_a^i \Po^i,\qquad \alpha_a^i \alpha_b^i\,.
\ee
Note that, if $P_\smp3^-$ \rf{20082017-man02-27} involves the bra-vector of continuous-spin field $\langle \phi(p_a,\alpha_a)|$, then in view of constraint \rf{19082017-man02-10}, the invariant $\alpha_a^i\alpha_a^i$ does not contribute to the $P_\smp3^-$, while, if $P_\smp3^-$ \rf{20082017-man02-27} involves bra-vector of massive field $\langle \phi(p_a,\alpha_a)|$, then in view of constraint \rf{19082017-man02-18}, the contribution of invariant $\alpha_a^i\alpha_a^i$, can be replaced by the $(-\zeta_a^2$).

To summarize the discussion of $J^{ij}$-symmetries, we note that general solution to parity invariant cubic vertices for the fields \rf{23082017-man02-01} and \rf{23082017-man02-02} that respect $J^{ij}$-symmetries and lead to the nontrivial $P_\smp3^-$ is given by the following respective expressions
\beq
\label{23082017-man02-06} && p_\smp3^-   = p_\smp3^-(\Po^i\Po^i, \beta_a, \alpha_a^i \Po^i, \alpha_{aa+1}\,,\zeta_1, \zeta_2,\ \upsilon_3)\,,
\\
\label{23082017-man02-07} && p_\smp3^- = p_\smp3^-(\Po^i\Po^i,  \beta_a, \alpha_a^i\Po^i\,, \alpha_{aa+1}\,,\upsilon_1\,, \upsilon_2,\ \zeta_3)\,.
\eeq
In \rf{23082017-man02-06}, \rf{23082017-man02-07} and below, the shortcut $p_\smp3^-(q_a)$ implies that $p_\smp3^-$ depends on $q_1$, $q_2$, $q_3$.

\noindent \iibf) We now analyse the restriction in \rf{21082017-man02-35-a2},\rf{21082017-man02-35-a3}.  One can demonstrate that, using field redefinitions, we can remove terms in \rf{23082017-man02-06}, \rf{23082017-man02-07} which  are proportional to $\Po^i\Po^i$ (see Appendix B in Ref.\cite{Metsaev:2005ar}). In other words, we can drop down the dependence on $\Po^i\Po^i$ in the vertices $p_\smp3^-$ \rf{23082017-man02-06}, \rf{23082017-man02-07}. Representation for the vertices in which they do not depend on $\Po^i\Po^i$ will be referred to minimal scheme in this paper. Obviously, in the minimal scheme, vertices satisfy Eqs.\rf{21082017-man02-35-a2},\rf{21082017-man02-35-a3} automatically. Thus, in the minimal scheme, vertices \rf{23082017-man02-06}, \rf{23082017-man02-07} take the form
\beq
\label{23082017-man02-08} && p_\smp3^-   = p_\smp3^-(\beta_a,B_a, \alpha_{aa+1}\,,\zeta_1, \zeta_2,\ \upsilon_3)\,,
\\
\label{23082017-man02-09} && p_\smp3^- = p_\smp3^-(\beta_a,B_a\,, \alpha_{aa+1}\,,\upsilon_1\,, \upsilon_2,\ \zeta_3)\,,
\eeq
where in \rf{23082017-man02-08}, \rf{23082017-man02-09} and below we use the notation
\be \label{23082017-man02-10}
B_a \equiv \frac{\alpha_a^i \Po^i}{\beta_a}\,, \qquad \alpha_{ab} \equiv \alpha_a^i\alpha_b^i\,.
\ee
Note also that, in place of the variables $\alpha_a^i \Po^i$ appearing in  \rf{23082017-man02-06}, \rf{23082017-man02-07}, we use re-scaled variables $B_a$ \rf{23082017-man02-10} in expressions \rf{23082017-man02-08}, \rf{23082017-man02-09}. We now see that dependence of vertices  \rf{23082017-man02-08}, \rf{23082017-man02-09} on the momentum $\Po^i$ enters through the variable $B_a$ \rf{23082017-man02-10}.

\noindent \iiibf) We now proceed with vertices in \rf{23082017-man02-08}, \rf{23082017-man02-09} and analyze the restrictions  given in \rf{21082017-man02-34}, \rf{21082017-man02-35-a1}. To this end, we compute action of the operator $\Jbf^{-i\dagger}$ \rf{21082017-man02-27} on the vertices \rf{23082017-man02-08}, \rf{23082017-man02-09},
\be  \label{23082017-man02-11}
\Jbf^{-i\dagger}|p_\smp3^-\rangle  = \Pbf^- \sum_{a=1,2,3} \frac{2\betach_a}{3\beta_a}\alpha_a^i \partial_{B_a} |p_\smp3^-\rangle + \Po^i G_\beta |p_\smp3^-\rangle + \sum_{a=1,2,3} \frac{\alpha_a^i}{\beta_a} G_{a,\Po^2} |p_\smp3^-\rangle\,,
\ee
where operators $G_{a,\Po^2}$, $G_\beta$ are given in the Appendices C,D. Using \rf{21082017-man02-34}, \rf{23082017-man02-11}, and explicit form of operators $G_{a,\Po^2}$, $G_\beta$, it is easy to see that requiring the density $|j_\smp3^{-i}\rangle$ to respect equations \rf{21082017-man02-34},\rf{21082017-man02-35-a3}, we get the equations (see Appendices C,D)
\beq
\label{23082017-man02-14} && G_a |p_\smp3^-\rangle =0\,,\qquad  a=1,2,3;
\\
\label{23082017-man02-15} && G_\beta |p_\smp3^-\rangle = 0\,.
\eeq
Using Eqs.\rf{21082017-man02-34}, \rf{23082017-man02-11}-\rf{23082017-man02-15} and \rf{09092017-man02-01}, \rf{11092017-man02-01}, we obtain the  representations for the densities $|j_\smp3^{-i}\rangle$ corresponding to the interaction vertices \rf{23082017-man02-08},\rf{23082017-man02-09},
\beq
\label{23082017-man02-16-a1} && |j_\smp3^{-i}\rangle = - \sum_{a=1,2,3} \frac{2 \betach_a}{3\beta_a} \alpha_a^i \partial_{B_a} |p_\smp3^-\rangle - \frac{2\beta}{\beta_3^3} \frac{ g_{\upsilon_3} \partial_{\upsilon_3} }{ 2 N_3  + d-2 } \partial_{B_3}^2 |p_\smp3^-\rangle\,,
\\
\label{23082017-man02-16-a2} && |j_\smp3^{-i}\rangle = - \sum_{a=1,2,3} \frac{2 \betach_a}{3\beta_a} \alpha_a^i \partial_{B_a} |p_\smp3^-\rangle - \sum_{a=1,2} \frac{2\beta}{\beta_a^3} \frac{ g_{\upsilon_a} \partial_{\upsilon_a} }{ 2 N_a  + d-2 } \partial_{B_a}^2 |p_\smp3^-\rangle\,,
\\
&& \hspace{2cm} N_a \equiv N_{B_a} + N_{\alpha_{aa+1}} + N_{\alpha_{a+2a}}\,, \quad a=1,2,3\,.
\eeq
From \rf{23082017-man02-16-a1},\rf{23082017-man02-16-a2}, we see that if vertices $|p_\smp3^-\rangle$ \rf{23082017-man02-08},\rf{23082017-man02-09} are expandable in $B_a$ \rf{23082017-man02-10} and satisfy Eqs.\rf{23082017-man02-14},\rf{23082017-man02-15}, then the respective densities $|j_\smp3^{-i}\rangle$ \rf{23082017-man02-16-a1},\rf{23082017-man02-16-a2} are also expandable in the $B_a$.

\bigskip
\noindent \ivbf) Finally, we analyze the restrictions of
$J^{+-}$-symmetry \rf{21082017-man02-32}. We note that, in
terms of vertices given in \rf{23082017-man02-08}, \rf{23082017-man02-09}, Eq.\rf{21082017-man02-32} is simplified as
\be \label{23082017-man02-17}
\sum_{a=1,2,3} \beta_a \partial_{\beta_a} p_\smp3^-=0\,.
\ee

{\bf Summary of analysis of Eqs.\rf{21082017-man02-32}-\rf{21082017-man02-35-a3}}.  To summarize the discussion in this Section, we note that, in the minimal scheme, the cubic vertices describing interactions of the fields given in \rf{23082017-man02-01} and \rf{23082017-man02-02} can be cast into the form given in \rf{23082017-man02-08} and  \rf{23082017-man02-09} respectively. Vertices \rf{23082017-man02-08} and  \rf{23082017-man02-09} should satisfy Eqs.\rf{23082017-man02-14}, \rf{23082017-man02-15}, \rf{23082017-man02-17}, while the respective densities $|j_\smp3^{-i}\rangle$ are expressed in terms of the cubic vertices $|p_\smp3^-\rangle$ as in \rf{23082017-man02-16-a1} and \rf{23082017-man02-16-a2}.

Thus all that remains is to solve Eqs.\rf{23082017-man02-14}, \rf{23082017-man02-15}, \rf{23082017-man02-17} for vertices \rf{23082017-man02-08} and  \rf{23082017-man02-09} which describe the respective interactions of the fields given in \rf{23082017-man02-01} and \rf{23082017-man02-02}.
From now on, we separately consider solutions of Eqs.\rf{23082017-man02-14}, \rf{23082017-man02-15}, \rf{23082017-man02-17} for vertices  \rf{23082017-man02-08} and  \rf{23082017-man02-09}.
We note also that, for vertices  \rf{23082017-man02-08}, there are two different cases: a) Two arbitrary spin massive fields have the same mass values; b) Two arbitrary spin massive fields have the different mass values. Structure of cubic vertices for these cases turns out to be different. We consider therefore these two cases in turn.

\newsection{ \large Parity invariant cubic vertices for one continuous-spin massless field and two massive fields with the same mass values}\label{sec-04}

We start with considering parity invariant cubic vertices for one continuous-spin massless field and two arbitrary spin massive fields having the same mass values.
Namely, using the shortcut $(0,\kappa)_\smCSF$ for a continuous-spin massless field and the shortcut $(m,s)$ for a mass-$m$ and spin-$s$ massive field, we consider a parity invariant cubic vertices for the following three fields:
\beq
\label{25082017-man02-01} && (m_1,s_1)\hbox{-}(m_2,s_2)\hbox{-}(0,\kappa_3)_\smCSF \hspace{0.8cm} \hbox{\small two massive fields and one continuous-spin massless field}\quad
\nonumber\\
&& m_1 = m_2 =m\,.
\eeq
Our notation in \rf{25082017-man02-01} implies that  the mass-$m_1$, spin-$s_1$ massive field and the mass-$m_2$, spin-$s_2$  massive field carry the respective external line indices $a=1,2$, while the continuous-spin massless field corresponds to $a=3$.

The general solution to cubic vertex for fields \rf{25082017-man02-01} takes the form (see Appendix C)
\beq
\label{30082017-man02-01}  p_\smp3^-  & = & U_{\upsilon_3} U_3 U_\beta U_\zeta U_B U_{z2} V^{(6)} \,,
\\
\label{30082017-man02-02} && p_\smp3^- = p_\smp3^-  (\beta_a, B_a, \alpha_{aa+1}\,,\zeta_1, \zeta_2,\ \upsilon_3)\,,
\\
\label{30082017-man02-03} && V^{(6)} = V^{(6)}(B_a, \alpha_{aa+1})\,,
\eeq
where, in solution \rf{30082017-man02-01}, we introduce new vertex $V^{(6)}$, while, in relations \rf{30082017-man02-02}, \rf{30082017-man02-03}, we show explicitly arguments of the generic vertex $p_\smp3^-$ and new vertex $ V^{(6)}$. The arguments $B_a$ and $\alpha_{ab}$  \rf{30082017-man02-02}, \rf{30082017-man02-03} are defined in \rf{23082017-man02-10}. Quantities denoted by $U$ in \rf{30082017-man02-01} are differential operators wrt the $B_a$ and $\alpha_{aa+1}$. Before presenting explicit expressions for the operators $U$ in \rf{30082017-man02-01}, we note that, for vertex $V^{(6)}$ \rf{30082017-man02-03}, we find two solutions given by
\beq
\label{30082017-man02-04}  V^{(6)} & = & \cosh (\Omega_3 z_3) V\,,
\\
\label{30082017-man02-05} V^{(6)} &  =  & \frac{\sinh(\Omega_3 z_3)}{\Omega_3} V\,,
\\
\label{30082017-man02-07} && V = V(B_1,B_2, \alpha_{12},\alpha_{23},\alpha_{31})\,,
\\
\label{30082017-man02-06}  && z_3 = \frac{B_3}{m}\,,
\eeq
where, in relations \rf{30082017-man02-04}, \rf{30082017-man02-05}, in place of $B_3$ \rf{23082017-man02-10}, we use a re-scaled variable $z_3$ \rf{30082017-man02-06} as a argument of the vertex $V^{(6)}$. In \rf{30082017-man02-07}, we show explicitly arguments of the vertex $V$. The quantity $\Omega_3$ is a differential operator independent of $z_3$. This operator is defined below.

From \rf{30082017-man02-01}-\rf{30082017-man02-07}, we see that general solution for the generic vertex $p_\smp3^-$, which depends on the twelve variables \rf{30082017-man02-02}, is expressed in terms of the vertex $V$ which is arbitrary function of the five variables \rf{30082017-man02-07}.  Note however that vertex $V$ \rf{30082017-man02-07} is restricted to be expandable in the five variables $B_1$, $B_2$, $\alpha_{12}$, $\alpha_{23}$, $\alpha_{31}$. Thus, the general solution for the generic vertex $p_\smp3^-$ \rf{30082017-man02-01} is expressed  in terms of the operators $U$, $\Omega_3$ and vertex $V$  \rf{30082017-man02-07}. Therefore all that remains to complete a description of the vertex $p_\smp3^-$ is to provide explicit expressions for the operators $U$, $\Omega_3$. The operators $U$, $\Omega_3$ entering our solution in \rf{30082017-man02-01} are given by
\beq
\label{30082017-man02-07-a1}&& \hspace{-1.5cm} U_{\upsilon_3} = \upsilon_3^{N_3}\,, \qquad  \qquad N_3 \equiv N_{B_3} + N_{ \alpha_{31} }  + N_{ \alpha_{23} }\,,
\\
\label{30082017-man02-07-a2} && \hspace{-1.5cm} U_3  =  \Bigl( \frac{2^{N_3} \Gamma(N_3 + \frac{d-2}{2})}{\Gamma( N_3 + 1)} \Bigr)^{1/2}\,,
\\
\label{30082017-man02-07-a3} && \hspace{-1.5cm} U_\beta   = \exp\bigl( - \frac{\betach_1}{2\beta_1}m\zeta_1 \partial_{B_1} - \frac{\betach_2}{2\beta_2} m\zeta_2 \partial_{B_2} - \frac{\betach_3}{2\beta_3} \kappa_3 \partial_{B_3} \bigr)\,,
\\
\label{30082017-man02-07-a4} && \hspace{-1.5cm} U_\zeta = \exp\bigl( - \half m \zeta_1\partial_{B_1} + \half m \zeta_2\partial_{B_2} \bigr)\,,
\\
\label{30082017-man02-07-a5} && \hspace{-1.5cm} U_B = \exp\Bigl( (\frac{\zeta_1}{m} B_2 - \frac{\zeta_2}{m} B_1 - \zeta_1 \zeta_2 )\partial_{\alpha_{12}}
- \frac{\zeta_1}{m }(B_3 + \half \kappa_3)\partial_{\alpha_{31}} + \frac{\zeta_2}{m }(B_3 - \half \kappa_3)\partial_{\alpha_{23}}\Bigr),
\\
\label{30082017-man02-07-a6} && \hspace{-1.5cm} U_{z2}= \exp\bigl( \half z_3 X + \frac{1}{4} z_3^2 Y \bigr)\,,
\\
\label{30082017-man02-07-a7} && \hspace{-1.5cm} X  =  -\frac{1}{m} ( B_2\partial_{\alpha_{23}} + B_1   \partial_{\alpha_{31}} ) \,,
\\
\label{30082017-man02-07-a8} && \hspace{-1.5cm} Y  =   \frac{2}{\kappa_3} (  B_2\partial_{\alpha_{23}} - B_1   \partial_{\alpha_{31}} )\,,
\\
\label{30082017-man02-07-a9} && \hspace{-1.5cm} Z  =  2 \alpha_{12} \partial_{\alpha_{31}}\partial_{\alpha_{23}}\,,
\\
\label{30082017-man02-07-a10} &&  \hspace{-1.5cm}  \Omega_3^2 = 1  + \half \{Y,\nu_3\}   + \frac{1}{4} X^2  +  Z\,,
\\
\label{30082017-man02-07-a11} &&  \hspace{-1.5cm}  \nu_3  =   N_{\alpha_{23}} +   N_{\alpha_{31}} + \frac{d-4}{2}\,,
\eeq
where $\betach_a$, $N_{B_a}$, $N_{\alpha_{ab}}$, $N_a$ are given in \rf{23092017-man02-04}-\rf{23092017-man02-06}, while $\Gamma$ \rf{30082017-man02-07-a2} stands for the Gamma-function.

Expressions \rf{30082017-man02-01}-\rf{30082017-man02-07-a11} provide the complete description of cubic vertices for one continuous-spin massless field and two arbitrary spin massive fields  \rf{25082017-man02-01}. More precisely, these cubic vertices describe an interaction of one continuous-spin massless field with two chains of totally symmetric massive fields. Each chain consists of every spin just once. Such chains of massive fields are described by ket-vectors given in \rf{19082017-man02-16-a1}. We now consider vertices for one continuous-spin massless field and two massive fields with arbitrary but fixed spin-$s_1$ and spin-$s_2$ values. Taking into account that the ket-vectors for massive fields $|\phi_{s_a}\rangle$  \rf{19082017-man02-16} are the respective degree-$s_a$ homogeneous polynomials in the oscillators $\alpha_a^i$, $\zeta_a$, $a=1,2$, \rf{19082017-man02-17}, it is easy to see that vertices we are interested in must satisfy the equations
\be \label{04092017-man02-01}
( N_{\alpha_a} + N_{\zeta_a} - s_a )|p_\smp3^-\rangle  = 0\,,\qquad a=1,2,
\ee
which tell us that the vertices should be degree-$s_1$ and degree-$s_2$ homogeneous polynomials in the respective oscillators $\alpha_1^i$, $\zeta_1$ and $\alpha_2^i$, $\zeta_2$. Using \rf{30082017-man02-01}, we verify that, in terms of $V$ \rf{30082017-man02-04}-\rf{30082017-man02-07}, Eqs.\rf{04092017-man02-01} take the form
\beq
\label{05092017-man02-01} && (N_{B_1} + N_{ \alpha_{12} } + N_{ \alpha_{31} } - s_1) V =0 \,,
\\
\label{05092017-man02-02} && (N_{B_2} + N_{ \alpha_{12} } + N_{ \alpha_{23} } - s_2) V =0 \,.
\eeq
As vertex $V$ \rf{30082017-man02-07} is considered to be expandable in the variables $B_1$, $B_2$, $\alpha_{12}$, $\alpha_{23}$, $\alpha_{31}$, a general solution of Eqs.\rf{05092017-man02-01}, \rf{05092017-man02-02} can be labelled by $s_1$, $s_2$ and by some three integers $n_1$, $n_2$, $n_3$. Using then  notation $V_{s_1,s_2}(n_1,n_2,n_3)$ for the vertex $V$ \rf{30082017-man02-07} that satisfies Eqs.\rf{05092017-man02-01}, \rf{05092017-man02-02}, we find the following general solution:
\beq
&& \hspace{-1.5cm} V  =   V_{s_1,s_2}(n_1,n_2,n_3) \,,
\nonumber\\
\label{05092017-man02-04} && \hspace{-1.5cm} V_{s_1,s_2}(n_1,n_2,n_3)  = B_1^{s_1 - n_1} B_2^{s_2 - n_2} \alpha_{12}^{l_3} \alpha_{23}^{l_1}  \alpha_{31}^{l_2} \,,
\\
\label{05092017-man02-05} && \hspace{-0.5cm} l_1 = \half (- n_1 + n_2 + n_3)\,, \quad l_2 = \half ( n_1 - n_2 +n_3)\,, \quad l_3 = \half ( n_1 + n_2 - n_3 )\,.
\eeq
Integers $n_1$, $n_2$, $n_3$ appearing in \rf{05092017-man02-04},\rf{05092017-man02-05} are the freedom of our solution for vertices, i.e., these integers label all possible
cubic vertices that can be constructed for three fields shown in \rf{25082017-man02-01}. In order for vertices \rf{05092017-man02-04} to be sensible, we should impose the following restrictions:
\be \label{05092017-man02-08}
0\leq n_1 \leq s_1\,, \quad 0\leq n_2 \leq s_2\,, \qquad \qquad l_1,l_2,l_3 \in \No_0\,.
\ee
Restrictions \rf{05092017-man02-08} amount to the requirement that the powers of all variables $B_1$, $B_2$, $\alpha_{12}$,  $\alpha_{23}$,  $\alpha_{31}$ in
\rf{05092017-man02-04} be non--negative integers. We note then that by using
relations \rf{05092017-man02-05}, we can rewrite restrictions
\rf{05092017-man02-08} as
\beq
\label{05092017-man02-09} && |n_1 - n_2| \leq n_3 \leq  n_1 + n_2 \,,
\\
\label{05092017-man02-10} && 0 \leq n_1 \leq s_1\,, \qquad 0 \leq n_2 \leq s_2\,,
\\
\label{05092017-man02-11} && n_1 + n_2 + n_3 \in 2\No_0\,, \qquad n_1,n_2,n_3 \in \No_0\,.
\eeq
Expressions for cubic vertices in \rf{30082017-man02-01},\rf{30082017-man02-04},\rf{30082017-man02-05},\rf{05092017-man02-04} supplemented by the restrictions on allowed values of the integers $n_1$, $n_2$, $n_3$ given in \rf{05092017-man02-09}-\rf{05092017-man02-11} provide the complete description and classification of cubic interaction vertices that can be constructed for the one continuous-spin massless field and the two spin-$s_1$ and spin-$s_2$ massive fields having the same mass parameter \rf{25082017-man02-01}.

The following remarks are in order.

\noindent \ibf)  From restrictions in \rf{05092017-man02-09}, \rf{05092017-man02-10}, we see that, given spin values $s_1$ and $s_2$, a number of cubic vertices that can be constructed for fields in \rf{25082017-man02-01} is finite.

\noindent \iibf)  From \rf{23082017-man02-10} and \rf{30082017-man02-06}, we see that the variable $z_3$ is a degree-1 homogeneous polynomial in the momentum $\Po^i$. Taking this into account and using expressions in \rf{30082017-man02-04}, \rf{30082017-man02-05}, we see that vertex $p_\smp3^-$ \rf{30082017-man02-01} involves all positive powers of the momentum $\Po^i$. Note also that vertex $V^{(6)}$ \rf{30082017-man02-04} involves all even positive powers of the momentum $\Po^i$, while vertex $V^{(6)}$ \rf{30082017-man02-05} involves all odd positive powers of the momentum $\Po^i$. On the other hand, from \rf{05092017-man02-04}, \rf{05092017-man02-10}, we see that vertex $V$ \rf{05092017-man02-04} is finite-order polynomial in the momentum $\Po^i$. Namely, vertex $V$ \rf{05092017-man02-04} is a degree-$(s_1+s_2-n_1 -n_2)$ homogeneous polynomial in the $\Po^i$.

\noindent \iiibf) From \rf{05092017-man02-04},\rf{05092017-man02-05}, we find the relation
\be \label{06092017-man02-01}
(N_{ \alpha_{23} } + N_{ \alpha_{31} } - n_3) V_{s_1,s_2}(n_1,n_2,n_3) =0 \,.
\ee
From \rf{05092017-man02-04},\rf{06092017-man02-01}, we see that the vertex $V_{s_1,s_2}(n_1,n_2,n_3)$  is a degree-$n_3$ monomial in the variables $\alpha_{23}$, $\alpha_{31}$. From relations \rf{30082017-man02-07-a7}-\rf{30082017-man02-07-a11}, we notice the commutators
\be \label{06092017-man02-02}
[\nu_3, X] = - X\,, \qquad [\nu_3,Y] = - Y\,, \qquad [\nu_3, Z] = - 2Z\,.
\ee
From \rf{06092017-man02-02}, it is clear that action of the operator $X^pY^qZ^n$ on the vertex $V_{s_1,s_2}(n_1,n_2,n_3)$ gives a degree-$(n_3-p-q-2n)$ homogeneous polynomial in the variables $\alpha_{23}$, $\alpha_{31}$. Taking this into account and using expressions for operators $U$ and $\Omega_3$ \rf{30082017-man02-07-a1}-\rf{30082017-man02-07-a11}, it is easy to see that given values $s_1$ and $s_2$ the cubic vertex $p_\smp3$ given by \rf{30082017-man02-01},\rf{30082017-man02-04},\rf{30082017-man02-05},\rf{05092017-man02-04} is a finite-order polynomial in the variables $B_1$, $B_2$, $\alpha_{12}$, $\alpha_{23}$, $\alpha_{31}$.

\noindent \ivbf) Two solutions for vertex $V^{(6)}$ \rf{30082017-man02-04},\rf{30082017-man02-05} appear as follows. Equations \rf{23082017-man02-14},\rf{23082017-man02-15} lead to the following 2nd-order differential equation for vertex $V^{(6)}$ (for details, see Appendix C)
\be \label{08092017-man02-01}
(\partial_{z_3}^2 - \Omega_3^2)V^{(6)}=0\,.
\ee
Differential equation \rf{08092017-man02-01} for the vertex $V^{(6)}$, which depends on the six variables $B_1$, $B_2$, $z_3$, $\alpha_{12}$, $\alpha_{23}$, $\alpha_{31}$ \rf{30082017-man02-03}, \rf{30082017-man02-06}, has two solutions presented in \rf{30082017-man02-04}, \rf{30082017-man02-05}, where the vertex $V$ depends on the five variables $B_1$, $B_2$, $\alpha_{12}$, $\alpha_{23}$, $\alpha_{31}$ \rf{30082017-man02-07}.

\noindent \vbf) {\bf Interaction of scalar massive fields and continuous-spin massless field}.%
\footnote{X. Bekaert informed us that, in collaboration with J.Mourad and
M.Najafizadeh, he described the minimal cubic coupling between a continuous-spin gauge field and scalar matter.}
By way of example and in order to demonstrate how to use our result we consider cubic vertex for two scalar massive fields and one continuous-spin massless field. For spin values of the scalar fields, we get $s_1=0$, $s_2=0$, while for mass values we set $m_1=m$, $m_2=m$ \rf{25082017-man02-01}. From \rf{05092017-man02-10}, we find $n_1=0$, $n_2=0$. Using this in \rf{05092017-man02-09}, we find $n_3=0$. Thus there is only one cubic vertex for two scalar massive fields and one continuous-spin massless field. Plugging values $n_1=0$, $n_2=0$, $n_3=0$ in \rf{05092017-man02-04},\rf{05092017-man02-05}, we find $V=1$. Plugging $V=1$ in \rf{30082017-man02-04}, \rf{30082017-man02-05} we find the following two vertices $V^{(6)}$:
\be \label{11092017-man02-45}
 V^{(6)} = \cosh z_3\,, \qquad V^{(6)}   =  \sinh z_3\,, \qquad z_3 = \frac{1}{m} B_3\,.
\ee
Plugging \rf{11092017-man02-45} into \rf{30082017-man02-01}, we get the following two interaction vertices $p_\smp3^-$:
\beq
\label{11092017-man02-46} && p_\smp3^- = U \cosh (\frac{1}{m} \upsilon_3 B_3  - \frac{\betach_3 \kappa_3}{2\beta_3 m} )\,,
\\
\label{11092017-man02-47} && p_\smp3^-   = U \sinh ( \frac{1}{m} \upsilon_3 B_3- \frac{\betach_3 \kappa_3}{2\beta_3 m} )\,,
\\
\label{11092017-man02-48} && \hspace{1cm} U = \Bigl( \frac{2^{N_{B_3}} \Gamma(N_{B_3} + \frac{d-2}{2})}{\Gamma( N_{B_3} + 1)} \Bigr)^{1/2}\,,\qquad N_{B_3} = B_3\partial_{B_3}\,.
\eeq
Vertex \rf{11092017-man02-46} is symmetric upon the replacement of external line indices of scalar fields, $1 \leftrightarrow 2$, and this vertex describes interaction of two scalar massive fields with one continuous-spin massless field. In the limit $\kappa_3\rightarrow 0$, vertex \rf{11092017-man02-46} is decomposed into a direct sum of vertices which describe interactions of two scalar massive fields with massless fields having even spin values.
Vertex \rf{11092017-man02-47} is anti-symmetric upon the replacement of external line indices of scalar fields, $1 \leftrightarrow 2$, and this vertex also describes interaction of two scalar massive fields with one continuous-spin massless field. In the limit $\kappa_3\rightarrow 0$, vertex \rf{11092017-man02-47} is decomposed into a direct sum of vertices which describe  interactions of two scalar massive fields with massless fields having odd spin values.%
\footnote{ In order to get non-trivial interaction for vertices in \rf{11092017-man02-47} one needs, as usually, to introduce internal symmetry. Incorporation of the internal symmetry into the game can be done via the Chan--Paton method in string theory \cite{Paton:1969je}, and could be performed as in Ref.\cite{Metsaev:1991nb}.}

\noindent \vibf) We see that vertices \rf{11092017-man02-46}, \rf{11092017-man02-47} are singular in the limit $m\rightarrow 0$. This implies that there are no cubic vertices describing consistent interaction of one continuous-spin massless field with two scalar massless fields. In Appendix  E, we demonstrate that, contrary to the cubic vertices for three arbitrary spin massless fields, cubic vertices for one  continuous-spin massless field and two arbitrary spin massless fields are not consistent.%
\footnote{ In light-cone gauge approach, parity invariant cubic vertices for three arbitrary spin massless fields in $R^{d-1,1}$, $d\geq 4$, were obtained in Ref.\cite{Metsaev:2005ar}. In BRST approach, such vertices were studied in Refs.\cite{Fotopoulos:2010ay,Metsaev:2012uy}, while, in metric-like approach, in Refs.\cite{Sagnotti:2010at}. In BRST approach, cubic vertices involving arbitrary spin massless and massive fields were derived in Ref.\cite{Metsaev:2005ar}. Interesting discussion of BRST approach may be found in Refs.\cite{Bekaert:2005jf}.
}

\noindent \viibf) We describe symmetry properties of various quantities and operators entering our solution \rf{30082017-man02-01}. Upon the replacement of the external line indices of arbitrary spin massive fields, $1 \leftrightarrow 2$, the quantities $\betach_a$, $\Po^i$ \rf{21082017-man02-09} and $B_a$ \rf{23082017-man02-10} are changed as
\be \label{12092017-man02-01}
\betach_1 \leftrightarrow - \betach_2\,,\qquad  \betach_3 \leftrightarrow - \betach_3\,,\qquad \Po^i \leftrightarrow -\Po^i\,,\qquad B_1 \leftrightarrow -B_2\,,\quad  B_3 \leftrightarrow - B_3\,.
\ee
Using \rf{12092017-man02-01}, we note then the behaviour  of quantities in \rf{30082017-man02-07-a7}-\rf{30082017-man02-07-a11} upon the replacement of the external line indices of arbitrary spin massive fields, $1 \leftrightarrow 2$,
\be \label{12092017-man02-02}
X \leftrightarrow X\,,\quad  Y \leftrightarrow - Y\,,\quad Z \leftrightarrow Z\,,\quad z_3 \leftrightarrow -z_3\,, \quad \Omega_3^2 \leftrightarrow \Omega_3^2\,,\quad \nu_3 \leftrightarrow \nu_3\,.\qquad
\ee
Relations \rf{12092017-man02-01},\rf{12092017-man02-02} imply that all operators $U$ \rf{30082017-man02-07-a1}-\rf{30082017-man02-07-a6} are symmetric upon the replacement of external line indices of arbitrary spin massive fields, $1 \leftrightarrow 2$,
\be \label{12092017-man02-03}
U_{\upsilon_3} \leftrightarrow U_{\upsilon_3}\,,\quad  U_3 \leftrightarrow U_3\,,\quad  U_\beta \leftrightarrow U_\beta\,,\quad  U_\zeta \leftrightarrow U_\zeta\,,\quad  U_B \leftrightarrow U_B\,, \quad U_{z2} \leftrightarrow U_{z2}\,.
\ee

\newsection{ \large Parity invariant cubic vertices for one continuous-spin massless field and two arbitrary spin massive fields with the different mass values}\label{sec-05}

In this Section, we consider parity invariant cubic vertices for one continuous-spin massless field and two arbitrary spin massive fields having different mass values.
Namely, using the shortcut $(0,\kappa)_\smCSF$ for continuous-spin massless field and the shortcut $(m,s)$ for mass-$m$ and spin-$s$ massive field, we consider  parity invariant cubic vertices for the following three fields:
\beq
\label{30082017-man02-08} && (m_1,s_1)\hbox{-}(m_2,s_2)\hbox{-}(0,\kappa_3)_\smCSF \hspace{0.8cm} \hbox{\small two massive fields and one continuous-spin massless field}\quad
\nonumber\\
&& m_1 \ne m_2\,.
\eeq
Our notation in \rf{30082017-man02-08} implies that mass-$m_1$, spin-$s_1$ and mass-$m_2$, spin-$s_2$  massive fields carry the respective external line indices $a=1,2$, while the  continuous-spin massless field corresponds to $a=3$.

The general solution to cubic vertex for fields in \rf{30082017-man02-08} takes the form (see Appendix C)
\beq
\label{02092017-man02-01}  p_\smp3^-  & = &  U_{\upsilon_3} U_3 U_\beta U_\zeta U_B U_{z1} U_{z\nu} U_W  V^{(8)}\,,
\\
\label{02092017-man02-02} && p_\smp3^- = p_\smp3^-  (\beta_a, B_a, \alpha_{aa+1}\,,\zeta_1, \zeta_2,\ \upsilon_3)\,,
\\
\label{02092017-man02-03} && V^{(8)}  = V^{(8)}(B_a,\alpha_{aa+1})\,,
\eeq
where, in the solution \rf{02092017-man02-01}, we introduce new vertex denoted by $V^{(8)}$, while, in relations \rf{02092017-man02-02}, \rf{02092017-man02-03}, we show explicitly arguments of the generic vertex $p_\smp3^-$ and the new vertex $ V^{(8)}$. Before presenting operators $U$ appearing in \rf{02092017-man02-01} we note that, for vertex $V^{(8)}$ \rf{02092017-man02-03}, we find two solutions which can be expressed in terms of the modified Bessel functions,
\beq
\label{02092017-man02-04} V^{(8)} &= &I_{\nu_3}(\sqrt{4z_3}) V\,,
\\
\label{02092017-man02-05} V^{(8)} & = &K_{\nu_3}(\sqrt{4z_3}) V\,,
\\
\label{02092017-man02-08} && V  = V(B_1,B_2, \alpha_{12},\alpha_{23},\alpha_{31}) \,,
\\
\label{02092017-man02-06} && z_3   =    \frac{\kappa_3^2(m_1^2 + m_2^2)}{2(m_1^2-m_2^2)^2} -   \frac{\kappa_3}{m_1^2-m_2^2} B_3\,,
\\
\label{02092017-man02-07} && \nu_3 = N_{\alpha_{23}} +  N_{\alpha_{31}} + \frac{d - 4}{2}\,,
\eeq
where $B_a$, $N_{\alpha_{ab}}$ are given in \rf{23092017-man02-04},\rf{23092017-man02-05}. In \rf{02092017-man02-04},\rf{02092017-man02-05}, in place of the $B_3$ \rf{23082017-man02-10}, we use a new variable $z_3$ \rf{02092017-man02-06} as a argument of the vertex $V^{(8)}$. For the modified Bessel functions $I_\nu$ and $K_\nu$ \rf{02092017-man02-04},\rf{02092017-man02-05}, we use conventions in Ref.\cite{NIST}.

From \rf{02092017-man02-02}, we see that the vertex $p_\smp3^-$ depends on twelve variables, while, from \rf{02092017-man02-08}, we learn that the vertex $V$ depends on five variables.  Vertex $V$ \rf{02092017-man02-08} is restricted to be expandable in the five variables $B_1$, $B_2$, $\alpha_{12}$, $\alpha_{23}$, $\alpha_{31}$. The general solution for vertex $p_\smp3^-$ \rf{02092017-man02-01} is expressed  in terms of the operators $U$, operator $\nu_3$ \rf{02092017-man02-07}, and vertex $V$  \rf{02092017-man02-08}. All that remains to complete the description of the solution for the vertex $p_\smp3^-$ in \rf{02092017-man02-01} is to provide explicit expressions for the operators $U$. The operators $U$ appearing in  \rf{02092017-man02-01} are given by
\beq
\label{02092017-man02-08-a1} && U_{\upsilon_3} = \upsilon_3^{N_3}\,, \hspace{3cm}  N_3 \equiv N_{B_3} + N_{ \alpha_{31} }  + N_{ \alpha_{23} }\,,
\\
\label{02092017-man02-08-a2} &&  U_3  =  \Bigl( \frac{2^{N_3} \Gamma(N_3 + \frac{d-2}{2})}{\Gamma( N_3 + 1)} \Bigr)^{1/2}\,,
\\
\label{02092017-man02-08-a3} && U_\beta   = \exp\bigl( - \frac{\betach_1}{2\beta_1}m_1\zeta_1 \partial_{B_1} - \frac{\betach_2}{2\beta_2} m_2\zeta_2 \partial_{B_2} - \frac{\betach_3}{2\beta_3} \kappa_3 \partial_{B_3} \bigr)\,,
\\
\label{02092017-man02-08-a4} &&  U_\zeta = \exp\bigl( - \frac{m_2^2}{2m_1} \zeta_1 \partial_{B_1} + \frac{m_1^2}{2m_2} \zeta_2 \partial_{B_2} \bigr)
\\
\label{02092017-man02-08-a5} &&  U_B = \exp\Bigl( \bigl(\frac{\zeta_1}{m_1} B_2 - \frac{\zeta_2}{m_2} B_1 - \frac{m_1^2 + m_2^2}{2m_1m_2} \zeta_1 \zeta_2 \bigr)\partial_{\alpha_{12}}
\nonumber\\
&& \hspace{1.8cm} - \frac{\zeta_1}{m_1}(B_3 + \half \kappa_3)\partial_{\alpha_{31}} + \frac{\zeta_2}{m_2}(B_3 - \half \kappa_3)\partial_{\alpha_{23}}\Bigr)\,,
\\
\label{02092017-man02-08-a6} && U_{z1} = \exp( - X + z_3^{\vphantom{1pt}} Y ) \,,
\\
\label{02092017-man02-08-a7} && U_{z\nu} = z_3^{-\nu_3/2}\,,
\\
\label{02092017-man02-08-a8} && U_W  = \sum_{n=0}^\infty \frac{\Gamma(\nu_3 + n)}{n!\Gamma(\nu_3 + 2n)} W^n\,,
\\
\label{02092017-man02-08-a9} && X  =  \frac{2\kappa_3}{(m_1^2-m_2^2)^2} ( m_1^2 B_1   \partial_{\alpha_{31}} - m_2^2 B_2\partial_{\alpha_{23}})\,,
\\
\label{02092017-man02-08-a10} && Y  =   \frac{2}{\kappa_3} (  B_2\partial_{\alpha_{23}} - B_1   \partial_{\alpha_{31}} )\,,
\\
\label{02092017-man02-08-a11} && Z  =  2 \alpha_{12} \partial_{\alpha_{31}}\partial_{\alpha_{23}}\,,
\\
\label{02092017-man02-08-a12} && W =    Z + X Y\,,
\eeq
where $\betach_a$, $N_{B_a}$, $N_{\alpha_{ab}}$, $\nu_a$, $N_a$ are defined in \rf{23092017-man02-04}-\rf{23092017-man02-06}, while the symbol $\Gamma$ \rf{02092017-man02-08-a2}, \rf{02092017-man02-08-a8}  stands for the Gamma-function.

Expressions in \rf{02092017-man02-01}-\rf{02092017-man02-08-a12} provide the complete description of the cubic vertex describing interaction of one continuous-spin  massless field with two infinite chains of arbitrary spin massive fields. We recall that the infinite chain of massive fields is described by ket-vector given in \rf{19082017-man02-16-a1}. In the chain of massive fields carrying external line index $a=1$, all fields have mass value $m_1$, while, in the chain of massive fields carrying external line index $a=2$, all fields have mass value $m_2$, $m_1 \ne m_2$. To consider vertices for one  continuous-spin massless field and arbitrary but fixed spin-$s_1$ and spin-$s_2$ massive fields \rf{30082017-man02-08} we note that the ket-vectors for massive spin-$s_a$ fields $|\phi_{s_a}\rangle$ are the respective degree-$s_a$ homogeneous polynomials in the oscillators $\alpha_a^i$, $\zeta_a$, $a=1,2$, \rf{19082017-man02-17}. This implies that the vertices we are interested in must satisfy the equations
\be \label{04092017-man02-01-ad}
( N_{\alpha_a} + N_{\zeta_a} - s_a )|p_\smp3^-\rangle  = 0\,,\qquad a=1,2,
\ee
which tell us that the $|p_\smp3^-\rangle$ should be degree-$s_1$ and degree-$s_2$ homogeneous polynomial in the respective oscillators $\alpha_1^i$, $\zeta_1$ and $\alpha_2^i$, $\zeta_2$. Using \rf{02092017-man02-01}, we verify that, in terms of vertex $V$ \rf{02092017-man02-04}, \rf{02092017-man02-05}, Eqs.\rf{04092017-man02-01-ad} take the form
\beq
\label{05092017-man02-01-ad} && (N_{B_1} + N_{ \alpha_{12} } + N_{ \alpha_{31} } - s_1) V =0 \,,
\\
\label{05092017-man02-02-ad} && (N_{B_2} + N_{ \alpha_{12} } + N_{ \alpha_{23} } - s_2) V =0 \,.
\eeq
Vertex $V$ \rf{02092017-man02-08} is restricted to be expandable in the variables $B_1$, $B_2$, $\alpha_{12}$, $\alpha_{23}$, $\alpha_{31}$. Therefore solution of Eqs.\rf{05092017-man02-01-ad}, \rf{05092017-man02-02-ad} can be labelled by spin values $s_1$, $s_2$ and by some three integers $n_1$, $n_2$, $n_3$. Using the notation $V_{s_1,s_2}(n_1,n_2,n_3)$ for vertex $V$ \rf{02092017-man02-08} that satisfies Eqs.\rf{05092017-man02-01-ad}, \rf{05092017-man02-02-ad}, we find the following general solution:
\beq
&& \hspace{-1cm} V  =   V_{s_1,s_2}(n_1,n_2,n_3) \,,
\nonumber\\
\label{05092017-man02-04-ad} && \hspace{-1cm} V_{s_1,s_2}(n_1,n_2,n_3)  =   B_1^{s_1 - n_1} B_2^{s_2 - n_2} \alpha_{12}^{l_3} \alpha_{23}^{l_1}  \alpha_{31}^{l_2} \,,
\\
\label{05092017-man02-05-ad} && l_1 = \half (- n_1 + n_2 + n_3)\,, \quad l_2 = \half ( n_1 - n_2 +n_3)\,, \quad  l_3 = \half ( n_1 + n_2 - n_3 )\,.\qquad
\eeq
Integers $n_1$, $n_2$, $n_3$ in \rf{05092017-man02-04-ad},\rf{05092017-man02-05-ad} are the freedom of our solution for vertices, i.e., these integers label all possible
cubic vertices that can be constructed for fields in \rf{30082017-man02-08}. In order for vertices \rf{05092017-man02-04-ad} to be sensible, we should impose the following restrictions:
\be \label{05092017-man02-08-ad}
0\leq n_1 \leq s_1\,, \quad 0\leq n_2 \leq s_2\,, \qquad \qquad l_1,l_2,l_3 \in \No_0\,,
\ee
which amount to the requirement that the powers of variables $B_1$, $B_2$, $\alpha_{12}$,  $\alpha_{23}$,  $\alpha_{31}$ in \rf{05092017-man02-04-ad} be non--negative integers. Using
relations \rf{05092017-man02-05-ad}, we represent restrictions
\rf{05092017-man02-08-ad} as
\beq
\label{05092017-man02-09-ad} && |n_1 - n_2| \leq n_3 \leq  n_1 + n_2 \,,
\\
\label{05092017-man02-10-ad} && 0 \leq n_1 \leq s_1\,, \qquad 0 \leq n_2 \leq s_2\,,
\\
\label{05092017-man02-11-ad} && n_1 + n_2 + n_3 \in 2\No_0\,, \qquad n_1,n_2,n_3 \in \No_0\,.
\eeq
Relations in \rf{02092017-man02-01},\rf{02092017-man02-04},\rf{02092017-man02-05},\rf{05092017-man02-04-ad}   and restrictions in \rf{05092017-man02-09-ad}-\rf{05092017-man02-11-ad} provide the complete description and classification of cubic vertices that can be constructed for one continuous-spin  massless field and two spin-$s_1$ and spin-$s_2$ massive fields having different mass parameters \rf{30082017-man02-08}.

The following remarks are in order.

\noindent \ibf)  From restrictions in \rf{05092017-man02-09-ad}, \rf{05092017-man02-10-ad}, we see that given spin values $s_1$ and $s_2$ a number of cubic vertices that can be constructed for fields in \rf{30082017-man02-08} is finite.

\noindent \iibf)  From \rf{23082017-man02-10} and \rf{02092017-man02-06}, we see that the variable $z_3$ is degree-1 polynomial in the momentum $\Po^i$. Taking this into account and using expressions in \rf{02092017-man02-04}, \rf{02092017-man02-05}, we see that vertex $p_\smp3^-$ \rf{02092017-man02-01} involves all positive powers of the momentum $\Po^i$. On the other hand, from \rf{05092017-man02-04-ad}, \rf{05092017-man02-10-ad}, we see that vertex $V$ \rf{05092017-man02-04-ad} is finite-order polynomial in the momentum $\Po^i$. Namely,  vertex $V$ \rf{05092017-man02-04-ad} is a degree-$(s_1+s_2-n_1 -n_2)$ homogeneous polynomial in the momentum $\Po^i$.

\noindent \iiibf) From \rf{05092017-man02-04-ad},\rf{05092017-man02-05-ad}, we find the relation
\be \label{06092017-man02-01-ad}
(N_{ \alpha_{23} } + N_{ \alpha_{31} } - n_3) V_{s_1,s_2}(n_1,n_2,n_3) =0 \,.
\ee
From \rf{05092017-man02-04-ad},\rf{06092017-man02-01-ad}, we see that the vertex $V_{s_1,s_2}(n_1,n_2,n_3)$ is a degree-$n_3$ monomial in the variables $\alpha_{23}$, $\alpha_{31}$. Using \rf{02092017-man02-07} and  \rf{02092017-man02-08-a9}-\rf{02092017-man02-08-a12}, we note the commutators
\be \label{06092017-man02-02-ad}
[\nu_3, X] = - X\,, \qquad [\nu_3,Y] = - Y\,, \qquad [\nu_3, Z] = - 2Z\,, \qquad [\nu_3, W] = - 2W\,.
\ee
From \rf{06092017-man02-02-ad}, it is clear that action of the operator $X^pY^qZ^nW^m$ on the vertex $V_{s_1,s_2}(n_1,n_2,n_3)$ gives a degree-$(n_3-p-q-2n-2m)$ homogeneous polynomial in the variables $\alpha_{23}$, $\alpha_{31}$. Taking this into account and using expressions for operators $U$  \rf{02092017-man02-08-a1}-\rf{02092017-man02-08-a12}, it is easy to see that given $s_1$ and $s_2$ the cubic vertex $p_\smp3$ given by  \rf{02092017-man02-01},\rf{02092017-man02-04},\rf{02092017-man02-05},\rf{05092017-man02-04-ad} is a finite-order polynomial in the variables $B_1$, $B_2$, $\alpha_{12}$, $\alpha_{23}$, $\alpha_{31}$.

\noindent \ivbf) Two solutions for vertex $V^{(8)}$ \rf{02092017-man02-04},\rf{02092017-man02-05} appear as follows. Equations \rf{23082017-man02-14},\rf{23082017-man02-15} lead to the following 2nd-order differential equation for the vertex $V^{(8)}$ (for details, see Appendix C)
\be \label{08092017-man02-02}
\bigl( 1 - (N_{z_3}+1)\partial_{z_3}  + \frac{\nu_3^2}{4z_3} \bigr) V^{(8)}=0\,, \qquad N_{z_3} \equiv z_3\partial_{z_3}\,.
\ee
Differential equation \rf{08092017-man02-02} for the vertex $V^{(8)}$, which depends on the six variables $B_1$, $B_2$, $z_3$, $\alpha_{12}$, $\alpha_{23}$, $\alpha_{31}$ \rf{02092017-man02-03}, \rf{02092017-man02-06}, has two solutions presented in \rf{02092017-man02-04}, \rf{02092017-man02-05}, where the vertex $V$ depends on the five variables $B_1$, $B_2$, $\alpha_{12}$, $\alpha_{23}$, $\alpha_{31}$ \rf{02092017-man02-08}.

\noindent \vbf) Using symmetry properties \rf{12092017-man02-01} of quantities $\betach_a$, $\Po^i$ \rf{21082017-man02-09} upon the replacement of external line indices of arbitrary spin massive fields, $1 \leftrightarrow 2$, we verify behavior of quantities in \rf{02092017-man02-06},\rf{02092017-man02-07} and \rf{02092017-man02-08-a9}-\rf{02092017-man02-08-a12} upon the replacement $1 \leftrightarrow 2$,
\be  \label{12092017-man02-04}
z_3 \leftrightarrow z_3\,, \quad  \nu_3 \leftrightarrow \nu_3\,, \quad X \leftrightarrow X\,,\quad  Y \leftrightarrow  Y\,,\quad Z \leftrightarrow Z\,, \quad W \leftrightarrow W\,.
\ee
Also, using \rf{12092017-man02-01} and \rf{12092017-man02-04}, we verify that all operators $U$ \rf{02092017-man02-08-a1}-\rf{02092017-man02-08-a8} are symmetric upon the replacement of external line indices of arbitrary spin massive fields,  $1 \leftrightarrow 2$,
\beq
\label{12092017-man02-05} & U_{\upsilon_3} \leftrightarrow U_{\upsilon_3}\,,\quad  U_3 \leftrightarrow U_3\,,\quad  U_\beta \leftrightarrow U_\beta\,,\quad  U_\zeta \leftrightarrow U_\zeta\,,\quad  U_B \leftrightarrow U_B\,,  &
\nonumber\\
& U_{z1} \leftrightarrow U_{z1}\,, \quad U_{z\nu} \leftrightarrow U_{z\nu}\,,\quad U_W \leftrightarrow U_W\,.  &
\eeq

\newsection{ \large Parity invariant cubic vertices for two continuous-spin  massless fields and one arbitrary spin massive field} \label{sec-06}

In this Section, we consider parity invariant cubic vertices for two continuous-spin  massless fields and one arbitrary spin massive field. Namely, using the shortcut $(0,\kappa)_\smCSF$ for continuous-spin massless field and the shortcut $(m,s)$ for spin-$s$ massive field with mass parameter $m$, we consider cubic interaction vertices for the following three fields:
\be \label{31082017-man02-01}
(0,\kappa_1)_\smCSF\hbox{-}(0,\kappa_2)_\smCSF\hbox{-}(m_3,s_3) \hspace{0.4cm}\hbox{ \small two  continuous-spin massless fields and one massive field}.\quad
\ee
Our notation in \rf{31082017-man02-01} implies that two continuous-spin massless fields denoted by $(0,\kappa_1)_\smCSF$ and $(0,\kappa_2)_\smCSF$ carry the respective external line indices $a=1,2$, while the mass-$m_3$, spin-$s_3$  massive field denoted by $(m_3,s_3)$ carries  external line index $a=3$.

The general solution to cubic vertex for fields in \rf{31082017-man02-01} takes the form (see Appendix D)
\beq
\label{03092017-man02-01} p_\smp3^-  & = & U_{\upsilon_1,\upsilon_2} U_{1,2} U_\beta U_B U_{z0} U_{z2} U_{z\nu} U_W V^{(8)}\,,
\\
\label{03092017-man02-02} && p_\smp3^-  = p_\smp3^- (\beta_a, B_a,\alpha_{aa+1}, \upsilon_1,\upsilon_2,\zeta_3)\,,
\\
\label{03092017-man02-03} && V^{(8)}  = V^{(8)}(B_a,\alpha_{aa+1})\,,
\eeq
where, in the solution \rf{03092017-man02-01}, we introduce new vertex denoted by $V^{(8)}$, while, in relations \rf{03092017-man02-02}, \rf{03092017-man02-03}, we show explicitly arguments of the generic vertex $p_\smp3^-$ and the new vertex $ V^{(8)}$. Before presenting operators $U$ appearing in \rf{03092017-man02-01}, we note that, for vertex $V^{(8)}$ \rf{03092017-man02-03}, we find four solutions which can be expressed in terms of the modified Bessel functions,
\beq
\label{03092017-man02-04} V^{(8)} &= & I_{\nu_1}(\sqrt{4z_1}) I_{\nu_2}(\sqrt{4z_2}) V\,,
\\
\label{03092017-man02-05} V^{(8)} &= & I_{\nu_1}(\sqrt{4z_1}) K_{\nu_2}(\sqrt{4z_2}) V\,,
\\
\label{03092017-man02-06} V^{(8)} &= & K_{\nu_1}(\sqrt{4z_1}) I_{\nu_2}(\sqrt{4z_2}) V\,,
\\
\label{03092017-man02-07} V^{(8)} &= & K_{\nu_1}(\sqrt{4z_1}) K_{\nu_2}(\sqrt{4z_2}) V\,,
\\
\label{03092017-man02-12} && V  = V(B_3,\alpha_{12},\alpha_{23},\alpha_{31}) \,,
\\
\label{03092017-man02-08} && z_1 = \frac{\kappa_1}{m_3^2} ( B_1 + \frac{\kappa_1}{2} )\,,
\\
\label{03092017-man02-09} && z_2 = \frac{\kappa_2}{m_3^2} ( - B_2 + \frac{\kappa_2}{2} )\,,
\\
\label{03092017-man02-10} && \nu_1 = N_{ \alpha_{12} } + N_{ \alpha_{31} } + \frac{d-4}{2}\,,
\\
\label{03092017-man02-11} && \nu_2 = N_{ \alpha_{12} } + N_{ \alpha_{23} } + \frac{d-4}{2}\,,
\eeq
where $B_a$, $N_{\alpha_{ab}}$ are defined in \rf{23092017-man02-04},\rf{23092017-man02-05}.  In \rf{03092017-man02-04}-\rf{03092017-man02-07}, in place of the $B_1$ and $B_2$, we use new respective variables $z_1$ and $z_2$ \rf{03092017-man02-08}, \rf{03092017-man02-09}. In relations \rf{03092017-man02-04}-\rf{03092017-man02-07}, the $I_\nu$ and $K_\nu$ stand for the modified Bessel functions. For the modified Bessel functions, we use conventions  in Ref.\cite{NIST}.

From \rf{03092017-man02-01}-\rf{03092017-man02-11}, we see that the general solution for the vertex $p_\smp3^-$, which depends on twelve variables \rf{03092017-man02-02}, is expressed in terms of the vertex $V$ which is arbitrary function of four variables \rf{03092017-man02-12}.  Note that vertex $V$ \rf{03092017-man02-12} is restricted to be expandable in the four variables $B_3$, $\alpha_{12}$, $\alpha_{23}$, $\alpha_{31}$. Thus, the general solution for vertex $p_\smp3^-$ \rf{03092017-man02-01} is expressed  in terms of the operators $U$, $\nu_1$, $\nu_2$ and the vertex $V$  \rf{03092017-man02-12}. Therefore all that remains to complete a description of the vertex $p_\smp3^-$ is to provide explicit expressions for the operators $U$. The operators $U$ entering our solution in \rf{03092017-man02-01} are given by
\beq
\label{03092017-man02-14} && U_{\upsilon_1,\upsilon_2} =  \upsilon_1^{N_1}\upsilon_2^{N_2}\,,
\\
\label{03092017-man02-15} && \hspace{1.3cm} N_1 = N_{B_1} + N_{\alpha_{12}} + N_{\alpha_{31}}\,,  \quad N_2 = N_{B_2} + N_{\alpha_{12}} + N_{\alpha_{23}}\,,
\\
\label{03092017-man02-16} && U_{1,2}  =  \Bigl( \frac{2^{N_1} \Gamma(N_1 + \frac{d-2}{2})}{\Gamma( N_1 + 1)} \frac{2^{N_2} \Gamma(N_2 + \frac{d-2}{2})}{\Gamma( N_2 + 1)} \Bigr)^{1/2}\,,
\\
\label{03092017-man02-17} && U_\beta  = \exp\bigl(  - \frac{\betach_1}{2\beta_1} \kappa_1 \partial_{B_1} - \frac{\betach_2}{2\beta_2} \kappa_2 \partial_{B_2} - \frac{\betach_3}{2\beta_3}  m_3 \zeta_3 \partial_{B_3} \bigr)\,,
\\
\label{03092017-man02-18} && U_\zeta  = \exp\Bigl( \frac{\zeta_3}{m_3} ( B_1 - \half \kappa_1  ) \partial_{\alpha_{31}} - \frac{\zeta_3}{m_3} ( B_2 + \half \kappa_2) \partial_{\alpha_{23}} \Bigr)
\\
\label{03092017-man02-19} && U_{z0} = \exp\bigl( \frac{2\kappa_1\kappa_2}{m_3^2}\partial_{\alpha_{12}} + \frac{2\kappa_1}{m_3^2} B_3\partial_{\alpha_{31}}  - \frac{2\kappa_2}{m_3^2} B_3\partial_{\alpha_{23}}\bigr) \,,
\\
\label{03092017-man02-20} && U_{z2} = \exp\bigl( \frac{2m_3^2}{\kappa_1\kappa_2} z_1 z_2 \partial_{\alpha_{12}} + \frac{2}{\kappa_1} z_1 B_3\partial_{\alpha_{31}}  - \frac{2}{\kappa_2} z_2 B_3\partial_{\alpha_{23}} \bigr) \,,
\\
\label{03092017-man02-21} && U_{z\nu} =  z_1^{-\nu_1 /2 } z_2^{- \nu_2/2 }\,,
\\
\label{03092017-man02-22} && U_W = \sum_{n=0}^\infty  U_n\,, \qquad  U_{n} = \sum_{m=0}^n U_{n-m,m} \,,
\\
\label{03092017-man02-23} && U_{n,m} = \frac{\Gamma(\nu_1+n)}{\Gamma(\nu_1+2n) } \sum_{k=0}^{\min(n,m)} \frac{1}{k!(n-k)!} W_{1,0}^{n-k} W_{1,1}^k  U_{0,m-k} \,,
\\
\label{03092017-man02-23-a1}&& U_{0,m} = \frac{\Gamma(\nu_2+m)}{ m!\Gamma(\nu_2+2m) } W_{0,1}^m \,,
\\
\label{03092017-man02-24} &&  \hspace{1.3cm}  W_{1,0} = 2 \alpha_{23} \partial_{\alpha_{12}} \partial_{\alpha_{31}}  - \frac{4}{m_3^2}B_3^2 \partial_{\alpha_{31}}^2\,,
\\
\label{03092017-man02-25} && \hspace{1.3cm}  W_{1,1} =  - 4 \partial_{\alpha_{12}}^2 \,,
\\
\label{03092017-man02-26} && \hspace{1.3cm}  W_{0,1}  =  2\alpha_{31} \partial_{\alpha_{12}} \partial_{\alpha_{23}}  - \frac{4}{m_3^2} B_3^2 \partial_{\alpha_{23}}^2\,,
\eeq
where $\betach_a$, $N_{B_a}$, $N_{\alpha_{ab}}$, $\nu_a$, $N_a$ are defined in \rf{23092017-man02-04}-\rf{23092017-man02-06}, while the $\Gamma$ \rf{03092017-man02-16}-\rf{03092017-man02-23-a1}  stands for the  Gamma-function.

Expressions given in \rf{03092017-man02-01}-\rf{03092017-man02-26} provide the complete description of the cubic vertex describing interaction of two continuous-spin massless fields with one infinite chain of massive fields. The chain of massive fields consists of every spin just once. Such chain of massive fields is described by ket-vector \rf{19082017-man02-16-a1}. To consider vertices for two continuous-spin massless fields and one arbitrary but fixed spin-$s_3$ massive field \rf{31082017-man02-01}, we note that the ket-vector for massive field $|\phi_{s_3}\rangle$ is a degree-$s_3$ homogeneous polynomial in the oscillators $\alpha_3^i$, $\zeta_3$ \rf{19082017-man02-17}. This implies that the vertices we are interested in must satisfy the equation
\be \label{04092017-man02-01-add}
( N_{\alpha_3} + N_{\zeta_3} - s_3 )|p_\smp3^-\rangle  = 0\,,
\ee
which tells us that the $|p_\smp3^-\rangle$ should be degree-$s_3$ homogeneous polynomial in the oscillators $\alpha_3^i$, $\zeta_3$. Using \rf{03092017-man02-01}, we verify that, in terms of vertex $V$ \rf{03092017-man02-04}-\rf{03092017-man02-12}, Eq.\rf{04092017-man02-01-add} takes the form
\be
\label{05092017-man02-01-add} (N_{B_3} + N_{ \alpha_{23} } + N_{ \alpha_{31} } - s_3) V =0 \,.
\ee
Vertex $V$ \rf{03092017-man02-12} is considered to be expandable in the variables $B_3$, $\alpha_{12}$, $\alpha_{23}$, $\alpha_{31}$. Therefore solution of Eq.\rf{05092017-man02-01-add} for the vertex $V$ can be labelled by $s_3$ and by three integers $n_1$, $n_2$, $n_3$. Using the notation $V_{s_3}(n_1,n_2,n_3)$ for vertex $V$ \rf{03092017-man02-12} that satisfies Eq.\rf{05092017-man02-01-add}, we find the general solution,
\beq
&& V =  V_{s_3}(n_1,n_2,n_3)\,,
\nonumber\\
\label{05092017-man02-03-add} && V_{s_3}(n_1,n_2,n_3)  = B_3^{s_3 - n_1 - n_2}  \alpha_{12}^{n_3} \alpha_{23}^{n_1} \alpha_{31}^{n_2} \,,
\\
&& n_1,n_2,n_3  =0,1,2,\ldots ,\infty\,.
\eeq
Integers $n_1$, $n_2$, $n_3$ in \rf{05092017-man02-03-add} are the freedom of our solution for vertices, i.e., these integers label all possible
cubic vertices that can be constructed for fields in \rf{31082017-man02-01}. In order for vertices \rf{05092017-man02-03-add} to be sensible, we should impose the following restrictions:
\be \label{05092017-man02-08-add}
n_1 + n_2 \leq s_3\,,  \qquad \qquad n_1,n_2,n_3 \in \No_0\,,
\ee
which amount to the requirement that the powers of variables $B_3$, $\alpha_{12}$,  $\alpha_{23}$,  $\alpha_{31}$ in \rf{05092017-man02-03-add} be non--negative integers.
Relations in \rf{03092017-man02-01},\rf{03092017-man02-04}-\rf{03092017-man02-12},\rf{05092017-man02-03-add}   and restrictions in \rf{05092017-man02-08-add} provide the complete description and classification of cubic vertices that can be constructed for two continuous-spin massless fields and one mass-$m_3$, spin-$s_3$ massive field.

The following remarks are in order.

\noindent \ibf)  Given $s_3$, the integers $n_1$ and $n_2$ take finite number of values \rf{05092017-man02-08-add}, while the integer $n_3$ takes infinite number of values, $n_3\in \No_0$ \rf{05092017-man02-08-add}. This implies that, given $s_3$, a number of cubic vertices that can be constructed for fields \rf{31082017-man02-01} is infinite.

\noindent \iibf)  From \rf{23082017-man02-10} and \rf{03092017-man02-08},\rf{03092017-man02-09}, we see that the variables $z_1$, $z_2$ are degree-1 polynomials in the momentum $\Po^i$. Taking this into account and using \rf{03092017-man02-04}-\rf{03092017-man02-07}, we see that vertex $p_\smp3^-$ \rf{03092017-man02-01} involves all positive powers of the momentum $\Po^i$. On the other hand, from \rf{05092017-man02-03-add}, \rf{05092017-man02-08-add}, we see that, given $s_3$,  vertex $V$ \rf{05092017-man02-03-add} is finite-order polynomial in the momentum $\Po^i$. Namely,  vertex $V$ \rf{05092017-man02-03-add} is a degree-$(s_3-n_1 -n_2)$ homogeneous polynomial in $\Po^i$.

\noindent \iiibf) Vertex $V$ \rf{05092017-man02-03-add} is a finite-order polynomial in the variables $B_3$, $\alpha_{12}$, $\alpha_{23}$, $\alpha_{31}$. We now show that vertex $p_\smp3^-$ \rf{03092017-man02-01} is also a finite-order polynomial in the variables $B_3$, $\alpha_{12}$, $\alpha_{23}$, $\alpha_{31}$. To this end we note that operators $\nu_1$, $\nu_2$ \rf{03092017-man02-10},\rf{03092017-man02-11} and operators $W_{1,0}$, $W_{1,1}$б $W_{0,1}$ \rf{03092017-man02-24}-\rf{03092017-man02-26} satisfy the following commutation relations
\beq
\label{08092017-man02-07} && [\nu_1,W_{1,0}] = - 2 W_{1,0}\,,
\\
\label{08092017-man02-08} && [\nu_2,W_{1,0}] = 0\,,
\\
\label{08092017-man02-09} && [\nu_1,W_{0,1}] = 0\,,
\\
\label{08092017-man02-10} && [\nu_2,W_{0,1}] = - 2 W_{0,1}\,,
\\
\label{08092017-man02-11} && [\nu_1,W_{1,1}] = - 2 W_{1,1}\,,
\\
\label{08092017-man02-12} && [\nu_2,W_{1,1}] = - 2 W_{1,1}\,,
\\
\label{08092017-man02-14} && [W_{1,0}, W_{0,1}] = (\nu_1-\nu_2) W_{1,1}\,.
\eeq
Using operator $U_{n,m}$ \rf{03092017-man02-23} and commutators  \rf{08092017-man02-07}-\rf{08092017-man02-12}, we find the commutators
\beq
\label{08092017-man02-15} && [\nu_1, U_{n,m}]  = -2n U_{n,m} \,,
\\
\label{08092017-man02-16} && [\nu_2, U_{n,m}]  = -2m U_{n,m} \,.
\eeq
Now using operator $U_n$ \rf{03092017-man02-22} and commutators \rf{08092017-man02-15}, \rf{08092017-man02-16}, we see that, for a sufficiently large number $N$, action of the operator $U_n$ on the vertex $V$ \rf{05092017-man02-03-add} gives zero for all $n\geq N$. This implies that an action of operator $U_W$ \rf{03092017-man02-22} on vertex $V$ \rf{05092017-man02-03-add} gives finite-order polynomial in the variables $B_3$, $\alpha_{12}$, $\alpha_{23}$, $\alpha_{31}$.
Taking this into account and using \rf{03092017-man02-14}-\rf{03092017-man02-21} it is clear that  vertex $p_\smp3^-$ \rf{03092017-man02-01} is also a finite-order polynomial in the variables $B_3$, $\alpha_{12}$, $\alpha_{23}$, $\alpha_{31}$. For the reader convenience, we note that operator $U_{n,m}$ \rf{03092017-man02-23} can equivalently be represented as
\beq
\label{08092017-man02-17} && U_{n,m} = \frac{\Gamma(\nu_2+m)}{\Gamma(\nu_2+2m) } \sum_{k=0}^{\min(n,m)} \frac{1}{k!(m-k)!} W_{0,1}^{m-k} W_{1,1}^k  U_{n-k,0} \,,
\\
\label{08092017-man02-18} && \hspace{1.3cm} U_{n,0} = \frac{\Gamma(\nu_1+n)}{ n!\Gamma(\nu_1+2n) } W_{1,0}^n\,.
\eeq

\noindent \ivbf) Four solutions for vertex $V^{(8)}$ \rf{03092017-man02-04}-\rf{03092017-man02-07} are obtained as follows. Equations \rf{23082017-man02-14},\rf{23082017-man02-15} lead to the following two second-order equations for the vertex $V^{(8)}$ (for details, see Appendix D):
\beq
\label{08092017-man02-03} && \bigl( 1 - (N_{z_1}+1)\partial_{z_1}  + \frac{\nu_1^2}{4z_1} \bigr) V^{(8)}=0\,, \qquad N_{z_1} \equiv z_1\partial_{z_1}\,,
\\
\label{08092017-man02-04} && \bigl( 1 - (N_{z_2}+1)\partial_{z_2}  + \frac{\nu_2^2}{4z_2} \bigr) V^{(8)}=0\,, \qquad N_{z_2} \equiv z_2\partial_{z_2}\,.
\eeq
Differential equations \rf{08092017-man02-03},\rf{08092017-man02-04} for the vertex $V^{(8)}$ depending on the six variables $z_1$, $z_2$, $B_3$, $\alpha_{12}$, $\alpha_{23}$, $\alpha_{31}$ \rf{03092017-man02-03}, \rf{03092017-man02-08}, \rf{03092017-man02-09} have four solutions presented in \rf{03092017-man02-04}, \rf{03092017-man02-05}, where the vertex $V$ depends on the four variables $B_3$, $\alpha_{12}$, $\alpha_{23}$, $\alpha_{31}$ \rf{03092017-man02-12}.

\noindent \vbf) Using symmetry properties \rf{12092017-man02-01} of quantities $\betach_a$, $\Po^i$ \rf{21082017-man02-09} upon the replacement of external line indices of continuous-spin massless fields, $1 \leftrightarrow 2$, we verify behavior of quantities in \rf{03092017-man02-08}-\rf{03092017-man02-11} and \rf{03092017-man02-24}-\rf{03092017-man02-26} upon the replacement $1 \leftrightarrow 2$,
\be  \label{12092017-man02-06}
z_1 \leftrightarrow z_2\,, \qquad \nu_1 \leftrightarrow \nu_2\,, \qquad W_{1,0} \leftrightarrow W_{0,1}\,,\qquad  W_{1,1} \leftrightarrow  W_{1,1}\,.\quad \qquad
\ee
Also, using \rf{12092017-man02-01} and \rf{12092017-man02-06}, we verify that all operators $U$ \rf{03092017-man02-14}-\rf{03092017-man02-22} are symmetric upon the replacement of external line indices of continuous-spin massless fields, $1 \leftrightarrow 2$,
\beq
\label{12092017-man02-07} & U_{\upsilon_1,\upsilon_2} \leftrightarrow U_{\upsilon_1,\upsilon_2}\,,\qquad  U_{1,2} \leftrightarrow U_{1,2}\,,\qquad  U_\beta \leftrightarrow U_\beta\,,\qquad  U_\zeta \leftrightarrow U_\zeta\,,\qquad  U_B \leftrightarrow U_B\,,  \qquad &
\nonumber\\
& U_{z1} \leftrightarrow U_{z1}\,, \qquad U_{z\nu} \leftrightarrow U_{z\nu}\,,\qquad U_W \leftrightarrow U_W\,. \qquad &
\eeq

\noindent \vibf) We see that vertices \rf{03092017-man02-04}-\rf{03092017-man02-09} are singular in the massless limit $m_3\rightarrow 0$, i.e., the massless limit of the cubic vertices describing interaction of two continuous-spin massless fields with one arbitrary spin massive field is problematic. In Appendix  F, by analysing equations for cubic vertices, we demonstrate explicitly that cubic vertices describing interaction of two continuous-spin massless fields with one arbitrary spin massless field are not consistent.

\newsection{ \large Conclusions}\label{sec-07}

In this paper, we applied the light-cone gauge formalism  to build
the parity invariant cubic interaction vertices for continuous-spin massless fields
and arbitrary spin massive fields. We considered two types of cubic vertices:
vertices describing interaction of one continuous-spin massless field with two arbitrary
spin massive fields and vertices describing interaction of two continuous-spin massless fields with one arbitrary spin massive field. We found the full lists of such vertices. Also we demonstrated that there are no cubic vertices describing consistent interaction of continuous-spin massless fields with arbitrary spin massless fields.  We expect that our results have the following interesting
applications and generalizations.

\noindent \ibf) In this paper, we studied interaction vertices for fields propagating in flat space and demonstrated that there are no cubic vertices describing consistent interaction of continuous-spin massless fields with arbitrary spin massless fields. Following ideas in Ref.\cite{Fradkin:1987ks}, we believe then that consistent interaction of continuous-spin fields with arbitrary spin massless fields can be constructed by considering fields propagating in AdS space. Metric-like gauge invariant Lagrangian formulation of continuous-spin free AdS field was developed in Refs.\cite{Metsaev:2016lhs,Metsaev:2017ytk}. We note however that, in the literature, there are many other interesting approaches which, upon a suitable generalization, could also be helpful for studying an interacting continuous-spin AdS field. For the reader convenience, we briefly review various approaches which could be used for the description of an interacting continuous-spin AdS field.
In the framework of frame-like approach, interacting higher-spin AdS fields are considered in  Refs.\cite{Vasilev:2011xf}, while, in the framework of ambient space metric-like approach, interacting higher-spin AdS fields are considered in Refs.\cite{Joung:2011ww}.%
\footnote{ In frame-like approach, equations of motion for continuous-spin field were studied in Ref.\cite{Ponomarev:2010st}.}
In this respect, recent interesting discussion of interacting higher-spin AdS fields may be found in Refs.\cite{Gelfond:2017wrh}.
In the frameworks of world-line particle and twistor-like approaches, higher-spin fields are considered, e.g., in Refs.\cite{Bonezzi:2017mwr} and \cite{Adamo:2016ple}. Interesting approach for analysing interacting conformal higher-spin fields in curved background is discussed in Ref.\cite{Grigoriev:2016bzl}. The light-cone gauge formulation of arbitrary spin free AdS fields was developed in Refs.\cite{Metsaev:1999ui,Metsaev:2002vr,Metsaev:2003cu}. It would be interesting to extend such formulation to the case of interacting continuous-spin fields. Extension of Hamiltonian form of AdS higher-spin fields dynamics \cite{Metsaev:2011iz}%
\footnote{ Alternative Hamiltonian formulations of AdS higher-spin field dynamics may be found in Refs.\cite{Vasiliev:1987hv}.
}
to the case of continuous-spin field could also be of some interest.%

\noindent \iibf) In this paper, we developed the light-cone gauge formulation for the bosonic continuous-spin fields. It would be interesting to extend a light-cone gauge formulation to the case of fermionic continuous-spin fields and apply such formulation for studying
vertices that describes interaction of fermionic continuous-spin fields with arbitrary spin massive fields as well as for studying supersymmetric continuous-spin field theories. Continuous-spin supermultiplets are studied Ref.\cite{Brink:2002zx,Zinoviev:2017rnj}. Recent interesting discussion of higher-spin supersymmetric theories may be found in Refs.\cite{Buchbinder:2015kca,Kuzenko:2016qwo}.

\noindent \iiibf) The methods for building $so(d-1,1)$ Lorentz covariant formulation of field dynamics  by using light-cone gauge $so(d-2)$ covariant formulation \cite{Siegel:1988yz} are most suitable for studying parity invariant vertices. Moreover such Lorentz covariant formulation turns out to be BRST gauge invariant. Therefore, we think that the parity invariant vertices obtained in this paper, can relatively straightforwardly be cast into BRST gauge invariant form. For example, the full list of the parity invariant light-cone gauge cubic vertices for arbitrary spin massless and massive fields in Ref.\cite{Metsaev:2005ar} has straightforwardly been cast into BRST gauge invariant form in Ref.\cite{Metsaev:2012uy}. BRST formulation of continuous-spin free field was discussed in Ref.\cite{Bengtsson:2013vra} and it was noted that such formulation has interesting interrelations with the formulations in terms of the unconstrained higher-spin gauge fields in Ref.\cite{Francia:2006hp}. Discussion of other interesting formulations in terms of unconstrained gauge fields can be found, e.g., in Refs.\cite{Sagnotti:2003qa}.
BRST formulations in terms of traceceless higher-spin gauge fields may be found in Refs.\cite{Metsaev:2015oza,Metsaev:2015yyv}. Use of BRST gauge fixed Lagrangian for the computation of partition functions of higher-spin fields is discussed in Refs.\cite{Metsaev:2014vda,Metsaev:2014iwa}.

\noindent \ivbf) In this paper, we studied totally symmetric continuous-spin fields.
As is known, string theory spectrum involves mixed-symmetry fields. Therefore from the perspective of studying interrelations between continuous-spin fields and string theory it seems reasonable to extend our study to the case of mixed-symmetry fields. We note, presently, even at the level of free fields, little is known about Lagrangian formulation of a continuous-spin mixed-symmetry fields. On the other hands, many interesting formulations for mixed-symmetry massless and massive higher-spin fields were developed (see, e.g., Refs.\cite{Boulanger:2008up}-\cite{Bastianelli:2014lia}). Extension of such formulations to continuous-spin fields could be also of some interest.

\noindent \vbf) In this paper, we studied cubic vertices for continuous-spin massless fields and arbitrary spin massive fields. Extension of our study to quartic vertices by using methods and approaches in  Refs.\cite{Metsaev:1991mt}-\cite{Dempster:2012vw} might contribute to better understanding of continuous-spin field dynamics.

\medskip

{\bf Acknowledgments}. Author cordially thanks Xavier Bekaert for his kind suggestion to use a light-cone gauge approach for the studying interacting continuous-spin fields. This work was supported by the Russian Science Foundation grant 14-42-00047.

\setcounter{section}{0}\setcounter{subsection}{0}
\appendix{ \large Notation and conventions  }

The vector indices of the $so(d-2)$ algebra take the values $i,j,k,l=1,\ldots ,d-2$. Creation operators $\alpha^i$, $\upsilon$, $\zeta$ and the respective annihilation operators $\alphab^i$, $\upsilonb$, $\zetab$  are referred to as oscillators in this paper. Commutation relations, the vacuum $|0\rangle $, and hermitian conjugation rules are fixed by the relations
\beq
\label{23092017-man02-01} && [ \bar\alpha^i,\alpha^j] = \delta^{ij}, \quad [\upsilonb,\upsilon]=1,  \quad [\zetab,\zeta]=1,  \quad \alphab^i |0\rangle = 0\,,\quad \upsilonb |0\rangle = 0\,,   \quad \zetab |0\rangle = 0\,,\qquad
\\
\label{23092017-man02-02} && \alpha^{i \dagger} = \alphab^i\,, \hspace{1.1cm} \upsilon^\dagger = \upsilonb\,, \hspace{1cm} \zeta^\dagger = \zetab\,.
\eeq
The oscillators $\alpha^i$, $\alphab^i$ and $\upsilon$, $\upsilonb$, $\zeta$, $\zetab$, transform under the respective vector and scalar representations of the $so(d-2)$ algebra. We use the following notation for the scalar product of the oscillators
\be \label{23092017-man02-03}
\alpha^2 \equiv \alpha^i\alpha^i\,,\qquad \alphab^2 \equiv \alphab^i \alphab^i\,,\qquad N_\alpha  \equiv \alpha^i \alphab^i\,, \qquad N_\zeta  \equiv \zeta \zetab\,,\qquad N_\upsilon  \equiv \upsilon \upsilonb\,.
\ee
Throughout this paper we use the following definitions for momentum $\Po^i$ and quantities $B_a$, $\alpha_{ab}$
\be \label{23092017-man02-04}
\Po^i \equiv \frac{1}{3}\sum_{a=1,2,3} \betach_a p_a^i\,, \qquad
\betach_a\equiv \beta_{a+1}-\beta_{a+2}\,, \qquad B_a \equiv \frac{\alpha_a^i \Po^i}{\beta_a}\,, \qquad \alpha_{ab} \equiv \alpha_a^i\alpha_b^i\,,
\ee
where $\beta_a\equiv \beta_{a+3}$. Various quantities constructed out of the $B_a$, $\alpha_{ab}$ and  derivatives of the $B_a$, $\alpha_{ab}$  are defined as
\beq
\label{23092017-man02-05} && N_{B_a} = B_a\partial_{B_a}\,,\quad N_{\alpha_{ab}} = \alpha_{ab}\partial_{\alpha_{ab}}\,,\quad \partial_{B_a} = \partial/\partial B_a\,, \qquad \partial_{\alpha_{ab}} = \partial/\partial \alpha_{ab}\,,
\\
\label{23092017-man02-06} && N_a = N_{B_a} + N_{\alpha_{aa+1}} + N_{\alpha_{a+2a}}\,,\qquad \nu_a =  N_{\alpha_{aa+1}} + N_{\alpha_{a+2a}} + \frac{d-4}{2}\,.
\eeq

\appendix{ \large Continuous-spin field in helicity basis}

Throughout this paper, we use the $so(d-2)$ covariant basis of light-cone gauge fields propagating in $R^{d-1,1}$ with arbitrary $d\geq 4$. For $d=4$, light-cone gauge fields can also be considered in a helicity basis. As a helicity basis is popular in many studies we decided, for the reader convenience, to work out light-cone gauge formulation of continuous-spin field in such basis. To discuss continuous-spin field in a helicity basis we introduce complex coordinates $x^\Rsm$, $x^\Lsm$ defined by the relations
\be \label{15092017-man02-01}
x^\Rsm
\equiv
\frac{1}{\sqrt{2}} (x^1 + \irm x^2)\,, \qquad
x^\Lsm
\equiv
\frac{1}{\sqrt{2}} (x^1 - \irm x^2)\,.
\ee
In the frame of the complex coordinates, a vector of the $so(2)$ algebra $X^i$ is decomposed as $X^i = X^\Rsm, X^\Lsm$, while a scalar product of $so(2)$ algebra vectors $X^i$, $Y^i$ is represented as $X^i Y^i = X^\Rsm Y^\Lsm + X^\Lsm Y^\Rsm$. We decompose the vector oscillators as $\alpha^i=\alpha^\Rsm,\alpha^\Lsm$ and, in place of the $so(2)$ covariant form of ket-vector $\phik$  \rf{19082017-man02-08}, we use a helicity basis ket-vector given by
\beq
\label{15092017-man02-02} && \hspace{-1cm} |\phi\rangle = |\phi_0\rangle + |\phi^\Rsm\rangle + |\phi^\Lsm\rangle \,,
\\
\label{15092017-man02-02-a1} && \hspace{-1cm} |\phi_0\rangle \equiv \phi_0(p)|0\rangle\,, \qquad |\phi^\Rsm\rangle = \sum_{n=1}^\infty \frac{ \upsilon^n \alpha_\Lsm^n }{ n! }\phi_n(p) |0\rangle \,,\qquad
|\phi^\Lsm \rangle = \sum_{n=1}^\infty \frac{ \upsilon^n \alpha_\Rsm^n }{ n! }\phi_{-n}(p) |0\rangle \,,\qquad
\eeq
$\alpha_\Lsm = \alpha^\Rsm$, $\alpha_\Rsm = \alpha^\Lsm$, where $\phi_0(p)$ is a scalar field, while $\phi_{\pm n}(p)$ are fields having helicities $\pm n$, $n=1,2,\ldots,\infty$.  We assume the following hermitian conjugation rules for the fields: $\phi_0^\dagger(p)=\phi_0(-p)$, $\phi_n^\dagger(p) = \phi_{-n}(-p)$. The oscillators and a vacuum $|0\rangle$ satisfy the relations
\be  \label{15092017-man02-03}
[\alphab^\Lsm,\alpha^\Rsm] = 1\,, \qquad [\alphab^\Rsm,\alpha^\Lsm] = 1\,, \qquad \alphab^\Rsm |0\rangle = 0 \qquad \alphab^\Lsm|0\rangle = 0 \,.
\ee
In the frame of the complex coordinates, the spin operator $M^i$ is decomposed as $M^i = M^\Rsm, M^\Lsm$, while the $so(2)$ algebra generator $M^{ij}=-M^{ji}$ is represented as $M^{\Rsm\Lsm}$. We now note that, in the frame of the complex coordinates, a realization of kinematical generators and dynamical generators \rf{20082017-man02-01}-\rf{20082017-man02-05} in terms of differential operators acting on ket-vector $|\phi\rangle$ \rf{15092017-man02-02} is given by
\beq
&& \hspace{-4cm}  \hbox{\it Kinematical generators}:
\nonumber\\
\label{15092017-man02-04}  && P^\Rsm=p^\Rsm\,,  \hspace{1.4cm}P^\Lsm=p^\Lsm\,,  \hspace{1.6cm} P^+=\beta\,,
\\
\label{15092017-man02-05} && J^{+\Rsm}=\partial_{p^\Lsm} \beta\,, \hspace{0.8cm} J^{+\Lsm}=\partial_{p^\Rsm} \beta\,,  \hspace{0.9cm} J^{+-} = \partial_\beta \beta\,,
\\
\label{15092017-man02-06} && J^{\Rsm\Lsm}=p^\Rsm \partial_{p^\Rsm} - p^\Lsm \partial_{p^\Lsm} + M^{\Rsm\Lsm}\,,
\\
&& \hspace{-4cm} \hbox{\it Dynamical generators}:
\nonumber\\
\label{15092017-man02-07} && P^- =  -\frac{2p^\Rsm p^\Lsm + m^2}{2\beta}\,,
\\
\label{15092017-man02-08} && J^{-\Rsm} = - \partial_\beta p^\Rsm + \partial_{p^\Lsm} P^- + \frac{1}{\beta}M^{\Rsm\Lsm} p^\Rsm + \frac{1}{\beta} M^\Rsm\,,
\\
\label{15092017-man02-09} && J^{-\Lsm} = - \partial_\beta p^\Lsm + \partial_{p^\Rsm} P^- - \frac{1}{\beta}M^{\Rsm\Lsm} p^\Lsm + \frac{1}{\beta} M^\Lsm\,,
\eeq
where we use the notation
\be \label{15092017-man02-10}
\beta\equiv p^+\,,\qquad
\partial_\beta\equiv \partial/\partial \beta\,, \quad
\partial_{p^\Rsm}\equiv \partial/\partial p^\Rsm\,, \qquad \partial_{p^\Lsm}\equiv \partial/\partial p^\Lsm\,.
\ee
The spin operators $M^{\Rsm\Lsm}$, $M^\Lsm$, and $M^\Lsm$ appearing in \rf{15092017-man02-06}-\rf{15092017-man02-09} are given by
\beq
\label{15092017-man02-11} && M^{\Rsm\Lsm} = N_R - N_L\,, \qquad N_\Rsm =  \alpha^\Rsm \alphab^\Lsm\,,\qquad N_\Lsm =  \alpha^\Lsm \alphab^\Rsm\,,
\\
\label{15092017-man02-12} &&  M^\Rsm  = g_\Lsm \alphab^\Rsm \upsilonb - \upsilon \alpha^\Rsm g_\Rsm \Pi^{\Rsm\Lsm}\,,
\\
\label{15092017-man02-14} &&  M^\Lsm  = g_\Rsm \alphab^\Lsm \upsilonb - \upsilon \alpha^\Lsm g_\Lsm \Pi^{\Lsm\Rsm}\,,
\eeq
where quantities $g_{\Rsm,\Lsm}$, $\Pi^{\Rsm\Lsm}$, and $\Pi^{\Lsm\Rsm}$ are defined by the relations
\beq
\label{15092017-man02-15} && g_\Rsm = \Bigl(\frac{ F_\Rsm }{ 2(N_\Rsm +1)^2 } \Bigr)^{1/2}\,, \hspace{2.2cm} g_\Lsm = \Bigl( \frac{ F_\Lsm }{ 2(N_\Lsm +1)^2 } \Bigr)^{1/2}\,,
\\
\label{15092017-man02-16} && F_\Rsm  =    \kappa^2 -  N_\Rsm (N_\Rsm + 1) m^2\,, \hspace{1.6cm} F_\Lsm  =    \kappa^2 -  N_\Lsm (N_\Lsm + 1) m^2\,,\qquad
\\
\label{15092017-man02-17} && \Pi^{\Rsm\Lsm} = 1 - \alpha^\Lsm \frac{1}{N_\Lsm+1}\alphab^\Rsm\,, \hspace{2cm} \Pi^{\Lsm\Rsm} = 1 - \alpha^\Rsm \frac{1}{N_\Rsm+1}\alphab^\Lsm\,.
\eeq
Relations in \rf{15092017-man02-04}-\rf{15092017-man02-17} provide the complete description of a realization for the generators of the Poincar\'e algebra $iso(3,1)$ in terms of differential operators acting on the ket-vector $|\phi\rangle$ \rf{15092017-man02-02}. To our knowledge, the realization of the operators $M^\Rsm$, $M^\Lsm$  for continuous-spin field in $R^{3,1}$ with arbitrary $\kappa^2 > 0$ and $m^2 <  0$ given in \rf{15092017-man02-12}-\rf{15092017-man02-17} has not been discussed in earlier literature.%
\footnote{ For continuous-spin massless field, $m=0$, the operators $M^\Rsm$, $M^\Lsm$ were discussed in Sec.2 in Ref.\cite{Brink:2002zx}.}

For the reader convenience, we note the following commutators for the spin operators:
\be
[M^\Rsm, M^\Lsm] = - m^2  M^{\Rsm\Lsm} \,, \qquad [M^{\Rsm\Lsm},M^\Rsm] = M^\Rsm\,, \qquad
[M^{\Rsm\Lsm}, M^\Lsm] = - M^\Lsm\,.
\ee
Action of the operators $\Pi^{\Rsm\Lsm}$, $\Pi^{\Lsm\Rsm}$ \rf{15092017-man02-17} on various ket-vectors defined in \rf{15092017-man02-02-a1} is given by
\beq
\label{15092017-man02-18} && \Pi^{\Rsm\Lsm}|\phi_0\rangle = |\phi_0\rangle\,,  \qquad \Pi^{\Rsm\Lsm}|\phi^\Rsm\rangle = |\phi^\Rsm\rangle\,, \qquad \Pi^{\Rsm\Lsm}|\phi^\Lsm\rangle = 0\,,
\\
\label{15092017-man02-19} && \Pi^{\Lsm\Rsm}|\phi_0\rangle = |\phi_0\rangle\,, \qquad \Pi^{\Lsm\Rsm}|\phi^\Rsm\rangle = 0 \,, \hspace{1.4cm} \Pi^{\Lsm\Rsm} |\phi^\Lsm\rangle = |\phi^\Lsm\rangle \,.
\eeq
We note the following helpful relations for the operators $\Pi^{\Rsm\Lsm}$, $\Pi^{\Lsm\Rsm}$ \rf{15092017-man02-17} and the oscillators
\beq
\label{15092017-man02-20} && \hspace{-0.8cm} \Pi^{\Rsm\Lsm}\Pi^{\Rsm\Lsm} = \Pi^{\Rsm\Lsm}\,, \qquad \Pi^{\Lsm\Rsm}\Pi^{\Lsm\Rsm} = \Pi^{\Lsm\Rsm}\,, \qquad \Pi^{\Rsm\Lsm}\Pi^{\Lsm\Rsm} = \Pi^{\Lsm\Rsm}  \Pi^{\Rsm\Lsm}\,, \qquad
\\
\label{15092017-man02-21} && \hspace{-0.8cm} \Pi^{\Rsm\Lsm}\alpha^\Lsm  = 0\,, \hspace{1.6cm}  \Pi^{\Lsm\Rsm} \alpha^\Rsm  = 0\,, \hspace{1.6cm} \alphab^\Rsm \Pi^{\Rsm\Lsm} = 0\,, \hspace{1.6cm}  \alphab^\Lsm \Pi^{\Lsm\Rsm}   = 0\,. \qquad
\eeq
Hermitian conjugation rules for various quantities above-defined are given by
\beq
\label{15092017-man02-22} && \alpha^{\Rsm\dagger} =\alphab^\Lsm\,, \qquad \alpha^{\Lsm\dagger} = \alphab^\Rsm\,,\qquad
\\
\label{15092017-man02-23} && N_\Rsm^\dagger = N_\Rsm\,, \qquad N_\Lsm^\dagger = N_\Lsm\,,\qquad \Pi^{\Rsm\Lsm\dagger} = \Pi^{\Rsm\Lsm}\,, \qquad \Pi^{\Lsm\Rsm\dagger} =
\Pi^{\Lsm\Rsm}\,.
\eeq

To quadratic order in fields, a field representation for generators of the Poincar\'e  algebra takes the form
\be \label{15092017-man02-24}
G_\smpt=\int \beta d\beta d^2p\,
\langle\phi(p)|
G_\smpt |\phi(p)\rangle\,,\qquad \langle\phi(p)|   \equiv |\phi(p)\rangle^\dagger\,,
\ee
where $G$ in \rf{15092017-man02-24}  are given in \rf{15092017-man02-04}-\rf{15092017-man02-09}.
The Poisson-Dirac commutator for fields entering ket-vector takes the form
\be \label{15092017-man02-25}
[\phi_{k'}(p')\,, \phi_{k''}(p'')]
\bigl|_{{\rm equal} \, x^+} =  \frac{1}{2\beta'} \delta(\beta'+\beta'')\delta^2(p'+p'') \delta_{k'+k'',0}\,, \qquad k',k''\in \Zo\,.
\ee

\appendix{ Derivation of cubic vertices $p_\smp3^-$ \rf{30082017-man02-01}, \rf{02092017-man02-01}}

Our aim in this Appendix is to outline some details of the derivation of cubic vertices  given in \rf{30082017-man02-01}, \rf{02092017-man02-01}. We would like to divide our derivation in ten steps which we now discuss in turn.

\noindent {\bf Step 1. Realization on $p_\smp3^-$}. First, we find realization of operators $G_{a,\Po^2}$, $G_\beta$ \rf{23082017-man02-11} on vertex $p_\smp3^-$ \rf{30082017-man02-02}. To this end we use $\Jbf^{-i\dagger}$ \rf{21082017-man02-25} with the spin operators $M_a^i$, $a=1,2$, for the arbitrary spin massive fields \rf{20082017-man02-20} and the spin operator $M_a^i$, $a=3$, for the continuous-spin massless field \rf{20082017-man02-14}-\rf{20082017-man02-18}. Doing so, we find that action of $\Jbf^{-i\dagger}$ \rf{21082017-man02-25} on $p_\smp3^-$ \rf{30082017-man02-02} can be cast into the form \rf{23082017-man02-11} with the following expressions for the $G_{a,\Po^2}$, $G_\beta$:
\beq
\label{09092017-man02-01} &&   \hspace{-1.5cm} G_{1,\Po^2} =   G_1\,, \qquad G_{2,\Po^2} =   G_2\,, \qquad G_{3,\Po^2} =   G_3 +  \Pbf^- \frac{2 \beta}{\beta_3^3} \alpha_3^i \frac{ g_{\upsilon_3} \partial_{\upsilon_3} }{2N_3 + d-2 } \partial_{B_3}^2\,,
\\
\label{09092017-man02-02}  G_1 &  =  &    (B_3 - \frac{\beta_1}{\beta_3}  g_{\upsilon_3} \partial_{\upsilon_3} )\partial_{\alpha_{31}}  - ( B_2 + \frac{\beta_1}{\beta_2} m_2 \zeta_2   ) \partial_{\alpha_{12}}
+ \half ( \frac{\betach_1}{ \beta_1} m_1^2 +  m_2^2) \partial_{B_1} + m_1\partial_{\zeta_1}\,,\qquad
\\
\label{09092017-man02-03} G_2 &  =  &   (B_1 - \frac{\beta_2}{\beta_1}  \zeta_1 m_1)\partial_{\alpha_{12}}
- ( B_3 + \frac{\beta_2}{\beta_3} g_{\upsilon_3} \partial_{\upsilon_3} ) \partial_{\alpha_{23}}
+ \half ( \frac{\betach_2}{ \beta_2} m_2^2 - m_1^2 ) \partial_{B_2} + m_2\partial_{\zeta_2}\,,
\\
\label{09092017-man02-04} G_3 &  =  &    (B_2 - \frac{\beta_3}{\beta_2} m_2 \zeta_2   )\partial_{\alpha_{2 3}}
-  ( B_1 + \frac{\beta_3}{\beta_1} m_1 \zeta_1 ) \partial_{\alpha_{31}}
+ \half (m_1^2 - m_2^2) \partial_{B_3} +  \upsilon_3 g_{\upsilon_3}
\nonumber\\
& + & \frac{ g_{\upsilon_3} \partial_{\upsilon_3}}{2N_3+d-2} \Bigl( \frac{2\beta_1}{\beta_3} B_1 \partial_{B_3} \partial_{\alpha_{31}} + \frac{2 \beta_2}{\beta_3} B_2 \partial_{B_3} \partial_{\alpha_{23}}
\nonumber\\
& + & 2 \alpha_{12} \partial_{\alpha_{31}}\partial_{\alpha_{23}}  +   \alpha_{11} \partial_{\alpha_{31}}^2 + \alpha_{22}\partial_{\alpha_{23}}^2
+ \frac{\beta}{\beta_3^2} \sum_{b=1,2} \frac{m_b^2}{\beta_b} \partial_{B_3}^2 \Bigr)\,,
\\
\label{09092017-man02-05} G_\beta &  =  &    -  \frac{1}{\beta}  \No_\beta -  \frac{1}{\beta_1^2} m_1\zeta_1  \partial_{B_1} - \frac{1}{\beta_2^2} m_2\zeta_2  \partial_{B_2}  - \frac{1}{\beta_3^2} g_{\upsilon_3} \partial_{\upsilon_3} \partial_{B_3}\,,
\eeq
where $g_\upsilon$ and $\betach_a$, $\No_\beta$  are given in \rf{20082017-man02-17} and \rf{21082017-man02-29} respectively, while the $B_a$,  $\alpha_{ab}$, $N_a$ are given in \rf{23092017-man02-04}-\rf{23092017-man02-06}. Using \rf{21082017-man02-34}, \rf{23082017-man02-11}, and operators $G_{a,\Po^2}$, $G_\beta$ \rf{09092017-man02-01}-\rf{09092017-man02-05}, we see that requiring the density $|j_\smp3^{-i}\rangle$ to respect equations \rf{21082017-man02-35-a3} amounts to the Eqs.\rf{23082017-man02-14},\rf{23082017-man02-15}.

\noindent {\bf Step 2. Realization on $V^{(1)}$}. At this step, we fix dependence of $p_\smp3^-$ \rf{30082017-man02-02} on the oscillator $\upsilon_3$. To this end, we note that in view of constraint \rf{19082017-man02-09}, vertex $p_\smp3^-$ \rf{30082017-man02-02} should satisfy the constraint
\be \label{09092017-man02-06}
(N_{\alpha_3} - N_{\upsilon_3}) | p_\smp3^- \rangle = 0 \,.
\ee
Introducing a vertex $V^{(1)}$ by the relations
\be \label{09092017-man02-07}
p_\smp3^-  =  U_{\upsilon_3} V^{(1)}\,, \qquad  U_{\upsilon_3} = \upsilon_3^{N_3}\,,  \qquad N_3 \equiv N_{B_3} + N_{ \alpha_{31} }  + N_{ \alpha_{23} }\,,
\ee
and using \rf{09092017-man02-06}, we find that the $V^{(1)}$ is independent of the oscillator $\upsilon_3$, i.e., we get
\be \label{09092017-man02-07-a1}
V^{(1)} = V^{(1)}(\beta_a, B_a, \alpha_{aa+1}\,,\zeta_1, \zeta_2)\,.
\ee
Using \rf{09092017-man02-07}, we find that, on vertex $V^{(1)}$ \rf{09092017-man02-07-a1}, operators $G_a$, $G_\beta$ \rf{09092017-man02-02}-\rf{09092017-man02-05} are realized as
\beq
\label{09092017-man02-08} && \hspace{-1cm} G_1 =   (B_3 - \frac{\beta_1}{\beta_3}  g_3(N_3+1) )\partial_{\alpha_{31}}  - ( B_2 + \frac{\beta_1}{\beta_2} m_2 \zeta_2   ) \partial_{\alpha_{12}} + \half ( \frac{\betach_1}{ \beta_1} m_1^2 +  m_2^2) \partial_{B_1} + m_1\partial_{\zeta_1}\,,\qquad
\\
\label{09092017-man02-09} &&  \hspace{-1cm} G_2  =    (B_1 - \frac{\beta_2}{\beta_1}  \zeta_1 m_1)\partial_{\alpha_{12}}
- ( B_3 + \frac{\beta_2}{\beta_3} g_3(N_3+1) ) \partial_{\alpha_{23}}
+ \half ( \frac{\betach_2}{ \beta_2} m_2^2 - m_1^2 ) \partial_{B_2} + m_2\partial_{\zeta_2}\,,
\\
\label{09092017-man02-10} && \hspace{-1cm} G_3  =    (B_2 - \frac{\beta_3}{\beta_2} m_2 \zeta_2   )\partial_{\alpha_{2 3}}
-  ( B_1 + \frac{\beta_3}{\beta_1} m_1 \zeta_1 ) \partial_{\alpha_{31}}  +  \half (m_1^2 - m_2^2) \partial_{B_3} +    g_3
\nonumber\\
& + & \frac{ g_3^\smone (N_3+2) }{2N_3+d-2} \Bigl( \frac{2\beta_1}{\beta_3} B_1 \partial_{B_3} \partial_{\alpha_{31}} + \frac{2 \beta_2}{\beta_3} B_2 \partial_{B_3} \partial_{\alpha_{23}}
\nonumber\\
&& +  2 \alpha_{12} \partial_{\alpha_{31}}\partial_{\alpha_{23}} +   \alpha_{11} \partial_{\alpha_{31}}^2 + \alpha_{22}\partial_{\alpha_{23}}^2
+ \frac{\beta}{\beta_3^3} \sum_{b=1,2} \frac{m_b^2}{\beta_b} \partial_{B_3}^2 \Bigr)\,,
\\
\label{09092017-man02-11} && \hspace{-1cm} G_\beta  =     -  \frac{1}{\beta}  \No_\beta -  \frac{1}{\beta_1^2} m_1\zeta_1  \partial_{B_1} - \frac{1}{\beta_2^2} m_2\zeta_2  \partial_{B_2}  - \frac{1}{\beta_3^2} g_3(N_3+1) \partial_{B_3}\,,
\\
\label{09092017-man02-12} && g_3 \equiv g_{\upsilon_3}\bigr|_{N_{\upsilon_3} \rightarrow N_3}\,, \qquad g_3^\smone \equiv g_{\upsilon_3}\bigr|_{N_{\upsilon_3} \rightarrow N_3+1}\,,
\eeq
where $N_{\upsilon_3} =\upsilon_3\partial_{\upsilon_3}$, $N_3$ is defined in \rf{09092017-man02-07}, while $g_\upsilon$ is given by \rf{20082017-man02-17},\rf{20082017-man02-18} for $m^2=0$.

\noindent {\bf Step 3. Realization on $V^{(2)}$}. We find it convenient to introduce a vertex $ V^{(2)}$ by the relations
\be \label{09092017-man02-14}
V^{(1)} = U_3 V^{(2)}\,, \qquad U_3  =  \Bigl( \frac{2^{N_3} \Gamma(N_3 + \frac{d-2}{2})}{\Gamma( N_3 + 1)} \Bigr)^{1/2}.
\ee
We note that the vertex $V^{(2)}$ depends on the same variables as vertex $V^{(1)}$ \rf{09092017-man02-07},
\be \label{09092017-man02-15}
V^{(2)} = V^{(2)}(\beta_a, B_a, \alpha_{aa+1}\,,\zeta_1, \zeta_2)\,.
\ee
On the vertex $V^{(2)}$, realization of $G_a$, $G_\beta$ \rf{09092017-man02-08}-\rf{09092017-man02-11} takes more convenient form given by
\beq
\label{10092017-man02-01} G_1 &  =  &    (B_3 - \frac{\beta_1}{\beta_3} \kappa_3  )\partial_{\alpha_{31}}  - ( B_2 + \frac{\beta_1}{\beta_2} m_2 \zeta_2   ) \partial_{\alpha_{12}} +  \half ( \frac{\betach_1}{ \beta_1} m_1^2 +  m_2^2 ) \partial_{B_1} + m_1\partial_{\zeta_1}\,,
\\
\label{10092017-man02-02} G_2 &  =  &   (B_1 - \frac{\beta_2}{\beta_1}  \zeta_1 m_1)\partial_{\alpha_{12}}
- ( B_3 + \frac{\beta_2}{\beta_3} \kappa_3 ) \partial_{\alpha_{23}}
+  \half ( \frac{\betach_2}{ \beta_2} m_2^2 - m_1^2 ) \partial_{B_2} + m_2\partial_{\zeta_2}\,,
\\
\label{10092017-man02-03} G_3 &  =  &    (B_2 - \frac{\beta_3}{\beta_2} m_2 \zeta_2   )\partial_{\alpha_{2 3}}
-  ( B_1 + \frac{\beta_3}{\beta_1} m_1 \zeta_1 ) \partial_{\alpha_{31}} +  \half ( m_1^2 - m_2^2) \partial_{B_3} +    \frac{g_3^2 (N_3+1)}{\kappa_3}
\nonumber\\
& + & \frac{ \kappa_3 }{2N_3+d-2} \Bigl( \frac{2\beta_1}{\beta_3} B_1 \partial_{B_3} \partial_{\alpha_{31}} + \frac{2 \beta_2}{\beta_3} B_2 \partial_{B_3} \partial_{\alpha_{23}}
\nonumber\\
& + & 2 \alpha_{12} \partial_{\alpha_{31}}\partial_{\alpha_{23}} +   \alpha_{11} \partial_{\alpha_{31}}^2 + \alpha_{22}\partial_{\alpha_{23}}^2
+ \frac{\beta}{\beta_3^2} \sum_{b=1,2} \frac{m_b^2}{\beta_b} \partial_{B_3}^2 \Bigr)\,,
\\
\label{10092017-man02-04} G_\beta &  =  &    -  \frac{1}{\beta}  \No_\beta -  \frac{1}{\beta_1^2} m_1\zeta_1  \partial_{B_1} - \frac{1}{\beta_2^2} m_2\zeta_2  \partial_{B_2}  - \frac{\kappa_3}{\beta_3^2} \partial_{B_3}\,.
\eeq
Namely, as compared with operators \rf{09092017-man02-08}-\rf{09092017-man02-11}, operators \rf{10092017-man02-01}-\rf{10092017-man02-04} do not involve square roots of $N_3$ \rf{09092017-man02-07}. Note, it is the quantities $g_3$, $g_3^\smone$ \rf{09092017-man02-12} that enter square roots of the $N_3$.

\noindent {\bf Step 4. Realization on $V^{(3)}$}. At this step, we fix dependence of  $V^{(2)}$ \rf{09092017-man02-15} on the momenta $\beta_1$, $\beta_2$, $\beta_3$. To this end, we use the transformation
\be \label{10092017-man02-05}
V^{(2)}  =  U_\beta  V^{(3)}\,, \qquad U_\beta   = \exp\bigl( - \frac{\betach_1}{2\beta_1}m_1\zeta_1 \partial_{B_1} - \frac{\betach_2}{2\beta_2} m_2\zeta_2 \partial_{B_2} - \frac{\betach_3}{2\beta_3} \kappa_3 \partial_{B_3} \bigr).
\ee
On vertex $V^{(3)}$ \rf{10092017-man02-05}, operators $G_a$, $G_\beta$ \rf{10092017-man02-01}-\rf{10092017-man02-04} are realized as%
\footnote{ To analyse equations like $G_aV=0$ we find it convenient to use equivalence classes for the operators $G_a$. Namely, the operators $G_a$ and $(2N_a+d-2)G_a$ are considered to be equivalent.
}
\beq
\label{10092017-man02-06} G_1 &  =  &    (B_3 + \half \kappa_3  )\partial_{\alpha_{31}}  - ( B_2 - \half m_2 \zeta_2   ) \partial_{\alpha_{12}} +  \half m_2^2 \partial_{B_1} + m_1\partial_{\zeta_1}\,,
\\
\label{10092017-man02-07} G_2 &  =  &   (B_1 + \half \zeta_1 m_1)\partial_{\alpha_{12}}
- ( B_3 - \half \kappa_3 ) \partial_{\alpha_{23}} -  \half  m_1^2 \partial_{B_2} + m_2\partial_{\zeta_2}\,,
\\
\label{10092017-man02-08} G_3 &  =  &    (B_2 + \half m_2 \zeta_2   )\partial_{\alpha_{23}}
-  ( B_1 - \half m_1 \zeta_1 ) \partial_{\alpha_{31}} + \half ( m_1^2 - m_2^2) \partial_{B_3}
\nonumber\\
& + & \frac{\kappa_3 }{2N_3+d-2} \Bigl( 1 - (B_1 + \frac{3}{2}m_1\zeta_1) \partial_{B_3} \partial_{\alpha_{31}} - (B_2 - \frac{3}{2}m_2\zeta_2) \partial_{B_3} \partial_{\alpha_{23}}
\nonumber\\
&  +  &  2\alpha_{12} \partial_{\alpha_{31}}\partial_{\alpha_{23}} +   \alpha_{11} \partial_{\alpha_{31}}^2 + \alpha_{22}\partial_{\alpha_{23}}^2
- \half (m_1^2 +  m_2^2) \partial_{B_3}^2 \Bigr),\qquad
\\
\label{10092017-man02-09} G_\beta &  =  &    -  \frac{1}{\beta}  \No_\beta \,.
\eeq
We find then the following equations for the vertex $V^{(3)}$,
\be \label{10092017-man02-10}
\sum_{a=1,2,3}\beta_a \partial_{\beta_a}  V^{(3)} = 0 \,, \qquad  \No_\beta  V^{(3)} = 0 \,.
\ee
Namely, the first equation in \rf{10092017-man02-10} is obtained from \rf{23082017-man02-17}, while the second equation in \rf{10092017-man02-10} is obtained by using \rf{23082017-man02-15} and \rf{10092017-man02-09}. Equations \rf{10092017-man02-10} imply that the vertex $V^{(3)}$ does not
dependent of the momenta $\beta_1$, $\beta_2$, $\beta_3$. Thus we have the following representation for the vertex $V^{(3)}$:
\be \label{10092017-man02-11}
V^{(3)} = V^{(3)}(B_a, \alpha_{aa+1}\,,\zeta_1, \zeta_2)\,.
\ee

\noindent {\bf Step 5. Realization on $V^{(4)}$}. We find it convenient to introduce a vertex $V^{(4)}$ by the relations
\be \label{10092017-man02-12}
V^{(3)}  =  U_\zeta V^{(4)}\,, \qquad U_\zeta = \exp\bigl( - \frac{m_2^2}{2m_1}  \zeta_1 \partial_{B_1} + \frac{m_1^2}{2m_2} \zeta_2 \partial_{B_2} \bigr).
\ee
We note that the vertex $V^{(4)}$ depends on the same variables as vertex $V^{(3)}$ \rf{10092017-man02-11},
\be \label{10092017-man02-14}
V^{(4)} = V^{(4)}(B_a, \alpha_{aa+1}\,,\zeta_1, \zeta_2)\,.
\ee
On the vertex $V^{(4)}$, realization of $G_a$ \rf{10092017-man02-06}-\rf{10092017-man02-08} takes more convenient form given by
\beq
\label{10092017-man02-15} G_1 &  =  &    (B_3 + \half \kappa_3  )\partial_{\alpha_{31}}  - ( B_2 -  \frac{m_1^2+m_2^2}{2m_2} \zeta_2   ) \partial_{\alpha_{12}} +  m_1\partial_{\zeta_1}\,,
\\
\label{10092017-man02-16} G_2 &  =  &   (B_1 +  \frac{m_1^2 + m_2^2}{2m_1} \zeta_1 )\partial_{\alpha_{12}}
- ( B_3 - \half \kappa_3 ) \partial_{\alpha_{23}} + m_2\partial_{\zeta_2}\,,
\\
\label{10092017-man02-17} G_3 &  =  &    (B_2  + \frac{m_2^2 - m_1^2}{2m_2} \zeta_2     )\partial_{\alpha_{23}} -  ( B_1  +  \frac{m_2^2 - m_1^2}{2m_1} \zeta_1 ) \partial_{\alpha_{31}}
+  \half ( m_1^2 - m_2^2) \partial_{B_3}
\nonumber\\
& + & \frac{\kappa_3 }{2N_3+d-2} \Bigl( 1  - (B_1  + \frac{3m_1^2 + m_2^2}{2m_1} \zeta_1) \partial_{B_3} \partial_{\alpha_{31}} - (B_2  - \frac{3m_2^2 +  m_1^2}{2m_2} \zeta_2) \partial_{B_3} \partial_{\alpha_{23}}
\nonumber\\
&  +  &  2\alpha_{12} \partial_{\alpha_{31}}\partial_{\alpha_{23}} + \alpha_{11} \partial_{\alpha_{31}}^2 + \alpha_{22}\partial_{\alpha_{23}}^2
- \half ( m_1^2 + m_2^2) \partial_{B_3}^2 \Bigr).\qquad
\eeq
Namely, as compared with \rf{10092017-man02-06},\rf{10092017-man02-07}, the $G_1$, $G_2$ in \rf{10092017-man02-15},\rf{10092017-man02-16} are independent of $\partial_{B_1}$, $\partial_{B_2}$.

\noindent {\bf Step 6. Realization on $V^{(5)}$}.  At this step, we fix dependence of  $V^{(4)}$ \rf{10092017-man02-14} on the variables $\zeta_1$, $\zeta_2$. To this end, we use the transformation
\beq
\label{10092017-man02-18} V^{(4)} & = & U_B V^{(5)}\,, \hspace{0.7cm}  U_B = \exp\Bigl( (\frac{\zeta_1}{m_1} B_2 - \frac{\zeta_2}{m_2} B_1 - \frac{m_1^2 + m_2^2}{2m_1m_2} \zeta_1 \zeta_2 )\partial_{\alpha_{12}}
\nonumber\\
&& \hspace{3.8cm} - \frac{\zeta_1}{m_1}(B_3 + \half \kappa_3)\partial_{\alpha_{31}} + \frac{\zeta_2}{m_2}(B_3 - \half \kappa_3)\partial_{\alpha_{23}}\Bigr).\qquad
\eeq
On vertex $V^{(5)}$ \rf{10092017-man02-18}, operators $G_a$, $G_\beta$ \rf{10092017-man02-15}-\rf{10092017-man02-17} are realized as
\beq
\label{10092017-man02-19} G_1 &  =  &      m_1\partial_{\zeta_1}\,, \qquad G_2   =     m_2\partial_{\zeta_2}\,,
\\
\label{10092017-man02-20} G_3 &  =  &\!\!\!    B_2\partial_{\alpha_{23}}
-   B_1   \partial_{\alpha_{31}} +  \half ( m_1^2 - m_2^2) \partial_{B_3}
\nonumber\\
& + &\!\!\! \frac{\kappa_3 }{2N_3+d-2} \Bigl( 1 -  B_1  \partial_{B_3} \partial_{\alpha_{31}} -  B_2   \partial_{B_3} \partial_{\alpha_{23}}
\nonumber\\
&  +  &\!\!\!  2 \alpha_{12} \partial_{\alpha_{31}}\partial_{\alpha_{23}} + (\alpha_{11} +\zeta_1^2) \partial_{\alpha_{31}}^2 + (\alpha_{22}+\zeta_2^2)\partial_{\alpha_{23}}^2
  - \half ( m_1^2 +  m_2^2) \partial_{B_3}^2 \Bigr).\qquad
\eeq
Using \rf{10092017-man02-19} in \rf{23082017-man02-14} when $a=1,2$, we see that the vertex $V^{(5)}$ is independent of the oscillators $\zeta_1$, $\zeta_2$,
\be \label{10092017-man02-21}
V^{(5)} = V^{(5)}(B_a, \alpha_{aa+1})\,.
\ee
Thus all that remains is to solve the equation
\be \label{10092017-man02-22}
G_3 V^{(5)}= 0
\ee
with $G_3$ as in \rf{10092017-man02-20}. Up to this point our treatment has been applied on an equal footing to vertices for massive fields having the same masses \rf{25082017-man02-01} and for massive
fields having different masses \rf{30082017-man02-08}. From now on, we separately consider Eq.\rf{10092017-man02-22} for vertices \rf{25082017-man02-01} and \rf{30082017-man02-08}.

\noindent {\bf Step 7. Case $m_1=m_2$. Realization on $V^{(5)}$}. We  now cast operator $G_3$ \rf{10092017-man02-20} and Eq.\rf{10092017-man02-22} into more convenient form.  First, we set $m_1=m_2=m$ in \rf{10092017-man02-20}. Second, in place of the $B_3$, we use variable $z_3$ \rf{30082017-man02-06}. Third, we note that in view of constraint \rf{19082017-man02-18}, contribution to commutators \rf{21082017-man02-20} of $(\alpha_{aa}+\zeta_a^2)$-terms, $a=1,2$, appearing in \rf{10092017-man02-20} cancel out. Therefore we drop down the just mentioned terms in $G_3$ \rf{10092017-man02-20}. Also we multiply Eq.\rf{10092017-man02-22} by the factor $(2N_3 + d-2)/\kappa_3$. Doing so, and using notation in \rf{30082017-man02-06},\rf{30082017-man02-07-a7}-\rf{30082017-man02-07-a11}, we verify that equation for the $V^{(5)}$ takes the form as in \rf{10092017-man02-22} with the following expression for $G_3$:
\be \label{10092017-man02-23}
G_3  = 1 - \partial_{z_3}^2  + X \partial_{z_3}  + Y (N_{z_3} + \nu_3)  + Z\,, \qquad N_{z_3} \equiv z_3\partial_{z_3}\,.
\ee

\noindent {\bf Step 8. Case $m_1=m_2$. Realization on $V^{(6)}$ }. Operator $G_3$ \rf{10092017-man02-23} is a second-order differential operator with respect to the variable $z_3$. We now use the transformation
\be \label{10092017-man02-24}
V^{(5)} =  U_{z2} V^{(6)}\,, \qquad U_{z2}= \exp\bigl( \half z_3 X + \frac{1}{4} z_3^2 Y \bigr)\,,
\ee
and verify that, on the vertex $V^{(6)}$, operator $G_3$ \rf{10092017-man02-23} is realized as
\be \label{10092017-man02-25}
G_3  = - \partial_{z_3}^2  + \Omega_3^2 \,,
\ee
where $\Omega_3^2$ is given in \rf{30082017-man02-07-a10}. Remarkable feature of $G_3$    \rf{10092017-man02-25} is that operator $\Omega_3^2$ \rf{30082017-man02-07-a10} is independent of the $z_3$. Solution to equation for $V^{(6)}$ \rf{08092017-man02-01} can be presented as in \rf{30082017-man02-04}, \rf{30082017-man02-05}.

\noindent {\bf Step 7. Case $m_1\ne m_2$. Realization on $V^{(5)}$.}
Here we cast operator $G_3$ \rf{10092017-man02-20} and Eq.\rf{10092017-man02-22} to more convenient form.  First, in place of $B_3$, we introduce variable $z_3$ \rf{02092017-man02-06}. Second, we note that in view of \rf{19082017-man02-18}, contribution to commutators \rf{21082017-man02-20} of $(\alpha_{aa}+\zeta_a^2)$-terms, $a=1,2$ appearing in \rf{10092017-man02-20} cancel out. Therefore we drop down the just mentioned terms in $G_3$ \rf{10092017-man02-20}. Also we multiply Eq.\rf{10092017-man02-22} by overall factor $(2N_3 + d-2)/\kappa_3$. Doing so, and using notation in \rf{02092017-man02-07} and \rf{02092017-man02-08-a9}-\rf{02092017-man02-08-a12} we verify that equation for the $V^{(5)}$ takes the form as in \rf{10092017-man02-22} with the following $G_3$
\be  \label{10092017-man02-26}
G_3   = 1 - (N_{z_3} + \nu_3 + 1)\partial_{z_3} +   X \partial_{z_3}  + (N_{z_3} + \nu_3 + 1)Y  + Z\,.
\ee

\noindent {\bf Step 8. Case $m_1\ne m_2$. Realization on $V^{(6)}$}. Operator $G_3$ \rf{10092017-man02-26} is a second-order differential operator with respect to the variable $z_3$. To simplify the $G_3$, we use the transformation
\be \label{10092017-man02-27}
V^{(5)} = U_{z1} V^{(6)}\,, \qquad  U_{z1} = e^{- X + z_3^{\vphantom{1pt}} Y }\,,
\ee
and verify that, on the vertex $V^{(6)}$, operator $G_3$ \rf{10092017-man02-26} is realized as
\be \label{10092017-man02-28}
G_3 = 1 -  (N_{z_3} + \nu_3 +1)\partial_{z_3} + W\,, \qquad W   =    Z + X Y\,.
\ee
Remarkable feature of $G_3$ \rf{10092017-man02-28} is that operator $W$ \rf{10092017-man02-28} is independent of the $z_3$. Therefore, as we demonstrate below,  the equation $G_3V^{(6)}=0$ can be solved in terms of the Bessel functions.

\noindent {\bf Step 9. Case $m_1\ne m_2$. Realization on $V^{(7)}$}. To get more convenient form  for $G_3$ \rf{10092017-man02-28}, we use the transformation
\be \label{10092017-man02-29}
V^{(6)} = U_{z\nu} V^{(7)}\,, \qquad U_{z\nu} = z_3^{-\nu_3/2}\,,
\ee
and verify that, on the vertex $V^{(7)}$, operator $G_3$ \rf{10092017-man02-28} is realized as
\be \label{10092017-man02-29-a1}
G_3    =   1 - (N_{z_3} + 1)\partial_{z_3}   + \frac{\Omega_3^2}{4z_3}\,, \qquad \Omega_3^2 = \nu_3^2 + 4 W\,.
\ee
Solution to equation $G_3 V^{(7)} = 0$ with $G_3$ as in \rf{10092017-man02-29-a1} can be expressed as
\be  \label{10092017-man02-30}
V^{(7)} = I_{\Omega_3}(\sqrt{4z_3})V'\,, \qquad K_{\Omega_3}(\sqrt{4z_3})V'\,, \qquad V'= V'(B_1, B_2, \alpha_{12}, \alpha_{23}, \alpha_{31}).
\ee
where $I_{\Omega_3}$, $K_{\Omega_3}$ are the modified Bessel functions. Operator $\nu_3$  \rf{02092017-man02-07} entering $\Omega_3^2$ \rf{10092017-man02-29-a1} is diagonal on vertex $V'$ \rf{10092017-man02-30}, while the operator $\Omega_3$ is not diagonal.

\noindent {\bf Step 10. Case $m_1\ne m_2$. Realization on $V^{(8)}$.} Our aim is to diagonalize operator $\Omega_3$ \rf{10092017-man02-29-a1}. To this end we use the transformation
\be \label{10092017-man02-31}
V^{(7)} = U_W V^{(8)} \qquad U_W  = \sum_{m=0}^\infty \frac{\Gamma(\nu_3 + m)}{m!\Gamma(\nu_3 + 2m)} W^m\,.
\ee
Using the relation
\be \label{10092017-man02-32}
(\nu_3^2 + 4W)U_W = U_W\nu_3^2\,,
\ee
we see that, on the vertex $V^{(8)}$, operator $\Omega_3^2$ \rf{10092017-man02-29-a1} becomes diagonal and operator $G_3$ \rf{10092017-man02-29-a1} takes the form
\be \label{10092017-man02-33}
G_3 =  1 - (N_{z_3} + 1)\partial_{z_3}  + \frac{\nu_3^2}{4z_3}\,.
\ee
Solutions to equation $G_3V^{(8)}=0$ with $G_3$ as in \rf{10092017-man02-33} are given in \rf{02092017-man02-04}-\rf{02092017-man02-08}. Thus we see that the vertex $p_\smp3^-$ takes the form given in \rf{02092017-man02-01}-\rf{02092017-man02-08-a12}.

\appendix{ Derivation of cubic vertex $p_\smp3^-$ \rf{03092017-man02-01}}

In this Appendix, we outline some details of the derivation of cubic vertex \rf{03092017-man02-01}. We divide our derivation in ten steps which we now discuss in turn.

{\bf Step 1. Realization on $p_\smp3^-$}. First, we find realization of operators $G_{a,\Po^2}$, $G_\beta$ \rf{23082017-man02-11} on vertex $p_\smp3^-$ \rf{03092017-man02-01}. To this end we use $\Jbf^{-i\dagger}$ \rf{21082017-man02-25} with the spin operators $M_a^i$, $a=1,2$, related to the continuous-spin massless fields \rf{20082017-man02-14}-\rf{20082017-man02-18}, $m=0$, and the spin operator $M_a^i$, $a=3$, related to the arbitrary spin massive field \rf{20082017-man02-20}. Doing so, we find that action of $\Jbf^{-i\dagger}$ \rf{21082017-man02-25} on $p_\smp3^-$ \rf{03092017-man02-01} can be cast into the form \rf{23082017-man02-11} with the following expressions for the $G_{a,\Po^2}$, $G_\beta$:
\beq
\label{11092017-man02-01} &&   \hspace{-1.5cm} G_{a,\Po^2} =   G_a +  \Pbf^- \frac{2 \beta}{\beta_a^3} \alpha_a^i \frac{ g_{\upsilon_a} \partial_{\upsilon_a} }{2N_a + d-2 } \partial_{B_a}^2\,,\qquad a=1,2;
\qquad G_{3,\Po^2} =   G_3\,,
\\
\label{11092017-man02-02} \hspace{-0.7cm} G_1 &  =  & (B_3 - \frac{\beta_1}{\beta_3}  m_3 \zeta_3 ) \partial_{\alpha_{31}}  - (B_2  \partial_{\alpha_{12}} + \frac{\beta_1}{\beta_2} g_{\upsilon_2} \partial_{\upsilon_2}) \partial_{\alpha_{12}}  -  \half m_3^2 \partial_{B_1} + \upsilon_1 g_{\upsilon_1}
\nonumber\\
\hspace{-0.7cm} & + &     \frac{ g_{\upsilon_1} \partial_{\upsilon_1}}{2N_1+d-2} \Bigl(\frac{ 2 \beta_2}{\beta_1} B_2 \partial_{B_1} \partial_{\alpha_{12}} + \frac{ 2 \beta_3}{\beta_1} B_3 \partial_{B_1} \partial_{\alpha_{31}}  +   2\alpha_{23} \partial_{\alpha_{12}}\partial_{\alpha_{31}}
\nonumber\\
& + & \alpha_{33} \partial_{\alpha_{31}}^2 + \frac{\beta_2 m_3^2 }{\beta_1} \partial_{B_1}^2 \Bigr) \,,
\\
\label{11092017-man02-03} \hspace{-0.7cm} G_2 &  =  &   ( B_1 - \frac{\beta_2}{\beta_1} g_{\upsilon_1} \partial_{\upsilon_1} ) \partial_{\alpha_{12}} -  ( B_3 + \frac{\beta_2}{\beta_3}  m_3 \zeta_3 ) \partial_{\alpha_{23}}  + \half m_3^2\partial_{B_2} + \upsilon_2 g_{\upsilon_2}
\nonumber\\
\hspace{-0.7cm} & + &       \frac{g_{\upsilon_2} \partial_{\upsilon_2}}{2N_2+d-2} \Bigl(  \frac{ 2 \beta_3}{\beta_2} B_3 \partial_{B_2} \partial_{\alpha_{23}}  + \frac{ 2 \beta_1}{\beta_2} B_1 \partial_{B_2} \partial_{\alpha_{12}}   +  2 \alpha_{31} \partial_{\alpha_{12}}\partial_{\alpha_{23}}
\nonumber\\
& + & \alpha_{33} \partial_{\alpha_{23}}^2
+ \frac{\beta_1 m_3^2}{\beta_2} \partial_{B_2}^2 \Bigr)  \,,
\\
\label{11092017-man02-04} \hspace{-0.7cm} G_3 &  =  & ( B_2 - \frac{\beta_3}{\beta_2} g_{\upsilon_2} \partial_{\upsilon_2} ) \partial_{\alpha_{23}} -  ( B_1 + \frac{\beta_3}{\beta_1} g_{\upsilon_1} \partial_{\upsilon_1} ) \partial_{\alpha_{31}} +   \frac{\betach_3}{ 2 \beta_3} m_3^2  \partial_{B_3} + m_3 \partial_{\zeta_3}\,,
\\
\label{11092017-man02-05}  \hspace{-0.7cm} G_\beta &  =  &   - \frac{1}{\beta}\No - \frac{1}{\beta_1^2} g_{\upsilon_1} \partial_{\upsilon_1} \partial_{B_1} - \frac{1}{\beta_2^2} g_{\upsilon_2} \partial_{\upsilon_2} \partial_{B_2} - \frac{1}{\beta_3^2}  m_3 \zeta_3 \partial_{B_3} \,,
\eeq
where $g_\upsilon$ is given in \rf{20082017-man02-17},\rf{20082017-man02-18} for $m=0$, while $\betach_a$, $\No_\beta$ are given in  \rf{21082017-man02-29}. The $B_a$,  $\alpha_{ab}$, $N_a$ are defined in \rf{23092017-man02-04}-\rf{23092017-man02-06}.
Using \rf{21082017-man02-34}, \rf{23082017-man02-11}, and explicit form of operators $G_{a,\Po^2}$, $G_\beta$ \rf{11092017-man02-01}-\rf{11092017-man02-05}, we see that requiring the density $|j_\smp3^{-i}\rangle$ to respect equations \rf{21082017-man02-35-a3} amounts to the equations
\rf{23082017-man02-14},\rf{23082017-man02-15}.

\noindent {\bf Step 2. Realization on $V^{(1)}$}. Here, we find dependence of $p_\smp3^-$ \rf{03092017-man02-02} on the oscillators $\upsilon_1$ and $\upsilon_2$. To this end, we note that in view of constraint \rf{19082017-man02-09}, vertex $p_\smp3^-$ \rf{03092017-man02-02} should satisfy the constraints
\be \label{11092017-man02-06}
(N_{\alpha_a} - N_{\upsilon_a}) | p_\smp3^- \rangle = 0 \,, \qquad a=1,2.
\ee
Introducing a vertex $V^{(1)}$ by the relations
\be \label{11092017-man02-07}
p_\smp3^- = U_{\upsilon_1,\upsilon_2}  V^{(1)}\,, \quad U_{\upsilon_1,\upsilon_2} =  \upsilon_1^{N_1}\upsilon_2^{N_2}\,, \quad N_1 = N_{B_1} + N_{\alpha_{12}} + N_{\alpha_{31}}\,,  \quad N_2 = N_{B_2} + N_{\alpha_{12}} + N_{\alpha_{23}}\,,
\ee
we verify that Eqs.\rf{11092017-man02-06} imply that the vertex $V^{(1)}$ does not depend on the $\upsilon_1$, $\upsilon_2$, i.e., we get
\be \label{11092017-man02-07-a1}
V^{(1)} = V^{(1)}(\beta_a, B_a, \alpha_{aa+1}\,,\zeta_3)\,.
\ee
Using \rf{11092017-man02-07}, we find that, on vertex $V^{(1)}$ \rf{11092017-man02-07-a1}, operators $G_a$, $G_\beta$ \rf{11092017-man02-02}-\rf{11092017-man02-05} are realized as
\beq
\label{11092017-man02-08} \hspace{-0.7cm} G_1 &  =  & (B_3 - \frac{\beta_1}{\beta_3}  m_3 \zeta_3 ) \partial_{\alpha_{31}}  - (B_2   + \frac{\beta_1}{\beta_2} g_2 (N_2+1)) \partial_{\alpha_{12}} -  \half m_3^2 \partial_{B_1} + g_1
\nonumber\\
\hspace{-0.7cm} & + &     \frac{ g_1^\smone (N_1+2)}{2N_1+d-2} \Bigl( \frac{ 2 \beta_2}{\beta_1} B_2 \partial_{B_1} \partial_{\alpha_{12}} + \frac{ 2 \beta_3}{\beta_1} B_3 \partial_{B_1} \partial_{\alpha_{31}}  +   2\alpha_{23} \partial_{\alpha_{12}}\partial_{\alpha_{31}}
\nonumber\\
& + & \alpha_{33} \partial_{\alpha_{31}}^2 + \frac{\beta_2 m_3^2 }{\beta_1} \partial_{B_1}^2 \Bigr) \,,
\\
\label{11092017-man02-09} \hspace{-0.7cm} G_2 &  =  &   ( B_1 - \frac{\beta_2}{\beta_1} g_1 (N_1+1) ) \partial_{\alpha_{12}} -  ( B_3 + \frac{\beta_2}{\beta_3}  m_3 \zeta_3 ) \partial_{\alpha_{23}} +  \half m_3^2 \partial_{B_2} +  g_2
\nonumber\\
\hspace{-0.7cm} & + &       \frac{g_2^\smone (N_2+2)}{2N_2 + d-2} \Bigl(  \frac{ 2 \beta_3}{\beta_2} B_3 \partial_{B_2} \partial_{\alpha_{23}}  + \frac{ 2 \beta_1}{\beta_2} B_1 \partial_{B_2} \partial_{\alpha_{12}}   +  2 \alpha_{31} \partial_{\alpha_{12}}\partial_{\alpha_{23}}
\nonumber\\
& + & \alpha_{33} \partial_{\alpha_{23}}^2 + \frac{\beta_1 m_3^2 }{\beta_2} \partial_{B_2}^2 \Bigr)  \,,
\\
\label{11092017-man02-10} \hspace{-0.9cm} G_3 &  =  & ( B_2 - \frac{\beta_3}{\beta_2} g_2 (N_2+1) ) \partial_{\alpha_{23}} -  ( B_1 + \frac{\beta_3}{\beta_1} g_1 (N_1+1) ) \partial_{\alpha_{31}}
  +  \frac{\betach_3}{ 2\beta_3} m_3^2  \partial_{B_3} + m_3 \partial_{\zeta_3}\,,\qquad
\\
\label{11092017-man02-11} \hspace{-0.7cm} G _\beta &  =  &   - \frac{1}{\beta}\No - \frac{1}{\beta_1^2} g_1 (N_1+1) \partial_{B_1} - \frac{1}{\beta_2^2} g_2 (N_2+1) \partial_{B_2} - \frac{1}{\beta_3^2}  m_3 \zeta_3 \partial_{B_3} \,,
\\
\label{11092017-man02-12} && \hspace{-1.6cm}  g_1 \equiv g_{\upsilon_1}\bigr|_{N_{\upsilon_1} \rightarrow N_1}\,, \quad
g_2 \equiv g_{\upsilon_2}\bigr|_{N_{\upsilon_2} \rightarrow N_2}\,,
\quad g_1^\smone \equiv g_{\upsilon_1}\bigr|_{N_{\upsilon_1} \rightarrow N_1+1}\,, \quad g_2^\smone \equiv g_{\upsilon_2}\bigr|_{N_{\upsilon_2} \rightarrow N_2+1}\,.
\eeq

\noindent {\bf Step 3. Realization on $V^{(2)}$}. We find it convenient to introduce a vertex $ V^{(2)}$ by the relations
\beq
\label{11092017-man02-14} && V^{(1)}= U_{1,2} V^{(2)}\,, \qquad U_{1,2}  =  \Bigl( \frac{2^{N_1} \Gamma(N_1 + \frac{d-2}{2})}{\Gamma( N_1 + 1)} \frac{2^{N_2} \Gamma(N_2 + \frac{d-2}{2})}{\Gamma( N_2 + 1)} \Bigr)^{1/2}\,,
\\
\label{11092017-man02-15} && V^{(2)} = V^{(2)}(\beta_a, B_a, \alpha_{aa+1}\,,\zeta_3)\,,
\eeq
where, in \rf{11092017-man02-15}, we show that vertex $V^{(2)}$ depends on the same variables as vertex $V^{(1)}$ \rf{11092017-man02-07}. On the vertex $V^{(2)}$, realization of $G_a$, $G_\beta$ \rf{11092017-man02-08}-\rf{11092017-man02-11} takes more convenient form given by
\beq
\label{11092017-man02-16} \hspace{-0.7cm} G_1 &  =  & (B_3 - \frac{\beta_1}{\beta_3}  m_3 \zeta_3 ) \partial_{\alpha_{31}}  - (B_2   + \frac{\beta_1}{\beta_2} \kappa_2) \partial_{\alpha_{12}} - \half m_3^2 \partial_{B_1} + \frac{1}{\kappa_1} g_1^2(N_1+1)
\nonumber\\
\hspace{-0.7cm} & + &     \frac{ \kappa_1}{2N_1+d-2} \Bigl( \frac{ 2 \beta_2}{\beta_1} B_2 \partial_{B_1} \partial_{\alpha_{12}} + \frac{ 2 \beta_3}{\beta_1} B_3 \partial_{B_1} \partial_{\alpha_{31}}  +   2\alpha_{23} \partial_{\alpha_{12}}\partial_{\alpha_{31}}
\nonumber\\
& + & \alpha_{33} \partial_{\alpha_{31}}^2 + \frac{\beta_2 m_3^2}{\beta_1} \partial_{B_1}^2 \Bigr) \,,
\\
\label{11092017-man02-17} \hspace{-0.7cm} G_2 &  =  &   ( B_1 - \frac{\beta_2}{\beta_1} \kappa_1 ) \partial_{\alpha_{12}} -  ( B_3 + \frac{\beta_2}{\beta_3}  m_3 \zeta_3 ) \partial_{\alpha_{23}} + \half  m_3^2   \partial_{B_2} +  \frac{1}{\kappa_2} g_2^2 (N_2+1)
\nonumber\\
\hspace{-0.7cm} & + &       \frac{\kappa_2}{2N_2+d-2} \Bigl(  \frac{ 2 \beta_3}{\beta_2} B_3 \partial_{B_2} \partial_{\alpha_{23}}  + \frac{ 2 \beta_1}{\beta_2} B_1 \partial_{B_2} \partial_{\alpha_{12}}   +  2 \alpha_{31} \partial_{\alpha_{12}}\partial_{\alpha_{23}}
\nonumber\\
& + & \alpha_{33} \partial_{\alpha_{23}}^2 + \frac{\beta_1 m_3^2}{\beta_2} \partial_{B_2}^2 \Bigr)  \,,
\\
\label{11092017-man02-18} \hspace{-0.7cm} G_3 &  =  & ( B_2 - \frac{\beta_3}{\beta_2} \kappa_2) \partial_{\alpha_{23}} -  ( B_1 + \frac{\beta_3}{\beta_1} \kappa_1  ) \partial_{\alpha_{31}} +  \frac{\betach_3}{ 2 \beta_3} m_3^2  \partial_{B_3} + m_3 \partial_{\zeta_3}\,,
\\
\label{11092017-man02-19} \hspace{-0.7cm} G _\beta &  =  &   - \frac{1}{\beta}\No - \frac{1}{\beta_1^2} \kappa_1 \partial_{B_1} - \frac{1}{\beta_2^2} \kappa_2 \partial_{B_2} - \frac{1}{\beta_3^2}  m_3 \zeta_3 \partial_{B_3} \,.
\eeq
Namely, as compared with operators \rf{11092017-man02-08}-\rf{11092017-man02-11}, operators \rf{11092017-man02-16}-\rf{11092017-man02-19} do not involve square roots of the operators $N_1$, $N_2$ \rf{11092017-man02-07}. Note, it is the quantities $g_a$, $g_a^\smone$, $a=1,2$ \rf{11092017-man02-12} that enter square roots of the $N_1$, $N_2$.

\noindent {\bf Step 4. Realization on $V^{(3)}$}.  At this step, we fix dependence of  $V^{(2)}$ \rf{11092017-man02-15} on the momenta $\beta_1$, $\beta_2$, $\beta_3$. To this end, using notation in \rf{21082017-man02-09}, we introduce the transformation
\be \label{11092017-man02-19-a1}
V^{(2)}  = U_\beta  V^{(3)}\,, \qquad U_\beta  = \exp\bigl(  - \frac{\betach_1}{2\beta_1} \kappa_1 \partial_{B_1} - \frac{\betach_2}{2\beta_2} \kappa_2 \partial_{B_2} - \frac{\betach_3}{2\beta_3}  m_3 \zeta_3 \partial_{B_3} \bigr).
\ee
On vertex $V^{(3)}$ \rf{11092017-man02-19-a1}, operators $G_a$, $G_\beta$  \rf{11092017-man02-16}-\rf{11092017-man02-19} are realized as
\beq
\label{11092017-man02-20} \hspace{-0.7cm} G_1 &  =  & (B_3 + \half  m_3 \zeta_3 ) \partial_{\alpha_{31}}  - (B_2   - \half \kappa_2) \partial_{\alpha_{12}} - \half  m_3^2 \partial_{B_1}
\nonumber\\
\hspace{-0.7cm} & + &     \frac{ \kappa_1}{2N_1 + d-2} \Bigl( 1  - ( B_2 + \frac{3}{2}\kappa_2) \partial_{B_1} \partial_{\alpha_{12}} - (B_3 - \frac{3}{2}m_3\zeta_3) \partial_{B_1} \partial_{\alpha_{31}}  +   2\alpha_{23} \partial_{\alpha_{12}}\partial_{\alpha_{31}}
\nonumber\\
& + & \alpha_{33} \partial_{\alpha_{31}}^2 -\half  m_3^2 \partial_{B_1}^2 \Bigr),
\\
\label{11092017-man02-21} \hspace{-0.7cm} G_2 &  =  &   ( B_1 + \half \kappa_1 ) \partial_{\alpha_{12}} -  ( B_3 - \half  m_3 \zeta_3 ) \partial_{\alpha_{23}} + \half  m_3^2  \partial_{B_2}
\nonumber\\
\hspace{-0.7cm} & + &       \frac{\kappa_2}{2N_2 +d-2} \Bigl(  1 - ( B_3 + \frac{3}{2} m_3 \zeta_3) \partial_{B_2} \partial_{\alpha_{23}}  - ( B_1  - \frac{3}{2} \kappa_1) \partial_{B_2} \partial_{\alpha_{12}}   +  2 \alpha_{31} \partial_{\alpha_{12}}\partial_{\alpha_{23}}
\nonumber\\
& + & \alpha_{33} \partial_{\alpha_{23}}^2  -\half  m_3^2 \partial_{B_2}^2 \Bigr),
\\
\label{11092017-man02-22} \hspace{-0.7cm} G_3 &  =  & ( B_2 + \half \kappa_2) \partial_{\alpha_{23}} -  ( B_1 - \half \kappa_1  ) \partial_{\alpha_{31}}   + m_3 \partial_{\zeta_3}\,,
\\
\label{11092017-man02-23} \hspace{-0.7cm} G_\beta &  =  &   - \frac{1}{\beta}\No\,.
\eeq
We find then the following equations for the vertex $V^{(3)}$,
\be \label{11092017-man02-24}
\sum_{a=1,2,3}\beta_a \partial_{\beta_a}  V^{(3)} = 0 \,, \qquad  \No_\beta  V^{(3)} = 0 \,.
\ee
The first equation in \rf{11092017-man02-24} is obtained from \rf{23082017-man02-17}, while the second equation in \rf{11092017-man02-24} is obtained by using \rf{23082017-man02-15} and \rf{11092017-man02-23}. Equations \rf{11092017-man02-24} imply that the vertex $V^{(3)}$  \rf{11092017-man02-19-a1} is independent of the momenta $\beta_1$, $\beta_2$, $\beta_3$. Thus, we have the following representation for the $V^{(3)}$:
\be \label{11092017-man02-25}
V^{(3)} = V^{(3)}(B_a, \alpha_{aa+1}\,,\zeta_3)\,.
\ee

\noindent {\bf Step 5. Realization on $V^{(4)}$}.  At this step, we fix dependence of  $V^{(3)}$ \rf{11092017-man02-25} on the oscillator $\zeta_3$. To this end, we use the transformation
\be \label{11092017-man02-26}
V^{(3)} = U_\zeta  V^{(4)}\,, \qquad U_\zeta  = \exp\Bigl( \frac{\zeta_3}{m_3} ( B_1 - \half \kappa_1  ) \partial_{\alpha_{31}} - \frac{\zeta_3}{m_3} ( B_2 + \half \kappa_2) \partial_{\alpha_{23}}  \Bigr).
\ee
On vertex $V^{(4)}$ \rf{11092017-man02-26}, operators $G_a$, $G_\beta$ \rf{11092017-man02-20}-\rf{11092017-man02-22} are realized as
\beq
\label{11092017-man02-27} \hspace{-0.7cm} G_1 &  =  & B_3\partial_{\alpha_{31}}  - (B_2   - \half \kappa_2) \partial_{\alpha_{12}} - \half m_3^2 \partial_{B_1}
\nonumber\\
\hspace{-0.7cm} & + &     \frac{ \kappa_1}{2N_1 + d-2} \Bigl( 1 - ( B_2 + \frac{3}{2}\kappa_2) \partial_{B_1} \partial_{\alpha_{12}} - B_3 \partial_{B_1} \partial_{\alpha_{31}}  +   2\alpha_{23} \partial_{\alpha_{12}}\partial_{\alpha_{31}}
\nonumber\\
& + & (\alpha_{33} + \zeta_3^2) \partial_{\alpha_{31}}^2  - \half m_3^2 \partial_{B_1}^2 \Bigr),
\\
\label{11092017-man02-28} \hspace{-0.7cm} G_2 &  =  &   ( B_1 + \half \kappa_1 ) \partial_{\alpha_{12}} -   B_3 \partial_{\alpha_{23}} + \half m_3^2 \partial_{B_2}
\nonumber\\
\hspace{-0.7cm} & + &       \frac{\kappa_2}{2N_2 +d-2} \Bigl(  1 - B_3 \partial_{B_2} \partial_{\alpha_{23}}  - ( B_1  - \frac{3}{2} \kappa_1) \partial_{B_2} \partial_{\alpha_{12}}   +  2 \alpha_{31} \partial_{\alpha_{12}}\partial_{\alpha_{23}}
\nonumber\\
& + & (\alpha_{33} + \zeta_3^2) \partial_{\alpha_{23}}^2  - \half  m_3^2 \partial_{B_2}^2 \Bigr),
\\
\label{11092017-man02-29} \hspace{-0.7cm} G_3 & = &  m_3\partial_{\zeta_3}\,.
\eeq
Using \rf{11092017-man02-29} in \rf{23082017-man02-14} for $a=3$, we see that the vertex $V^{(4)}$ is independent of the oscillator $\zeta_3$,
\be \label{11092017-man02-30}
V^{(4)} = V^{(4)}(B_a, \alpha_{aa+1})\,.
\ee
Thus all that remains is to solve the equations
\be \label{11092017-man02-31}
G_1 V^{(4)}= 0\,, \qquad G_2 V^{(4)}= 0\,,
\ee
where $G_1$, $G_2$ take the form given in \rf{11092017-man02-27},\rf{11092017-man02-28}.

\noindent {\bf Step 6. Realization on $V^{(4)}$}. Here we cast $G_1$, $G_2$ \rf{11092017-man02-27},\rf{11092017-man02-28}, and Eq.\rf{11092017-man02-31} to more convenient form.  First, in place of $B_1$, $B_2$ \rf{23082017-man02-10}, we introduce variables $z_1$, $z_2$ \rf{03092017-man02-08}, \rf{03092017-man02-09}. Second, we note that in view of \rf{19082017-man02-18}, contribution of $(\alpha_{33}+\zeta_3^2)$-terms \rf{11092017-man02-27},\rf{11092017-man02-28} to commutators \rf{21082017-man02-20} cancel out. Therefore we drop down the just mentioned terms in $G_1$, $G_2$ \rf{11092017-man02-27}, \rf{11092017-man02-28}. Also we multiply first and second equations in \rf{11092017-man02-31} by the respective overall factors $(2N_1 + d-2)/\kappa_1$ and $(2N_2 + d-2)/\kappa_2$. Doing so, and using notation in \rf{03092017-man02-08}-\rf{03092017-man02-11}, we verify that equations for $V^{(4)}$ take the form as in \rf{11092017-man02-31} with the following $G_1$, $G_2$:
\beq
\label{11092017-man02-32} G_1 & = & 1 - (N_{z_1}  +  \nu_1 + 1)\partial_{z_1} +  \frac{2}{\kappa_1} ( N_{z_1} +  \nu_1 + 1 ) B_3\partial_{\alpha_{31}} - \frac{2\kappa_1}{m_3^2} B_3 \partial_{z_1} \partial_{\alpha_{31}}
\nonumber\\
& + & 2\alpha_{23} \partial_{\alpha_{12}} \partial_{\alpha_{31}} + \frac{2m_3^2}{\kappa_1\kappa_2}( N_{z_1}  +  \nu_1 + 1 )z_2  \partial_{\alpha_{12}} -  \frac{ 2 \kappa_1 \kappa_2}{m_3^2}  \partial_{z_1} \partial_{\alpha_{12}}\,,
\\
\label{11092017-man02-33} G_2 & = & 1 - (N_{z_2} +  \nu_2 + 1 )\partial_{z_2} -  \frac{2}{\kappa_2} ( N_{z_2}  +  \nu_2 + 1 ) B_3\partial_{\alpha_{23}} + \frac{ 2\kappa_2}{m_3^2} B_3 \partial_{z_2} \partial_{\alpha_{23}}
\nonumber\\
& + & 2\alpha_{31} \partial_{\alpha_{12}} \partial_{\alpha_{23}} + \frac{2m_3^2}{\kappa_1\kappa_2}( N_{z_2}  +  \nu_2 + 1 ) z_1 \partial_{\alpha_{12}}  - \frac{ 2\kappa_1 \kappa_2}{m_3^2} \partial_{z_2} \partial_{\alpha_{12}} \,,
\eeq
where $N_{z_a}=z_a\partial_{z_a}$. The $G_1$, $G_2$ \rf{11092017-man02-32},\rf{11092017-man02-33} are 2nd-order differential operators wrt $z_1$, $z_2$. Our next aim is to simplify operators $G_1$, $G_2$ \rf{11092017-man02-32},\rf{11092017-man02-33}.

\noindent {\bf Step 7. Realization on $V^{(5)}$ }.  To simplify $G_1$, $G_2$, \rf{11092017-man02-32},\rf{11092017-man02-33}, we use the transformation
\be \label{11092017-man02-34}
V^{(4)}  =   U_{z0}  V^{(5)}\,, \qquad
U_{z0} =  \exp\bigl( \frac{2\kappa_1\kappa_2}{m_3^2}\partial_{\alpha_{12}} + \frac{2\kappa_1}{m_3^2} B_3\partial_{\alpha_{31}}  - \frac{2\kappa_2}{m_3^2} B_3\partial_{\alpha_{23}}\bigr),
\ee
and verify that, on the vertex $V^{(5)}$, operators $G_1$, $G_2$ \rf{11092017-man02-32},\rf{11092017-man02-33} are realized as
\beq
 \label{11092017-man02-35} G_1 &= & 1 - (N_{z_1} +   \nu_1 + 1 )\partial_{z_1} +   \frac{2}{\kappa_1} ( N_{z_1} +  \nu_1 + 1 ) B_3\partial_{\alpha_{31}}  +   2\alpha_{23} \partial_{\alpha_{12}} \partial_{\alpha_{31}}
\nonumber\\
& + & \frac{2m_3^2}{\kappa_1\kappa_2}( N_{z_1} +  \nu_1 + 1 )z_2  \partial_{\alpha_{12}} -  4z_2 \partial_{\alpha_{12}}^2 - \frac{4}{\kappa_2} z_2 B_3 \partial_{\alpha_{12}}\partial_{\alpha_{31}}  - \frac{4}{m_3^2}B_3^2 \partial_{\alpha_{31}}^2\,,\qquad
\\
\label{11092017-man02-35-a1} G_2 &= & 1 - (N_{z_2}  +  \nu_2 + 1 )\partial_{z_2} +  \frac{2}{\kappa_2} ( N_{z_2} +  \nu_2 + 1 ) B_3\partial_{\alpha_{23}}  +   2\alpha_{31} \partial_{\alpha_{12}} \partial_{\alpha_{23}}
\nonumber\\
& + & \frac{2m_3^2}{\kappa_1\kappa_2}( N_{z_2} +  \nu_2 + 1 )z_1  \partial_{\alpha_{12}}
-  4z_1 \partial_{\alpha_{12}}^2 - \frac{4}{\kappa_1} z_1 B_3 \partial_{\alpha_{12}}\partial_{\alpha_{23}}  - \frac{4}{m_3^2} B_3^2 \partial_{\alpha_{23}}^2\,.
\eeq
As compared with \rf{11092017-man02-32}, \rf{11092017-man02-33}, operators $G_1$, $G_2$ in \rf{11092017-man02-35}, \rf{11092017-man02-35-a1} do  not involve $\partial_{z_a}\partial_{\alpha_{12}}$-terms and $B_3\partial_{z_a}\partial_{\alpha_{a3}}$-terms, $a=1,2$.

\noindent {\bf Step 8. Realization on $V^{(6)}$}. To get more simple form for $G_1$,$G_2$ \rf{11092017-man02-35},\rf{11092017-man02-35-a1}, we use the transformation
\be \label{11092017-man02-36}
V^{(5)}  =   U_{z2}  V^{(6)}\,, \qquad
U_{z2}  =   \exp\bigl( \frac{2m_3^2}{\kappa_1\kappa_2} z_1 z_2 \partial_{\alpha_{12}} + \frac{2}{\kappa_1} z_1 B_3\partial_{\alpha_{31}}  - \frac{2}{\kappa_2} z_2 B_3\partial_{\alpha_{23}} \bigr) \,,
\ee
and verify that, on the vertex $V^{(6)}$, operators $G_1$, $G_2$ \rf{11092017-man02-35},\rf{11092017-man02-35-a1} are realized as
\beq
\label{11092017-man02-37} G_1  & = &   1 - (N_{z_1} +  \nu_1 + 1 )\partial_{z_1} +   2\alpha_{23} \partial_{\alpha_{12}} \partial_{\alpha_{31}} - 4z_2 \partial_{\alpha_{12}}^2   - \frac{4}{m_3^2}B_3^2 \partial_{\alpha_{31}}^2\,,
\\
\label{11092017-man02-38} G_2  & = & 1 - (N_{z_2} +  \nu_2 + 1 )\partial_{z_2}
+ 2\alpha_{31} \partial_{\alpha_{12}} \partial_{\alpha_{23}}
- 4z_1 \partial_{\alpha_{12}}^2   - \frac{4}{m_3^2} B_3^2 \partial_{\alpha_{23}}^2\,.
\eeq
As compared with \rf{11092017-man02-35},\rf{11092017-man02-35-a1}, operators \rf{11092017-man02-37},\rf{11092017-man02-38} do  not involve $N_{z_a}\partial_{\alpha_{12}}$-terms, $a=1,2$.

\noindent {\bf Step 9. Realization on $V^{(7)}$}. In order to remove the dependence of the $G_1$ on the variable $z_2$ \rf{11092017-man02-37} and the dependence of the $G_2$ on the variable $z_1$ \rf{11092017-man02-38} we use the transformation
\be \label{11092017-man02-39}
V^{(6)} =   U_{z\nu}  V^{(7)}\,, \qquad U_{z\nu}  =   z_1^{-\nu_1 /2 } z_2^{- \nu_2/2 }\,,
\ee
and verify that, on the vertex $V^{(7)}$, operators $G_1$, $G_2$ \rf{11092017-man02-37},\rf{11092017-man02-38} are realized as
\beq
\label{11092017-man02-40} && G_1   =  1 - (N_{z_1}+1)\partial_{z_1}   + \frac{\Omega_1^2}{4z_1}\,, \qquad   \Omega_1^2 = \nu_1^2 + 4  W_{1,0} + 4 W_{1,1}\,,
\\
\label{11092017-man02-41} && G_2   =  1 - ( N_{z_2} +1)\partial_{z_2}  + \frac{\Omega_2^2}{4z_2}\,, \hspace{1cm} \Omega_2^2 = \nu_2^2 + 4 W_{0,1} + 4 W_{1,1}\,,
\eeq
where $W_{1,0}$, $W_{0,1}$, $W_{1,1}$ are defined in \rf{03092017-man02-24}-\rf{03092017-man02-26}.

\noindent {\bf Step 10. Realization on $V^{(8)}$}. Using commutators \rf{08092017-man02-07}-\rf{08092017-man02-14}, we verify that $\Omega_1^2$, $\Omega_2^2$ \rf{11092017-man02-40},\rf{11092017-man02-41} are commuting, $[\Omega_1^2,\Omega_2^2] = 0$. We then find a transformation that casts the $\Omega_1^2$, $\Omega_2^2$ into a diagonal form. Namely, we introduce  the transformation
\be \label{11092017-man02-42}
V^{(7)}  =  U_W  V^{(8)}\,,
\ee
where operator $U_W$ is defined in \rf{03092017-man02-22}-\rf{03092017-man02-26}.  Using the relations
\be \label{11092017-man02-43}
\Omega_1^2 U_W = U_W \nu_1^2 \,, \qquad \Omega_2^2 U_W = U_W \nu_2^2 \,,
\ee
we see that, on the vertex $V^{(8)}$, operators $\Omega_1^2$, $\Omega_2^2$ \rf{11092017-man02-40}, \rf{11092017-man02-41} become diagonal, i.e., on the vertex $V^{(8)}$, operators $G_1$, $G_2$ \rf{11092017-man02-40},\rf{11092017-man02-41} are realized as
\be  \label{11092017-man02-44}
G_1   =   1 - (N_{z_1}+1)\partial_{z_1}    + \frac{\nu_1^2}{4z_1}\,, \qquad  G_2  =  1 - ( N_{z_2} +1)\partial_{z_2}   + \frac{\nu_2^2}{4z_2}\,.
\ee
Solutions to equations $G_1V^{(8)}=0$, $G_2V^{(8)}=0$ with $G_1$, $G_2$ as in \rf{11092017-man02-44}, are given in \rf{03092017-man02-04}-\rf{03092017-man02-07}. Thus, we see that the vertex $p_\smp3^-$ takes the form given in \rf{03092017-man02-01}-\rf{03092017-man02-26}.

\appendix{ \large Analysis of equations to vertices for one continuous-spin massless field and two massless fields}

In this Appendix, we prove that one continuous-spin massless field and two arbitrary spin massless fields have no consistent cubic interaction vertices. We divide our proof into the five steps.

\noindent {\bf Step 1. Realization on $p_\smp3^-$}. Using the shortcut $(0,\kappa)_\smCSF$ for  continuous-spin massless field and the shortcut $(0,s)$ for spin-$s$  massless field we consider the cubic vertex for the following three fields:
\be  \label{14092017-man02-45}
(0,s_1)\hbox{-}(0,s_2)\hbox{-}(0,\kappa_3)_\smCSF \hspace{0.8cm} \hbox{\small one continuous-spin massless field and two massless fields}.\quad
\ee
Our notation in \rf{14092017-man02-45} implies that spin-$s_1$ and spin-$s_2$ massless fields carry external line indices $a=1,2$, while the continuous-spin massless field corresponds to $a=3$. Using notation in \rf{23082017-man02-10}, we note that a general form of parity invariant vertex that respect $J^{ij}$-symmetries is given  by
\be  \label{14092017-man02-46}
p_\smp3^- = p_\smp3^-(\beta_a, B_a\,, \alpha_{aa+1}\,,\upsilon_3)\,.
\ee
To find realization of $G_{a,\Po^2}$, $G_\beta$ \rf{23082017-man02-11} on $p_\smp3^-$ \rf{14092017-man02-46} we use $\Jbf^{-i\dagger}$ \rf{21082017-man02-25}, where operators $M_a^i$, $a=1,2$, for arbitrary spin massless fields are equal to zero, \rf{20082017-man02-21}, while operator $M_a^i$, $a=3$, for continuous-spin  massless field can be read from \rf{20082017-man02-14}-\rf{20082017-man02-18} for $m^2=0$. Doing so, we find that action of $\Jbf^{-i\dagger}$ \rf{21082017-man02-25} on $p_\smp3^-$ \rf{14092017-man02-46} can be cast into the form \rf{23082017-man02-11} with the following $G_{a,\Po^2}$, $G_\beta$:
\beq
\label{13092017-man02-01} &&   \hspace{-1.5cm} G_{1,\Po^2} =   G_1\,, \qquad G_{2,\Po^2} =   G_2\,, \qquad G_{3,\Po^2} =   G_3 +  \Pbf^- \frac{2 \beta}{\beta_3^3} \alpha_3^i \frac{ g_{\upsilon_3} \partial_{\upsilon_3} }{2N_3 + d-2 } \partial_{B_3}^2\,,
\\
\label{13092017-man02-02}  G_1 &  =  &    (B_3 - \frac{\beta_1}{\beta_3}  g_{\upsilon_3} \partial_{\upsilon_3} )\partial_{\alpha_{31}}  -   B_2  \partial_{\alpha_{12}}\,,
\\
\label{13092017-man02-03} G_2 &  =  &   B_1  \partial_{\alpha_{12}}
- ( B_3 + \frac{\beta_2}{\beta_3} g_{\upsilon_3} \partial_{\upsilon_3} ) \partial_{\alpha_{23}}\,,
\\
\label{13092017-man02-04} G_3 &  =  &    B_2 \partial_{\alpha_{2 3}}
-   B_1  \partial_{\alpha_{31}} +  \upsilon_3 g_{\upsilon_3}
\nonumber\\
& + & \frac{ g_{\upsilon_3} \partial_{\upsilon_3}}{2N_3+d-2} \Bigl( \frac{2\beta_1}{\beta_3} B_1 \partial_{B_3} \partial_{\alpha_{31}} + \frac{2 \beta_2}{\beta_3} B_2 \partial_{B_3} \partial_{\alpha_{23}} + 2 \alpha_{12} \partial_{\alpha_{31}}\partial_{\alpha_{23}}   \Bigr)\,,
\\
\label{13092017-man02-05} G_\beta &  =  &    -  \frac{1}{\beta}  \No_\beta  - \frac{1}{\beta_3^2} g_{\upsilon_3} \partial_{\upsilon_3} \partial_{B_3}\,,
\eeq
where $g_\upsilon$ is given in \rf{20082017-man02-17},\rf{20082017-man02-18} for $m=0$, while $\betach_a$, $\No_\beta$ are given in  \rf{21082017-man02-29}. The $B_a$,  $\alpha_{ab}$, $N_a$ are defined in \rf{23092017-man02-04}-\rf{23092017-man02-06}.
Using \rf{21082017-man02-34}, \rf{23082017-man02-11}, and operators $G_{a,\Po^2}$, $G_\beta$ \rf{13092017-man02-01}-\rf{13092017-man02-05}, we see that requiring the density $|j_\smp3^{-i}\rangle$ to respect equations \rf{21082017-man02-35-a3} amounts to the Eqs.\rf{23082017-man02-14},\rf{23082017-man02-15}.

\noindent {\bf Step 2. Realization on $V^{(1)}$}. At this step, we fix dependence of $p_\smp3^-$ \rf{14092017-man02-46} on the oscillator $\upsilon_3$. To this end, we note that in view of constraint \rf{19082017-man02-09}, vertex $p_\smp3^-$ \rf{14092017-man02-46} should satisfy the constraint
\be \label{13092017-man02-06}
(N_{\alpha_3} - N_{\upsilon_3}) | p_\smp3^- \rangle = 0 \,.
\ee
Introducing a vertex $V^{(1)}$ by the relations
\be \label{13092017-man02-07}
p_\smp3^-  =  U_{\upsilon_3} V^{(1)}\,, \qquad  U_{\upsilon_3} = \upsilon_3^{N_3}\,,  \qquad N_3 \equiv N_{B_3} + N_{ \alpha_{31} }  + N_{ \alpha_{23} }\,,
\ee
we obtain from Eq.\rf{13092017-man02-06} that the vertex $V^{(1)}$ is independent of the oscillator $\upsilon_3$, i.e., we get
\be \label{13092017-man02-07-a1}
V^{(1)} = V^{(1)}(\beta_a, B_a, \alpha_{aa+1})\,.
\ee
Using \rf{13092017-man02-07}, we find that, on vertex $V^{(1)}$ \rf{13092017-man02-07-a1}, operators $G_a$, $G_\beta$ \rf{13092017-man02-02}-\rf{13092017-man02-05} are realized as
\beq
\label{13092017-man02-08} && \hspace{-1cm} G_1 =   (B_3 - \frac{\beta_1}{\beta_3}  g_3(N_3+1) )\partial_{\alpha_{31}}  -   B_2  \partial_{\alpha_{12}}\,, \qquad
\\
\label{13092017-man02-09} &&  \hspace{-1cm} G_2  =     B_1  \partial_{\alpha_{12}}
- ( B_3 + \frac{\beta_2}{\beta_3} g_3(N_3+1) ) \partial_{\alpha_{23}} \,,
\\
\label{13092017-man02-10} && \hspace{-1cm} G_3  =     B_2 \partial_{\alpha_{2 3}}
-    B_1  \partial_{\alpha_{31}}   +    g_3
\nonumber\\
& + & \frac{ g_3^\smone (N_3+2) }{2N_3+d-2} \Bigl( \frac{2\beta_1}{\beta_3} B_1 \partial_{B_3} \partial_{\alpha_{31}} + \frac{2 \beta_2}{\beta_3} B_2 \partial_{B_3} \partial_{\alpha_{23}}
  +  2 \alpha_{12} \partial_{\alpha_{31}}\partial_{\alpha_{23}}  \Bigr)\,,
\\
\label{13092017-man02-11} && \hspace{-1cm} G_\beta  =     -  \frac{1}{\beta}  \No_\beta    - \frac{1}{\beta_3^2} g_3(N_3+1) \partial_{B_3}\,,
\\
\label{13092017-man02-12} && g_3 \equiv g_{\upsilon_3}\bigr|_{N_{\upsilon_3} \rightarrow N_3}\,, \qquad g_3^\smone \equiv g_{\upsilon_3}\bigr|_{N_{\upsilon_3} \rightarrow N_3+1}\,,
\eeq
where $N_{\upsilon_3} =\upsilon_3\partial_{\upsilon_3}$, $N_3$ is defined in \rf{13092017-man02-07}, while $g_\upsilon$ is given by \rf{20082017-man02-17},\rf{20082017-man02-18} for $m^2=0$.

\noindent {\bf Step 3. Realization on $V^{(2)}$}. We find it convenient to introduce a vertex $ V^{(2)}$ by the relations
\be \label{13092017-man02-14}
V^{(1)} = U_3 V^{(2)}\,, \qquad V^{(2)} = V^{(2)}(\beta_a, B_a, \alpha_{aa+1})\,, \qquad U_3  =  \Bigl( \frac{2^{N_3} \Gamma(N_3 + \frac{d-2}{2})}{\Gamma( N_3 + 1)} \Bigr)^{1/2}.
\ee
Note that vertex $V^{(2)}$ \rf{13092017-man02-14} depends on the same variables as vertex $V^{(1)}$ \rf{13092017-man02-07-a1}.
On vertex $V^{(2)}$ \rf{13092017-man02-14} , realization of $G_a$, $G_\beta$ \rf{13092017-man02-08}-\rf{13092017-man02-11} takes more convenient form given by
\beq
\label{13092017-man02-16} G_1 &  =  &    (B_3 - \frac{\beta_1}{\beta_3} \kappa_3  )\partial_{\alpha_{31}}  -  B_2 \partial_{\alpha_{12}} \,,
\\
\label{13092017-man02-17} G_2 &  =  &   B_1 \partial_{\alpha_{12}} - ( B_3 + \frac{\beta_2}{\beta_3} \kappa_3 ) \partial_{\alpha_{23}} \,,
\\
\label{13092017-man02-18} G_3 &  =  &    B_2 \partial_{\alpha_{2 3}}
-   B_1  \partial_{\alpha_{31}}  +    \frac{g_3^2 (N_3+1)}{\kappa_3}
\nonumber\\
& + & \frac{ \kappa_3 }{2N_3+d-2} \Bigl( \frac{2\beta_1}{\beta_3} B_1 \partial_{B_3} \partial_{\alpha_{31}} + \frac{2 \beta_2}{\beta_3} B_2 \partial_{B_3} \partial_{\alpha_{23}}
 + 2 \alpha_{12} \partial_{\alpha_{31}}\partial_{\alpha_{23}}  \Bigr)\,,
\\
\label{13092017-man02-19} G_\beta &  =  &    -  \frac{1}{\beta}  \No_\beta    - \frac{\kappa_3}{\beta_3^2} \partial_{B_3}\,.
\eeq
Namely, as compared with operators \rf{13092017-man02-08}-\rf{13092017-man02-11}, operators \rf{13092017-man02-16}-\rf{13092017-man02-19} do not involve square roots of $N_3$ \rf{13092017-man02-07}. Note, it is the quantities $g_3$, $g_3^\smone$ \rf{13092017-man02-12} that enter square roots of the $N_3$.

\noindent {\bf Step 4. Realization on $V^{(3)}$}. At this step, we remove the dependence of operators $G_a$ \rf{13092017-man02-16}-\rf{13092017-man02-18} on the momenta $\beta_1$, $\beta_2$, $\beta_3$. To this end we use the transformation
\be \label{13092017-man02-20}
V^{(2)}  =  U_\beta  V^{(3)}\,, \qquad V^{(3)} = V^{(3)}(\beta_a, B_a, \alpha_{aa+1})\,, \qquad U_\beta   = \exp\bigl(  - \frac{\betach_3}{2\beta_3} \kappa_3 \partial_{B_3} \bigr).
\ee
On vertex $V^{(3)}$ \rf{13092017-man02-20}, operators $G_a$, $G_\beta$ \rf{13092017-man02-16}-\rf{13092017-man02-19} are realized as
\beq
\label{13092017-man02-21}  \hspace{-1.2cm} && G_1  =     (B_3 + \half \kappa_3  )\partial_{\alpha_{31}}  - B_2 \partial_{\alpha_{12}}\,,
\\
\label{13092017-man02-22}  \hspace{-1.2cm} && G_2  =    B_1 \partial_{\alpha_{12}} -  ( B_3 - \half \kappa_3 ) \partial_{\alpha_{23}}\,,
\\
\label{13092017-man02-23} \hspace{-1.2cm} && G_3  =     B_2\partial_{\alpha_{23}} -   B_1\partial_{\alpha_{31}} + \frac{\kappa_3 }{2N_3+d-2} \Bigl( 1 - B_1 \partial_{B_3} \partial_{\alpha_{31}} - B_2 \partial_{B_3} \partial_{\alpha_{23}}   +  2\alpha_{12} \partial_{\alpha_{31}}\partial_{\alpha_{23}}  \Bigr),\qquad
\\
\label{13092017-man02-24}  \hspace{-1.2cm} && G_\beta  =   -  \frac{1}{\beta}  \No_\beta \,.
\eeq

\noindent {\bf Step 5. Realization on $V^{(4)}$}. Solution to equations for vertex $V^{(3)}$ \rf{13092017-man02-20}
\be \label{13092017-man02-25}
G_1 V^{(3)} = 0 \,, \qquad G_2 V^{(3)}  = 0\,,
\ee
with $G_1$, $G_2$ as in \rf{13092017-man02-21},\rf{13092017-man02-22} is given by
\beq
\label{13092017-man02-26}  && V^{(3)} = V^{(4)} (B_a,Z)\,,
\\
\label{13092017-man02-27}  && Z = B_3 (B_1\alpha_{23} + B_2\alpha_{31} + B_3 \alpha_{12}) + \half \kappa_3 (B_1 \alpha_{23} - B_2 \alpha_{31}) - \frac{1}{4}\kappa_3^2 \alpha_{12}\,.
\eeq
Multiplying \rf{13092017-man02-23} by the factor $(2N_3 + d-2)/\kappa_3$,  we verify that equation $G_3 V^{(3)} = 0$ with $G_3$ and $V^{(3)}$ as in \rf{13092017-man02-23},\rf{13092017-man02-26} amounts to the following equation for $V^{(4)}$ \rf{13092017-man02-26}:
\be \label{13092017-man02-28}
\bigl( B_1 B_2 (2N_Z + d-4) \partial_Z + 1\bigr) V^{(4)} = 0 \,.
\ee
Solutions to Eq.\rf{13092017-man02-28} are given by
\beq
\label{13092017-man02-29}  && V^{(4)} = W^{ \frac{6-d}{2} } I_{ \frac{d-6}{2} } (\sqrt{ 4 W })V^{(5)}\,,  \qquad V^{(4)} = W^{ \frac{6-d}{2} } K_{ \frac{d-6}{2} } (\sqrt{ 4 W })V^{(5)}\,,
\\
\label{13092017-man02-30} &&  \hspace{1.3cm} V^{(5)} =  V^{(5)}(B_a)\,, \hspace{1.8cm} W = - \frac{Z}{2B_1B_2}\,.
\eeq
Taking into account expression for $Z$ \rf{13092017-man02-27}, we see that a power series expansion of vertices $V^{(4)}$ \rf{13092017-man02-29} in the $B_1$, $B_2$ involves negative powers of the $B_1$, $B_2$. Therefore vertices \rf{13092017-man02-29} are not consistent. To summarize, in the flat space, there are no cubic vertices describing consistent interaction of two arbitrary spin massless fields with  one continuous-spin massless field.

\appendix{ \large Analysis of equations to vertices for two continuous-spin massless fields and one  massless field}

In this Appendix, we prove that two continuous-spin massless fields and one arbitrary spin massless field have no consistent cubic interaction vertices. We divide our proof into five steps.

\noindent {\bf Step 1. Realization on $p_\smp3^-$}. Using the shortcut $(0,\kappa)_\smCSF$ for  continuous-spin massless field and the shortcut $(0,s)$ for spin-$s$ massless field we consider a cubic vertex for the following fields
\be  \label{14092017-man02-43}
(0,\kappa_1)_\smCSF\hbox{-}(0,\kappa_2)_\smCSF\hbox{-}(0,s_3) \hspace{0.8cm} \hbox{\small two  continuous-spin massless fields and one massless field}.\quad
\ee
Our notation in \rf{14092017-man02-43} implies that continuous-spin massless fields carry external line indices $a=1,2$, while the spin-$s_3$ massless field corresponds to $a=3$.
Using notation in \rf{23082017-man02-10}, we note that a general form of parity invariant vertex that respect $J^{ij}$-symmetries is given by
\be  \label{14092017-man02-44}
p_\smp3^- = p_\smp3^-(\beta_a, B_a\,, \alpha_{aa+1}\,,\upsilon_1\,, \upsilon_2)\,.
\ee
To find realization of $G_{a,\Po^2}$, $G_\beta$ \rf{23082017-man02-11} on $p_\smp3^-$ \rf{14092017-man02-44} we use $\Jbf^{-i\dagger}$ \rf{21082017-man02-25}, where operators $M_a^i$, $a=1,2$ for continuous-spin massless fields can be read from \rf{20082017-man02-14}-\rf{20082017-man02-18} for $m^2=0$, while operator $M_a^i$, $a=3$, for arbitrary spin massless field is equal to zero \rf{20082017-man02-21}. Doing so, we find that action of $\Jbf^{-i\dagger}$ \rf{21082017-man02-25} on $p_\smp3^-$ \rf{14092017-man02-44} can be cast into the form \rf{23082017-man02-11} with the following $G_{a,\Po^2}$, $G_\beta$:
\beq
\label{14092017-man02-01} &&   \hspace{-1.5cm} G_{a,\Po^2} =   G_a +  \Pbf^- \frac{2 \beta}{\beta_a^3} \alpha_a^i \frac{ g_{\upsilon_a} \partial_{\upsilon_a} }{2N_a + d-2 } \partial_{B_a}^2\,,\qquad a=1,2;
\qquad G_{3,\Po^2} =   G_3\,,
\\
\label{14092017-man02-02} \hspace{-0.7cm} G_1 &  =  &  B_3   \partial_{\alpha_{31}}  - (B_2  \partial_{\alpha_{12}} + \frac{\beta_1}{\beta_2} g_{\upsilon_2} \partial_{\upsilon_2}) \partial_{\alpha_{12}}    + \upsilon_1 g_{\upsilon_1}
\nonumber\\
\hspace{-0.7cm} & + &     \frac{ g_{\upsilon_1} \partial_{\upsilon_1}}{2N_1+d-2} \Bigl(\frac{ 2 \beta_2}{\beta_1} B_2 \partial_{B_1} \partial_{\alpha_{12}} + \frac{ 2 \beta_3}{\beta_1} B_3 \partial_{B_1} \partial_{\alpha_{31}}  +   2\alpha_{23} \partial_{\alpha_{12}}\partial_{\alpha_{31}} \Bigr) \,,
\\
\label{14092017-man02-03} \hspace{-0.7cm} G_2 &  =  &   ( B_1 - \frac{\beta_2}{\beta_1} g_{\upsilon_1} \partial_{\upsilon_1} ) \partial_{\alpha_{12}} -    B_3  \partial_{\alpha_{23}}   + \upsilon_2 g_{\upsilon_2}
\nonumber\\
\hspace{-0.7cm} & + &       \frac{g_{\upsilon_2} \partial_{\upsilon_2}}{2N_2+d-2} \Bigl(  \frac{ 2 \beta_3}{\beta_2} B_3 \partial_{B_2} \partial_{\alpha_{23}}  + \frac{ 2 \beta_1}{\beta_2} B_1 \partial_{B_2} \partial_{\alpha_{12}}   +  2 \alpha_{31} \partial_{\alpha_{12}}\partial_{\alpha_{23}} \Bigr)  \,,
\\
\label{14092017-man02-04} \hspace{-0.7cm} G_3 &  =  & ( B_2 - \frac{\beta_3}{\beta_2} g_{\upsilon_2} \partial_{\upsilon_2} ) \partial_{\alpha_{23}} -  ( B_1 + \frac{\beta_3}{\beta_1} g_{\upsilon_1} \partial_{\upsilon_1} ) \partial_{\alpha_{31}}  \,,
\\
\label{14092017-man02-05}  \hspace{-0.7cm} G_\beta &  =  &   - \frac{1}{\beta}\No - \frac{1}{\beta_1^2} g_{\upsilon_1} \partial_{\upsilon_1} \partial_{B_1} - \frac{1}{\beta_2^2} g_{\upsilon_2} \partial_{\upsilon_2} \partial_{B_2}\,,
\eeq
where $g_\upsilon$ is given in \rf{20082017-man02-17},\rf{20082017-man02-18} for $m=0$, while $\betach_a$, $\No_\beta$ are given in  \rf{21082017-man02-29}. The $B_a$,  $\alpha_{ab}$, $N_a$ are defined in \rf{23092017-man02-04}-\rf{23092017-man02-06}.
Using operators $G_{a,\Po^2}$, $G_\beta$ \rf{14092017-man02-01}-\rf{14092017-man02-05} in \rf{21082017-man02-34}, \rf{23082017-man02-11}, we see that requiring the density $|j_\smp3^{-i}\rangle$ to respect equations \rf{21082017-man02-35-a3} amounts to equations
\rf{23082017-man02-14},\rf{23082017-man02-15}.

\noindent {\bf Step 2. Realization on $V^{(1)}$}. We find dependence of the vertex $p_\smp3^-$ on the oscillators $\upsilon_1$ and $\upsilon_2$. To this end, we note that in view of constraint \rf{19082017-man02-09}, vertex $p_\smp3^-$ \rf{30082017-man02-02} should satisfy the constraints
\be \label{14092017-man02-06}
(N_{\alpha_a} - N_{\upsilon_a}) | p_\smp3^- \rangle = 0 \,, \qquad a=1,2.
\ee
Introducing a vertex $V^{(1)}$ by the relations
\be \label{14092017-man02-07}
p_\smp3^- = U_{\upsilon_1,\upsilon_2}  V^{(1)}\,, \quad U_{\upsilon_1,\upsilon_2} =  \upsilon_1^{N_1}\upsilon_2^{N_2}\,, \quad N_1 = N_{B_1} + N_{\alpha_{12}} + N_{\alpha_{31}}\,,  \quad N_2 = N_{B_2} + N_{\alpha_{12}} + N_{\alpha_{23}}\,,
\ee
and using Eqs.\rf{14092017-man02-06}, we find that the vertex $V^{(1)}$ does not depend on the $\upsilon_1$, $\upsilon_2$, i.e., we get
\be \label{14092017-man02-07-a1}
V^{(1)} = V^{(1)}(\beta_a, B_a, \alpha_{aa+1} )\,.
\ee
Using \rf{14092017-man02-07}, we find that, on vertex $V^{(1)}$ \rf{14092017-man02-07-a1}, operators $G_a$, $G_\beta$ \rf{14092017-man02-02}-\rf{14092017-man02-05} are realized as
\beq
\label{14092017-man02-08} \hspace{-0.7cm} G_1 &  =  &  B_3 \partial_{\alpha_{31}}  - (B_2   + \frac{\beta_1}{\beta_2} g_2 (N_2+1)) \partial_{\alpha_{12}}   + g_1
\nonumber\\
\hspace{-0.7cm} & + &     \frac{ g_1^\smone (N_1+2)}{2N_1+d-2} \Bigl( \frac{ 2 \beta_2}{\beta_1} B_2 \partial_{B_1} \partial_{\alpha_{12}} + \frac{ 2 \beta_3}{\beta_1} B_3 \partial_{B_1} \partial_{\alpha_{31}}  +   2\alpha_{23} \partial_{\alpha_{12}}\partial_{\alpha_{31}}
  \Bigr) \,,
\\
\label{14092017-man02-09} \hspace{-0.7cm} G_2 &  =  &   ( B_1 - \frac{\beta_2}{\beta_1} g_1 (N_1+1) ) \partial_{\alpha_{12}} -    B_3  \partial_{\alpha_{23}} +  g_2
\nonumber\\
\hspace{-0.7cm} & + &       \frac{g_2^\smone (N_2+2)}{2N_2 + d-2} \Bigl(  \frac{ 2 \beta_3}{\beta_2} B_3 \partial_{B_2} \partial_{\alpha_{23}}  + \frac{ 2 \beta_1}{\beta_2} B_1 \partial_{B_2} \partial_{\alpha_{12}}   +  2 \alpha_{31} \partial_{\alpha_{12}}\partial_{\alpha_{23}}
\Bigr)  \,,
\\
\label{14092017-man02-10} \hspace{-0.9cm} G_3 &  =  & ( B_2 - \frac{\beta_3}{\beta_2} g_2 (N_2+1) ) \partial_{\alpha_{23}} -  ( B_1 + \frac{\beta_3}{\beta_1} g_1 (N_1+1) ) \partial_{\alpha_{31}} \,,\qquad
\\
\label{14092017-man02-11} \hspace{-0.7cm} G _\beta &  =  &   - \frac{1}{\beta}\No - \frac{1}{\beta_1^2} g_1 (N_1+1) \partial_{B_1} - \frac{1}{\beta_2^2} g_2 (N_2+1) \partial_{B_2}  \,,
\\
\label{14092017-man02-12} && \hspace{-1.6cm}  g_1 \equiv g_{\upsilon_1}\bigr|_{N_{\upsilon_1} \rightarrow N_1}\,, \quad
g_2 \equiv g_{\upsilon_2}\bigr|_{N_{\upsilon_2} \rightarrow N_2}\,,
\quad g_1^\smone \equiv g_{\upsilon_1}\bigr|_{N_{\upsilon_1} \rightarrow N_1+1}\,, \quad g_2^\smone \equiv g_{\upsilon_2}\bigr|_{N_{\upsilon_2} \rightarrow N_2+1}\,.
\eeq

\noindent {\bf Step 3. Realization on $V^{(2)}$}. We find it convenient to introduce a vertex $ V^{(2)}$ by the relations
\beq
\label{14092017-man02-14} && V^{(1)}= U_{1,2} V^{(2)}\,, \qquad U_{1,2}  =  \Bigl( \frac{2^{N_1} \Gamma(N_1 + \frac{d-2}{2})}{\Gamma( N_1 + 1)} \frac{2^{N_2} \Gamma(N_2 + \frac{d-2}{2})}{\Gamma( N_2 + 1)} \Bigr)^{1/2}\,,
\\
\label{14092017-man02-15} && V^{(2)} = V^{(2)}(\beta_a, B_a, \alpha_{aa+1})\,,
\eeq
where, in \rf{14092017-man02-15}, we show that the vertex $V^{(2)}$ depends on the same variables as vertex $V^{(1)}$ \rf{14092017-man02-07-a1}. On the vertex $V^{(2)}$, the realization of $G_a$, $G_\beta$ \rf{14092017-man02-08}-\rf{14092017-man02-11} takes more convenient form given by
\beq
\label{14092017-man02-16} \hspace{-0.7cm} G_1 &  =  & B_3  \partial_{\alpha_{31}}  - (B_2   + \frac{\beta_1}{\beta_2} \kappa_2) \partial_{\alpha_{12}}  + \frac{1}{\kappa_1} g_1^2(N_1+1)
\nonumber\\
\hspace{-0.7cm} & + &     \frac{ \kappa_1}{2N_1+d-2} \Bigl( \frac{ 2 \beta_2}{\beta_1} B_2 \partial_{B_1} \partial_{\alpha_{12}} + \frac{ 2 \beta_3}{\beta_1} B_3 \partial_{B_1} \partial_{\alpha_{31}}  +   2\alpha_{23} \partial_{\alpha_{12}}\partial_{\alpha_{31}}
\Bigr) \,,
\\
\label{14092017-man02-17} \hspace{-0.7cm} G_2 &  =  &   ( B_1 - \frac{\beta_2}{\beta_1} \kappa_1 ) \partial_{\alpha_{12}} -   B_3  \partial_{\alpha_{23}}  +  \frac{1}{\kappa_2} g_2^2 (N_2+1)
\nonumber\\
\hspace{-0.7cm} & + &       \frac{\kappa_2}{2N_2+d-2} \Bigl(  \frac{ 2 \beta_3}{\beta_2} B_3 \partial_{B_2} \partial_{\alpha_{23}}  + \frac{ 2 \beta_1}{\beta_2} B_1 \partial_{B_2} \partial_{\alpha_{12}}   +  2 \alpha_{31} \partial_{\alpha_{12}}\partial_{\alpha_{23}} \Bigr)  \,,
\\
\label{14092017-man02-18} \hspace{-0.7cm} G_3 &  =  & ( B_2 - \frac{\beta_3}{\beta_2} \kappa_2) \partial_{\alpha_{23}} -  ( B_1 + \frac{\beta_3}{\beta_1} \kappa_1  ) \partial_{\alpha_{31}} \,,
\\
\label{14092017-man02-19} \hspace{-0.7cm} G _\beta &  =  &   - \frac{1}{\beta}\No - \frac{1}{\beta_1^2} \kappa_1 \partial_{B_1} - \frac{1}{\beta_2^2} \kappa_2 \partial_{B_2} \,.
\eeq
Namely, as compared with operators \rf{14092017-man02-08}-\rf{14092017-man02-11}, operators \rf{14092017-man02-16}-\rf{14092017-man02-19} do not involve square roots of the operators $N_1$, $N_2$ \rf{14092017-man02-07}. Note, it is the quantities $g_a$, $g_a^\smone$, $a=1,2$ \rf{14092017-man02-12} that enter square roots of the $N_1$, $N_2$.

\noindent {\bf Step 4. Realization on $V_{Z\beta}^{(2)}$}. We study dependence of $V^{(2)}$ \rf{14092017-man02-15} on the $\alpha_{23}$ and $\alpha_{31}$. Namely, we note that solution to equation $G_3 V^{(2)}=0$ with $V^{(2)}$, $G_3$ as in \rf{14092017-man02-15}, \rf{14092017-man02-18}, is given by
\be \label{14092017-man02-20}
V^{(2)} = V_{Z\beta}^{(2)}(\beta_a,B_a,\alpha_{12},Z)\,,\qquad Z  =  (B_1 + \frac{\beta_3}{\beta_1} \kappa_1) \alpha_{23} + (B_2 - \frac{\beta_3}{\beta_2} \kappa_2) \alpha_{31}\,.
\ee
Using \rf{14092017-man02-20}, we find that, on vertex $V_{Z\beta}^{(2)}$ \rf{14092017-man02-20}, operators $G_1$, $G_2$ \rf{14092017-man02-16}, \rf{14092017-man02-17} are realized as
\beq
\label{14092017-man02-21} \hspace{-0.7cm} G_1 &  = & B_3 (B_2 -\frac{\beta_3}{\beta_2}\kappa_2) \partial_Z  - (B_2   + \frac{\beta_1}{\beta_2} \kappa_2) \partial_{\alpha_{12}}
\nonumber\\
\hspace{-0.7cm} & + &     \frac{ \kappa_1}{2N_{1Z} + d-2} \Bigl(1+ \frac{ 2 \beta_2}{\beta_1} B_2 \partial_{B_1} \partial_{\alpha_{12}} + \frac{ 2 \beta_3}{\beta_1} B_3 (B_2 -\frac{\beta_3}{\beta_2}\kappa_2) \partial_{B_1} \partial_Z    \Bigr) \,,
\\
\label{14092017-man02-22} \hspace{-0.7cm} G_2 & = &   ( B_1 - \frac{\beta_2}{\beta_1} \kappa_1 ) \partial_{\alpha_{12}} -   B_3 (B_1 + \frac{\beta_3}{\beta_1} \kappa_1) \partial_Z
\nonumber\\
\hspace{-0.7cm} & + &       \frac{\kappa_2}{2N_{2Z} + d-2} \Bigl( 1 +  \frac{ 2 \beta_3}{\beta_2} B_3 (B_1 + \frac{\beta_3}{\beta_1} \kappa_1) \partial_{B_2} \partial_Z  + \frac{ 2 \beta_1}{\beta_2} B_1 \partial_{B_2} \partial_{\alpha_{12}}  \Bigr)  \,,
\\
\label{14092017-man02-23} && N_{1Z}  =  N_{B_1} + N_{\alpha_{12}} + N_Z\,, \qquad N_{2Z}  =  N_{B_2} + N_{\alpha_{12}} + N_Z\,.
\eeq

\noindent {\bf Step 5. Analysis of equations for $V_{Z\beta}^{(2)}$}. We now analyse equations for vertex $V_{Z\beta}^{(2)}$ \rf{14092017-man02-20},
\be \label{14092017-man02-23-a1}
G_1 V_{Z\beta}^{(2)} = 0 \,, \qquad G_2 V_{Z\beta}^{(2)} = 0 \,, \qquad V_{Z\beta}^{(2)} = V_{Z\beta}^{(2)}(\beta_a,B_a,\alpha_{12},Z)\,,
\ee
where $G_1$, $G_2$ are given in \rf{14092017-man02-21}-\rf{14092017-man02-23}. As the vertex  $V_{Z\beta}^{(2)}$ is considered to be expandable in the $B_3$ we use the following power series expansion:
\be
\label{14092017-man02-24} V_{Z\beta}^{(2)} = \sum_{n=0}^\infty V_n\,, \qquad N_{B_3} V_n = n V_n\,, \qquad  V_n = V_n(\beta_a,B_a,\alpha_{12},Z)\,,
\ee
where $V_n$ is a degree-$n$ monomial in $B_3$. Equations \rf{14092017-man02-23-a1} then imply
\beq
\label{14092017-man02-26} && \hspace{-1.6cm} G_1 V_0  = 0 \,, \qquad G_2 V_0 = 0 \,,
\\
\label{14092017-man02-27}  G_1 & = &     - ( B_2 + \frac{\beta_1}{\beta_2} \kappa_2) \partial_{\alpha_{12}}
+ \frac{\kappa_1}{2N_{1Z}+d-2}  \bigl(1 + \frac{2\beta_2}{\beta_1} B_2 \partial_{B_1} \partial_{\alpha_{12}} \bigr),\qquad
\\
\label{14092017-man02-28} G_2 & = &  (B_1  - \frac{\beta_2}{\beta_1} \kappa_1) \partial_{\alpha_{12}}  +        \frac{\kappa_2}{2N_{2Z}+d-2} \bigl( 1 + \frac{2\beta_1}{\beta_2} B_1 \partial_{B_2} \partial_{\alpha_{12}} \bigr),
\eeq
where $N_{1Z}$, $N_{2Z}$ are given in \rf{14092017-man02-23}. As $V_0$ is considered to be expandable in the $B_1$, $B_2$, we use the following power series expansion in the $B_1$, $B_2$:
\be
\label{14092017-man02-29} V_0 = \sum_{n=0}^\infty \Vb_n\,,  \qquad (N_{B_1} + N_{B_2})\Vb_n = n \Vb_n\,,\qquad \Vb_n = \Vb_n(\beta_a,B_1,B_2,\alpha_{12},Z)\,.
\ee
From \rf{14092017-man02-26}-\rf{14092017-man02-29}, we get the following equations for the vertex $\Vb_0$:
\beq
\label{14092017-man02-30} && \Bigl( - (2N_{1Z} + d-2) \partial_{\alpha_{12}}     + \frac{\kappa_1 \beta_2}{ \kappa_2 \beta_1}\Bigr)\Vb_0 = 0\,,
\\
\label{14092017-man02-31} && \Bigl( - (2 N_{2Z} +d-2) \partial_{\alpha_{12}}     + \frac{\kappa_2 \beta_1}{ \kappa_1 \beta_2}\Bigr) \Vb_0 = 0\,,
\eeq
where $N_{1Z}$, $N_{2Z}$ are given in \rf{14092017-man02-23}. Subtracting \rf{14092017-man02-31} from \rf{14092017-man02-30}, we get the equation
\be  \label{14092017-man02-32}
\Bigl(\frac{\kappa_1 \beta_2}{ \kappa_2 \beta_1} -    \frac{\kappa_2 \beta_1}{ \kappa_1 \beta_2}\Bigr) \Vb_0 = 0\,,
\ee
which implies
\be \label{14092017-man02-33}
\Vb_0 = 0 \,.
\ee
Using \rf{14092017-man02-33}  in \rf{14092017-man02-26}-\rf{14092017-man02-29}, we get the following equations for the vertex $\Vb_1$:
\beq
\label{14092017-man02-34}  && \Bigl( - (2N_{1Z}+d-2) \partial_{\alpha_{12}}  +   \frac{2\kappa_1\beta_2^2}{\kappa_2 \beta_1^2} B_2 \partial_{B_1} \partial_{\alpha_{12}} + \frac{\kappa_1 \beta_2}{ \kappa_2 \beta_1} \Bigr)\Vb_1 = 0\,,
\\
\label{14092017-man02-35}  && \Bigl( - (2N_{2Z}+d-2) \partial_{\alpha_{12}}  +   \frac{2\kappa_2\beta_1^2}{\kappa_1 \beta_2^2} B_1 \partial_{B_2} \partial_{\alpha_{12}}  + \frac{\kappa_2 \beta_1}{ \kappa_1 \beta_2} \Bigr)\Vb_1 = 0\,,
\eeq
where $N_{1Z}$, $N_{2Z}$ are given in \rf{14092017-man02-23}. Subtracting \rf{14092017-man02-35}  from \rf{14092017-man02-34}, we get the equation
\be
\label{14092017-man02-36}   \Bigl( (-2N_{B_1} + 2 N_{B_2}) \partial_{\alpha_{12}}  +   \frac{2\kappa_1\beta_2^2}{\kappa_2 \beta_1^2} B_2 \partial_{B_1} \partial_{\alpha_{12}} - \frac{2\kappa_2\beta_1^2}{\kappa_1 \beta_2^2} B_1 \partial_{B_2} \partial_{\alpha_{12}}
+ \frac{\kappa_1 \beta_2}{ \kappa_2 \beta_1} -    \frac{\kappa_2 \beta_1}{ \kappa_1 \beta_2} \Bigr)\Vb_1 = 0\,.
\ee
Now, in place of the $B_1$, $B_2$, we introduce new variables $B_\pm$,
\be \label{14092017-man02-37}
B_\pm  =  \frac{\beta_1^2}{\kappa_1} B_1 \pm  \frac{\beta_2^2}{\kappa_2} B_2\,,
\ee
and note that, in terms of the $B_\pm$, equation \rf{14092017-man02-36} takes the form
\be \label{14092017-man02-38}
\Bigl( - 4 B_- \partial_{B_+}\partial_{\alpha_{12}}  +  \frac{\kappa_1 \beta_2}{ \kappa_2 \beta_1} -    \frac{\kappa_2 \beta_1}{ \kappa_1 \beta_2} \Bigr) \Vb_1  = 0\,.
\ee
As the vertex $\Vb_1$ is a degree-1 homogeneous polynomial in the $B_+$, $B_-$ \rf{14092017-man02-29}, we use the expansion
\be  \label{14092017-man02-38-a1}
\Vb_1 = B_+ \Vb_{1+} + B_- \Vb_{1-}\,,
\ee
and verify that Eq.\rf{14092017-man02-38} amounts to the equations
\be
\label{14092017-man02-39}   - 4 \partial_{\alpha_{12}}\Vb_{1+}   +  \Bigl( \frac{\kappa_1 \beta_2}{ \kappa_2 \beta_1} -    \frac{\kappa_2 \beta_1}{ \kappa_1 \beta_2} \Bigr) \Vb_{1-} = 0\,,\qquad \Bigl( \frac{\kappa_1 \beta_2}{ \kappa_2 \beta_1} -    \frac{\kappa_2 \beta_1}{ \kappa_1 \beta_2} \Bigr) \Vb_{1+} = 0\,.
\ee
The 2nd equation in \rf{14092017-man02-39} implies $\Vb_{1+} = 0$. Setting then $\Vb_{1+}=0$ in the 1st equation in \rf{14092017-man02-39}, we find $\Vb_{1-}=0$, i.e., vertex $\Vb_1$ \rf{14092017-man02-38-a1} is trivial,
\be \label{14092017-man02-40}
\Vb_1 = 0\,.
\ee
Thus we have demonstrated that Eqs.\rf{14092017-man02-26} and relations in \rf{14092017-man02-27}-\rf{14092017-man02-29} lead to $\Vb_0=0$ and $\Vb_1=0$. We now use the induction method. Suppose Eqs.\rf{14092017-man02-26}-\rf{14092017-man02-29} lead to trivial $\Vb_k=0$ for $k=0,1,\ldots, n-1$. Then we are going to prove that $\Vb_n=0$. To this end we note that if $\Vb_k=0$, for $k=0,1,\ldots n-1$ then Eqs.\rf{14092017-man02-26} and relations in \rf{14092017-man02-27}-\rf{14092017-man02-29} lead
to the equation for $\Vb_n$ which is obtained by replacement $\Vb_1\rightarrow \Vb_n$ in \rf{14092017-man02-34}, \rf{14092017-man02-35}. Therefore repeating analysis we used above for the vertex $\Vb_1$, we obtain the following equation for the vertex $\Vb_n$:
\be  \label{14092017-man02-41}
\Bigl( - 4 B_- \partial_{B_+}\partial_{\alpha_{12}}  +  \frac{\kappa_1 \beta_2}{ \kappa_2 \beta_1} -    \frac{\kappa_2 \beta_1}{ \kappa_1 \beta_2} \Bigr) \Vb_n = 0\,.
\ee
By definition \rf{14092017-man02-29}, the vertex $\Vb_n$ is a degree-$n$ homogeneous polynomial in variables $B_\pm$ \rf{14092017-man02-37}. Therefore, we can present the vertex $\Vb_n$ as
\be  \label{14092017-man02-42}
\Vb_n = \sum_{k=0}^n B_+^k B_-^{n-k} \Vb_{n,k}\,,
\ee
where $\Vb_{n,k}$ are independent of the $B_\pm$. Plugging \rf{14092017-man02-42} into \rf{14092017-man02-41}, we verify that $\Vb_{n,k}=0$ for all $k=0,1,\ldots n$, i.e., $\Vb_n=0$.
Using the induction method, we conclude then that $\Vb_n=0$ for all $n=0,1,\ldots \infty$. In other words, $V_0 =0$ \rf{14092017-man02-29}. Using the relation $V_0=0$ and applying the induction method to the vertices $V_n$ in \rf{14092017-man02-24} we find therefore that $V_n=0$ for all $n=0,1,\ldots,\infty$. Thus we conclude that $V_{Z\beta}^{(2)} =0 $ \rf{14092017-man02-24}. This implies that vertex \rf{14092017-man02-44} is trivial, $p_\smp3^- =0$.


\end{document}